\documentclass[runningheads]{llncs}
\usepackage{color}

\usepackage{amsmath,amssymb,amscd,stmaryrd
}

\usepackage{comment} 
\usepackage{color}

\usepackage{multirow}

\usepackage{times}

\usepackage{wrapfig}
\usepackage{listings}
\usepackage{tikz}
\usetikzlibrary{automata,positioning,arrows,shapes,snakes,backgrounds,fit,calc}

\tikzstyle{dem}=[shape=diamond,draw,inner sep=0pt,minimum size=6mm]
\tikzstyle{ran}=[shape=circle,draw,inner sep=0pt,minimum size=5mm]
\tikzstyle{det}=[shape=rectangle,draw,inner sep=0pt,minimum size=5mm]
\tikzstyle{up}=[shape=regular polygon,draw,inner sep=0pt,minimum size=6mm]
\tikzstyle{tran}=[draw,->,>=stealth, rounded corners]

\newif\ifignore 
\ignorefalse
\newcommand{\auxproof}[1]{
\ifignore\mbox{}\newline
\color{gray}
\textbf{BEGIN: AUX-PROOF} \dotfill\newline
{#1}\mbox{}\newline
\textbf{END: AUX-PROOF}\dotfill\newline
\color{black}
\fi}

\newcommand{\linit}{l_{\mathrm{init}}}
\newcommand{\xinit}{\vec{x}_{\mathrm{init}}}
\newcommand{\trrel}{\mapsto}
\newcommand{\myPr}{\mathrm{Pr}}

\newcommand{\preach}{\mathbb{P}^{\mathrm{reach}}}
\newcommand{\preachN}[1][N]{\mathbb{P}^{\mathrm{reach}\le #1}}
\newcommand{\preachEN}[1][N]{\mathbb P^{\mathrm{reach}=#1}}
\newcommand{\preachNN}[1][N]{\mathbb P^{\lnot\mathrm{reach}\le #1}}
\newcommand{\upreach}{\overline{\mathbb{P}}^{\mathrm{reach}}}
\newcommand{\upreachN}[1][N]{\overline{\mathbb{P}}^{\mathrm{reach}\le #1}}
\newcommand{\lpreach}{\underline{\mathbb{P}}^{\mathrm{reach}}}
\newcommand{\lpreachN}[1][N]{\underline{\mathbb{P}}^{\mathrm{reach}\le #1}}

\newcommand{\esteps}{\mathbb{E}^{\mathrm{steps}}}
\newcommand{\estepsN}[1][N]{\mathbb{E}^{\mathrm{steps}\le #1}}
\newcommand{\uesteps}{\overline{\mathbb{E}}^{\mathrm{steps}}}
\newcommand{\uestepsN}[1][N]{\overline{\mathbb{E}}^{\mathrm{steps}\le #1}}
\newcommand{\lesteps}{\underline{\mathbb{E}}^{\mathrm{steps}}}
\newcommand{\lestepsN}[1][N]{\underline{\mathbb{E}}^{\mathrm{steps}\le #1}}

\newcommand{\dist}{\mathcal{D}}
\newcommand{\borel}{\mathcal{B}}
\newcommand{\real}{\mathbb{R}}
\newcommand{\nat}{\mathbb{N}}

\newcommand{\true}{\mathsf{true}}

\newcommand{\Sch}{\mathrm{Sch}}

\newcommand{\FarkasInput}{\mathrm{Fml}^{\Rightarrow}}
\newcommand{\FarkasInputExp}{\mathrm{Fml}^{\Rightarrow e^{\place}}}

\newcommand{\vars}{\mathcal{V}}

\newcommand{\pre}{\mathbb{X}}
\newcommand{\upre}{\overline{\mathbb{X}}}
\newcommand{\lpre}{\underline{\mathbb{X}}}

\newcommand{\config}{L \times \mathbb R^V}

\newcommand{\sem}[1]{\llbracket#1\rrbracket}

\newcommand{\prob}{\mathbf{prob}}
\newcommand{\ndet}{\mathbf{ndet}}
\newcommand{\sample}{\mathbf{sample}}

\newcommand{\Real}{\mathbf{Real}}

\newcommand{\pc}[1]{\langle\mathit{#1}\rangle}

\newcommand{\app}[1]{\mathbf{#1}}

\newcommand{\place}{\underline{\phantom{n}}\,} 

\newcommand{\myUp}{\mathrm{Up}} 
\newcommand{\x}{\vec{x}} 

\newcommand{\mysucc}{\mathop{\mathrm{succ}}}

\makeatletter
  \def\@thmcountersep{.}
\makeatother

\begin{document}
\spnewtheorem{mytheorem}{Theorem}[section]{\bfseries}{\itshape} 
\spnewtheorem{mylemma}[mytheorem]{Lemma}{\bfseries}{\itshape}
\spnewtheorem{myproposition}[mytheorem]{Proposition}{\bfseries}{\itshape}
\spnewtheorem{mysublemma}[mytheorem]{Sublemma}{\bfseries}{\itshape}
\spnewtheorem{mycorollary}[mytheorem]{Corollary}{\bfseries}{\itshape}
\spnewtheorem{myfact}[mytheorem]{Fact}{\bfseries}{\itshape}
\spnewtheorem{mynotation}[mytheorem]{Notation}{\bfseries}{\rmfamily}
\spnewtheorem{myremark}[mytheorem]{Remark}{\bfseries}{\rmfamily}
\spnewtheorem{myexample}[mytheorem]{Example}{\bfseries}{\rmfamily}
\spnewtheorem{myassumption}[mytheorem]{Assumption}{\bfseries}{\rmfamily}
\spnewtheorem{mydefinition}[mytheorem]{Definition}{\bfseries}{\rmfamily}
\spnewtheorem{myrequirements}[mytheorem]{Requirements}{\bfseries}{\rmfamily}
\spnewtheorem{myproblem}[mytheorem]{Problem}{\bfseries}{\rmfamily}

\title{Ranking and Repulsing Supermartingales for  Reachability  in Probabilistic Programs
}
\titlerunning{Ranking and Repulsing Supermartingales for Approximating Reachability}
%
\author{Toru Takisaka\inst{1} \and
Yuichiro Oyabu\inst{1,2} \and
Natsuki Urabe\inst{3} \and
Ichiro Hasuo\inst{1,2}}
\authorrunning{T. Takisaka et al.}
%
\institute{National Institute of Informatics, Japan \and
The Graduate University for Advanced Studies (SOKENDAI), Japan \and
University of Tokyo, Japan
}
\maketitle              
\begin{abstract}
Computing reachability probabilities is a fundamental problem in the
 analysis of probabilistic programs. This paper aims at a comprehensive and comparative account of various \emph{martingale-based methods} for over- and underapproximating reachability probabilities. Based on the existing works that stretch across different communities (formal verification, control theory, etc.), we offer a unifying account. In particular, we emphasize the role of order-theoretic fixed points---a classic topic in computer science---in the analysis of probabilistic programs. This  leads us to two new martingale-based techniques, too. 
 We also make an experimental comparison using our implementation of template-based synthesis algorithms for those martingales.
%
\end{abstract}

\section{Introduction}\label{sec:intro}

\begin{wrapfigure}[12]{r}{4.4cm}
\hspace*{8mm}
\scriptsize
\begin{minipage}{6cm}
\begin{lstlisting}[numbers=left]
x := 2; y := 2; t := 0;
while t <= 100 do
  t := t + 1;
  z := Unif (-2,1);
  if * then 
    x := x + z
  else
   y := y + z
  fi
  
\end{lstlisting}
\end{minipage}
\caption{An example of 
probabilistic programs. The line 4 means that the value of $z$ is randomly sampled from the interval $[-2,1]$.}\label{fig:APP}
\end{wrapfigure}

Computing \emph{reachability probabilities} is a fundamental problem in
the analysis of
probabilistic systems. It is known
that probabilistic model checking titleems can be  solved via
reachability probabilities~\cite{BaierK08}, 
much like 
nondeterministic model checking problems are reduced to emptiness and
hence to reachability~\cite{Vardi95}. While the computation of reachability
probabilities for finite-state systems is effectively solved 
 by linear programming, the problem becomes much more challenging for
 \emph{probabilistic programs}---a paradigm that
 attracts growing attention as a  programming language foundation for
 machine learning~\cite{GordonHNR14probprog}---because  their transition
 graphs are infinite in general.

 Reachability probabilities of probabilistic programs with while loops
 are clearly not computable, because the problem encompasses
 termination of (non-probabilistic) while programs. Therefore the existing research efforts
 have focused on sound \emph{approximation methods} for reachability
 probabilities. An approach that is widely used in the literature is to
 use  \emph{ranking supermartingales}---a probabilistic analogue of
 \emph{ranking functions}---as a witness for the qualitative question of
 almost-sure reachability. Ranking supermartingales are amenable to
 template-based synthesis~\cite{chakarovS13probprog,ChatterjeeFG16,chakarovVS16TACAS}, making them
 appealing from the automatic analysis point of view.
 Recently, methods for quantitatively
 underapproximating reachability probabilities are also proposed
 in~\cite{ChatterjeeNZ17,urabeHH17LICS}.

The dual question of \emph{overapproximating} reachability
 probabilities, which can then be used to qualitatively \emph{refute} almost-sure
 reachability, is also considered. In the control theory, supermartingales are used
 as a probabilistic counterpart of \emph{barrier certificates}~\cite{PrajnaJ04,steinhardtT12IJRR}.
 A similar idea is recently used for the purpose of synthesizing 
 \emph{stochastic} invariants for probabilistic programs~\cite{ChatterjeeNZ17}.
 Here an overapproximation of reachability probability serves as 
quantitative verification for safety: it gives an upper bound for the probability that the system or the program reaches a bad state.

\begin{table}[tbp]
\centering
\caption{Martingale-based techniques for approximation of reachability probabilities. MC stands for Markov chains, and PP stands for probabilistic programs}
\label{table:overview}
\begin{tabular}{c|c|c|c}
&& 
\shortstack{certificate for}
&from\\\hline\hline
\multirow{2}{*}{\shortstack{ranking \\(super- and sub-)\\ martingale for\\ 
 \emph{under}-approximation
}
}   & 
\shortstack{additive\\ supermartingale\\(ARnkSupM, \S{}\ref{sec:ARnkSupM})}       
&
\shortstack{
\\
$\mathbb{E}(\text{steps to $C$})\le {?}$
}  & 
\cite{DBLP:conf/psse/McIverM04,chakarovS13probprog}
 \\\cline{2-4}
& 
\shortstack{$\gamma$-scaled \\ submartingale\\ ($\gamma$-SclSubM, \S{}\ref{sec:gammaSclSubM})} 
&
\shortstack{$\mathbb{P}(\text{reach $C$})\ge {?}$}
  & 
\shortstack{this paper for PP, \\following categorical\\ observations in~\cite{urabeHH17LICS} for MC}
 \\\hline\hline
\multirow{2}{*}{\shortstack{repulsing\\ supermartingale for\\
 \emph{over}-approximation 
}} 
&      
\shortstack{$\varepsilon$-decreasing\\ supermartingale\\($\varepsilon$-RepSupM, \S{}\ref{sec:EpsRepSupM})
}          &
\shortstack{$\mathbb{P}(\text{reach $C$})\le{?}$
\\
$\mathbb{P}(\text{reach $C$})<^{?}1$
}
  &
\shortstack{\cite{ChatterjeeNZ17}, derived from\\ Azuma's martingale\\ concentration inequality}
  \\\cline{2-4}
&\shortstack{nonnegative\\ supermartingale\\ (NNRepSupM, \S{}\ref{sec:NNRepSupM})}                           &    
$\mathbb{P}(\text{reach $C$})\le{?}$
            &  
\shortstack{
this paper, derived from\\ the Knaster--Tarski theorem
\\(\cite{PrajnaJ04,steinhardtT12IJRR}, without nondeterminism,\\
 derived from Markov's\\concentration inequality)}
\end{tabular}
\end{table}

Table~\ref{table:overview} lists four supermartingale-based techniques for over- and underapproximating reachability probabilities. 
The table is not meant to be exhaustive---still, it shows that multiple methods have been introduced and studied, in different communities (formal verification, control theory, etc.) and with different mathematical backgrounds (ranking functions, martingale concentration inequalities, etc.).

%

The current work aims at a comprehensive and comparative account of those martingale-based techniques in Table~\ref{table:overview}. Central to our account is the role of \emph{order-theoretic fixed points},  a classic topic in theoretical computer science. More specifically, we characterize our objectives---namely reachability probability and expected reaching time---as suitable least fixed points. It turns out that a large part of the theory of martingale-based methods can be developed based on this order-theoretic characterization, without using mathematical gadgets unique to probabilistic settings such as martingale concentration inequalities. 
Our contributions are summarized  as follows.

\begin{itemize}
 \item A comprehensive and comparative account of different martingale-based techniques for approximating reachability probabilities. We identify their key mathematical principles to be order-theoretic fixed points and martingale concentration inequalities, and we emphasize the role of the former. 
 \item We introduce two martingale-based techniques that seem to be new, namely $\gamma$-SclSubM and NNRepSupM in Table~\ref{table:overview}. Their purely probabilistic versions have been in the literature: $\gamma$-SclSubM  is from a category-theoretic account in~\cite{urabeHH17LICS}, and NNRepSupM is from control theory~\cite{steinhardtT12IJRR}. We extend them to probabilistic programs that additionally have nondeterminism. Moreover, completeness of ARnkSupM for probabilistic programs with real-valued variables seems to be new.
 \item We formalize those techniques, taking probabilistic programs (with nondeterminism) as the target of analyses. We investigate soundness and completeness of  the  techniques in Table~\ref{table:overview}. While  the order-theoretic fixed-point foundation gives us clear theoretical guidance, additional nondeterminism requires us to carefully establish measure-theoretic arguments. 
 \item We implemented template-based automated synthesis algorithms for $\gamma$-SclSubM, $\varepsilon$-RepSupM and NNRepSupM, following~\cite{chakarovS13probprog,ChatterjeeFG16}. Our experimental results suggest the advantage of $\gamma$-SclSubM in quantitative reasoning, and the comparative advantage of NNRepSupM over $\varepsilon$-RepSupM in the quality of bounds.
\end{itemize}

\vspace{.3em}
The paper is organized as follows. Preliminaries are in~\S{}\ref{prelim}, where we introduce our system models (pCFGs) for operational semantics of probabilistic programs, and review the theory of order-theoretic fixed points (the Knaster--Tarski and Cousot--Cousot theorems). In~\S{}\ref{sec:EpsRepSupM}--\ref{sec:gammaSclSubM} we discuss the four techniques in Table~\ref{table:overview}, offering a unifying account based on order-theoretic fixed points, and providing some new techniques and results. 
In~\S{}\ref{sec:implexp} we give implementations and experiment results of template-based synthesis. 
After discussion of related work in~\S{}\ref{sec:relwork}, we conclude in~\S{}\ref{sec:concfut}. 

\section{Preliminaries}\label{prelim}
We first fix some notations.
We write $\mathbb N$ and $\mathbb R$ for the set of all natural numbers (i.e.\ nonnegative integers) and reals, respectively. 
We use subscripts to denote subsets of $\mathbb N$ and $\mathbb R$; for example, $\real_{\geq 0}$ denotes the set of all nonnegative reals. 
%
We write $X^*, X^+, X^\omega$ for the sets of all finite, nonempty finite, and infinite sequences of elements of $X$, respectively.

 We use the Borel measurable structure $\mathcal{B}(\mathbb{R})$ of the set $\mathbb{R}$ of real numbers. This induces the measurable structures of all the other sets used in this paper:  $\mathbb{R}^{k}$ where $k\in \mathbb{N}$, $X\times\mathbb{R}^{k}$ where $X$ is finite, 
 and so on. The induced measurable structures are defined in a standard manner: for example, 
$X\times\mathbb{R}^{k}$ where $X$ is finite, it is  given by $\mathcal{B}(X\times\mathbb{R}^{k})=\bigl(\mathcal{B}(\mathbb{R}^{k})\bigr)^{X}$.
The set of 
probability distributions on 
$(X,\mathcal B(X))$ is denoted by $\dist(X)$.
The Dirac measure on $x\in X$ is denoted by $\delta_x$.
The \emph{support} $\mathrm{supp}(d)$ of $d \in \dist(X)$ is defined by
$\mathrm{supp}(d) = \bigl\{x \in X \mid \text{for any $A\in\mathcal{B}(X)$, } x\in A \text{ implies }d(A)>0\bigr\}$. 
The set of all Borel measurable function from $X$ to $Y$ is denoted by $\mathcal B (X,Y)$. The functions $\mathbf{0}$ and $\mathbf{1}$ are the real-valued constant function of which coefficient is $0$ and $1$, respectively.

\subsection{Probabilistic Control Flow Graphs (pCFGs)}\label{sec:pCFG}



We take the notion of pCFG from~\cite{AgrawalC018} and use it as our model of probabilistic systems. pCFGs 
can be thought of as a subclass of Markov decision processes (MDPs), but tailored for 
operational semantics of probabilistic programs (\S{}\ref{sec:pp}).
\auxproof{In particular, the partition $L=L_N+L_P+L_D+L_A$ allows one to simplify the formalization of  measurability conditions on the transition relation. 
}

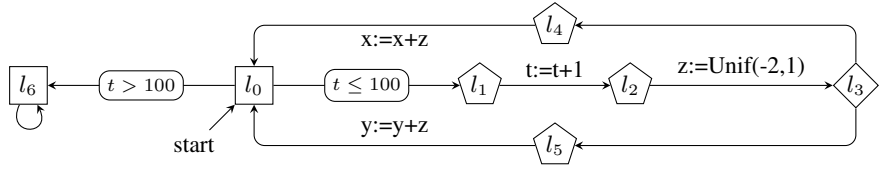
\begin{figure}[t]
\begin{center}
\begin{tikzpicture}[x = 2cm]
\node(init) at (1.1,-0.8){start};
\node[det] (det1) at (1.5,0)  {$l_0$};
\node[up] (up1) at (3,0) {$l_1$};
\node[up] (up2) at (4,0) {$l_2$};
\node[dem] (ndet) at (5.5,0) {$l_3$};
\node [up](up3) at (3.5,0.8) {$l_4$};
\node [up](up4) at (3.5,-0.8) {$l_5$};
\node[det] (det2) at (0,0) {$l_6$};
\draw[tran] (init) -- (det1);
\draw[tran] (det1) to node[font=\scriptsize,draw, fill=white, 
rectangle,pos=0.5] {$t>100$} (det2);
\draw[tran, loop, looseness = 5, in =-65, out = -115] (det2) to (det2);
\draw[tran] (det1) to node[font=\scriptsize,draw, fill=white, 
rectangle,pos=0.5] {$t\leq 100$} (up1);
\draw[tran] (up1) to node[auto] {t:=t+1}  (up2);
\draw[tran] (up2) to node[auto] {z:=Unif(-2,1)}  (ndet);
\draw[tran] (ndet) -- (ndet|-up3) -- (up3);
\draw[tran](up3) --node[auto] {x:=x+z}(det1|-up3)--(det1);
\draw[tran] (ndet) -- (ndet|-up4) -- (up4);
\draw[tran](up4) -- node[auto,swap] {y:=y+z} (det1|-up4)--(det1);
\end{tikzpicture}
\caption{The pCFG that models the probabilistic program in Fig.~\ref{fig:APP}. 
Rectangles, diamonds, and pentagons represent 
deterministic, nondeterministic and assignment locations, respectively.
The variables are initially set $x:=2$, $y:=2$ and $t:=0$.}
\end{center}
\end{figure}

\begin{mydefinition}[pCFG, \cite{AgrawalC018}]\label{def:pCFG}
A \emph{probabilistic control flow graph (pCFG)} is a tuple $\Gamma=(L,V,\linit,\xinit,\trrel,\myUp,\myPr,G)$
consisting of the following components.
\begin{itemize}
\item A finite set $L$ of \emph{locations},  equipped with a partition $L=L_N+L_P+L_D+L_A$ 
into
 \emph{nondeterministic}, \emph{probabilistic}, \emph{deterministic} and \emph{assignment} locations.

\item A finite set $V=\{x_1,\ldots,x_{|V|}\}$ 
of \emph{program variables}.

\item An \emph{initial location} $\linit\in L$, and 
 an \emph{initial valuation vector} $\xinit\in\mathbb{R}^V$.

\item A \emph{transition relation} ${\trrel}\subseteq L\times L$ which is total (each location has a successor).
For $l\in L\setminus L_A$, we write $\mysucc(l)$ to denote the set of all successors of $l$, i.e. $\mysucc(l) = \{l'\in L \mid l\trrel l'\}$ .
We require that each assignment location $l \in L_A$ has a unique successor; in this case, $\mysucc(l)$ denotes this unique location.

\item An \emph{update function} $\myUp: L_A \to
V\times \mathcal{U}$, where $\mathcal U = \borel(\real^V,\real)\cup\dist(\real)\cup\borel(\real)$. Here, three components of $\mathcal U$ represent deterministic, probabilistic and nondeterministic assignment, respectively.


\item A family $\myPr=\bigl(\myPr_l\in \dist(\mysucc(l)) \bigr)_{l\in L_P}$ of probability distributions.

\item A \emph{guard function} $G: {L_D\times L} \to \mathcal B (\mathbb R^V)$ such that, for each $l \in L_D$, the following hold: (collective exhaustion) $\bigcup_{l\trrel l'}G(l,l') =  \mathbb R^V$; and (mutual exclusion) $l\trrel l'$, $l\trrel l''$ and $l' \neq l''$ imply
$G(l,l') \cap G(l,l'') = \emptyset$. 
We write $\vec{x} \models G(l,l')$ if $\vec{x} \in G(l,l')$. 
\end{itemize}
%

\end{mydefinition}
A \emph{configuration} of a pCFG $\Gamma$ is a pair $(l,\vec{x})\in L\times\mathbb{R}^V$ of a location and a vector. 
%
%
A \emph{successor} $(l',\vec{x}')$ of a configuration $(l,\vec{x})$
 is a one such that $l \mapsto l'$ and 

\noindent
\begin{minipage}{\textwidth}
\begin{itemize}
\item if $l \in L_{N} \cup L_{P}$ then ${\vec{x}' = \vec{x}}$;
\item if $l \in L_{D}$ then ${\vec{x}' = \vec{x}}$ and $\vec{x} \models G(l,l')$; and 
\item if $l \in L_A$ and  $\myUp(l) = (x_{j},u)$, then ${\vec{x}}' = {\vec{x}}(x_{j} \leftarrow a)$. Here ${\vec{x}}(x_{j} \leftarrow a)$ denotes an update of the vector (the $x_{j}$-component of $\vec{x}$  is replaced by $a$), and 
i) $a = u(\vec{x})$ if $u\in \borel(\config,\real)$, 
ii) $a \in \mathrm{supp}(u)$ if $u\in\dist(\real)$, and 
iii) $a \in u$ if $u\in\borel(\real)$. 
\end{itemize}
\end{minipage}

 A \emph{finite path} of $\Gamma$ is a finite sequence $c_0,c_1,\ldots, c_k$ 
 of configurations where $c_i$ is a successor of $c_{i-1}$ for each $i$. 
Similarly, A {\it run} of $\Gamma$ is an infinite sequence $c_0,c_1,\ldots$ of configurations such that each $c_i$ is a successor of $c_{i-1}$.

\medskip

\emph{Schedulers} resolve nondeterminism. 
Given a history $c_0\ldots c_i$ of configurations, it gives a distribution of the successor's location or valuation vector.
We assume that a scheduler is \emph{universally measurable}, which is standard in control theory (see e.g.~\cite{BertsekasS07stochastic}). 
\medskip

If a pCFG $\Gamma$ and a scheduler $\sigma$ for $\Gamma$ are given, then the behavior of $\Gamma$ is determined for each initial configuration $c_0$; we represent it by the map $\mathcal \mu_{\_}^{\sigma}: (L \times \mathbb R^V)^+ \to \mathcal D(L \times \mathbb R^V)$. For each nonempty sequence $c_0\ldots c_i$ the distribution $\mathcal \mu_{c_0\ldots c_i}^{\sigma}$ is, intuitively, the distribution of the next configuration given a current history $c_0\ldots c_i$ of configurations under the scheduler $\sigma$. 
For the set $\Sch_{\Gamma}$ of all schedulers for $\Gamma$ we define the following.
\begin{mydefinition}[reachability probabilities $\preach_{C,\sigma}, \upreach_{C},\lpreach_{C}$]\label{def:reachabilityprob}
Let $\Gamma$ 
be a pCFG.
The \emph{reachability probability} $\preach_{C,\sigma}(c)$ from a configuration $c_0\in L\times\mathbb{R}^V$ to a region $C \in \mathcal B(L\times\mathbb{R}^V)$ under a scheduler $\sigma \in \Sch_{\Gamma}$
is defined 
by 
\[
 \preach_{C,\sigma}(c_0) = \sum_{i \geq 1} \int_{L\times\mathbb{R}^V\setminus C} \!\!\!\mu^\sigma_{c_0}(\mathrm{d}c_1) 
 \ldots 
 \int_{L\times\mathbb{R}^V\setminus C} \!\!\!\mu^\sigma_{c_0\ldots c_{i-2}}(\mathrm{d}c_{i-1})
 \int_{C}{\mathbf 1} \mu^\sigma_{c_0\ldots c_{i-1}}(\mathrm{d}c_{i})
\]
for the case of $c_0 \not\in C$, and $\preach_{C,\sigma}(c_0) = 1$ otherwise.
The \emph{upper reachability probability} $\upreach_{C}(c)$ from $c$ to $C$ 
is defined by 
\begin{math}
 \upreach_{C}(c) = \sup_{\sigma\in\Sch_{\Gamma}} \preach_{C,\sigma}(c)
\end{math}; the \emph{lower reachability probability}  $\lpreach_{C}(c)$ is defined 
by
\begin{math}
 \lpreach_{C}(c) = \inf_{\sigma\in\Sch_{\Gamma}} \preach_{C,\sigma}(c)
\end{math}.
\end{mydefinition}


\begin{mydefinition}[reaching times $\esteps_{C,\sigma},\uesteps_C,\lesteps_C$]
Let $\Gamma$ be a pCFG.
The \emph{expected reaching time} of $\Gamma$ from a configuration $c_0\in \config$ to $C\in\borel(\config)$ under a scheduler $\sigma \in \Sch_{\Gamma}$ is defined by 
$\esteps_{C,\sigma}(c_0)=0$ for $c_0\in C$; for $c_0 \not\in C$ it is defined by
\[
\esteps_{C,\sigma}(c_0) = \sum_{i \geq 1} i\cdot\int_{L\times\mathbb{R}^V\setminus C} \!\!\!\mu^\sigma_{c_0}(\mathrm{d}c_1) 
\ldots 
\int_{L\times\mathbb{R}^V\setminus C} \!\!\!\mu^\sigma_{c_0\ldots c_{i-2}}(\mathrm{d}c_{i-1})
\int_{C}{\mathbf 1} \mu^\sigma_{c_0\ldots c_{i-1}}(\mathrm{d}c_{i})
\]
if $\preach_{C,\sigma}(c_0) = 1$, or $\esteps_{C,\sigma}(c_0) = \infty$ otherwise.
%
The \emph{upper expected reaching time}
$\uesteps_C(c)$ of $\Gamma$ from $c$ to $C$ is
\begin{math}
\uesteps_C(c) = \sup_{\sigma \in \Sch_\Gamma}\esteps_{C,\sigma}(c) 
\end{math}, 
 and the \emph{lower expected reaching time}
is
\begin{math}
\lesteps_C(c) = \inf_{\sigma \in \Sch_\Gamma}\esteps_{C,\sigma}(c)
\end{math}.
\end{mydefinition}

\auxproof{
\begin{myremark}
In many papers these characteristics are defined using the probability space $(\Omega_{\Gamma}, \mathcal{F}_{\Gamma},\mathbb{P}_{\Gamma}^{\sigma})$, where $(\Omega_{\Gamma}, \mathcal{F}_{\Gamma})$ is a measure space of infinite runs of $\Gamma$ and $\mathbb{P}_{\Gamma}^{\sigma}$ is a canonical distribution that is constructed from $\mu_{\_}^\sigma$. Our definitions are equivalent to the ones that are defined in this way.
\end{myremark}
}




\subsection{Probabilistic Programs: APP and PPP}\label{sec:pp}

The goal of this paper is the reachability analysis of imperative programs with probabilistic and nondeterministic branching. 
We consider two languages taken from~\cite{ChatterjeeNZ17,ChatterjeeFG16}, called
\emph{affine probabilistic programs} (APP) and
\emph{polynomial probabilistic programs} (PPP). 
%
The two languages differ only in the arithmetic expressions  allowed in the assignment commands and Boolean expressions. For example, the assignment command $x:= xy+x+1$ is allowed in PPP but not in APP;  $x:=3x+2y-1$  is allowed in both since its right-hand side is an affine expression. 

Both APP and PPP have the standard control structure in imperative languages---such as if-branches and while-loops.
APP and PPP additionally have nondeterministic and probabilistic if-branches (\lstinline{if $\star$ then $\dotsc$} and \lstinline{if $\prob(p)$ then $\dotsc$}, respectively, where $p\in [0,1]$). They also have nondeterministic and probabilistic assignment commands:
\lstinline{$x:=\ndet\,A$} where a value is chosen from a set $A\subseteq \mathbb{R}$; and \lstinline{$x:=d$} where a value is sampled from a probability distribution $d$ over $\mathbb{R}$. 

The definition of the semantical model pCFG (\S{}\ref{sec:pCFG}) mirrors the structure of these languages. The translation from APP/PPP
to pCFGs is straightforward and  omitted.



\subsection{Order-Theoretic Foundation of Fixed Points}\label{subsec:fixedPt}
\auxproof{(*** Toru's version, 2018-05-11 ***)
In this section we illustrate how we use results of fixed points in Lattice theory, using ranking functions for a simple Kripke frame as the toy model. First recall the following:

\begin{mylemma}[Ranking function for liveness]\label{rfunction}
 Let $(S,\,{\mapsto}\subseteq S\times S)$ be a Kripke frame, $s_{0}\in S$ and $C\subseteq S$. 
 Let $\eta\colon S\to \overline\nat$ be a \emph{ranking function} for $C$. That is, 1) for each $s\in S\setminus C$, there is a successor $s'$ such that $\eta(s)\ge \eta(s')+1$; and 2) for each $s\in S$, $\eta(s)=0$ implies $s\in C$. Then, $\eta(s_{0})\neq\infty$ implies that there is a path from $s_{0}$ to $C$. \qed
\end{mylemma}

Let $F(S,\overline\nat)$ be the set of all functions from $S$ to $\overline\nat$, and endow a partial order $\sqsubseteq$ on it by the pointwise majorization, i.e. $f \sqsubseteq g \iff \forall s \in S. \ f(s) \leq g(s)$. 
Now define a function $\Phi:F(S,\overline\nat)\to F(S,\overline\nat)$ by $\Phi(f)(s) = \min_{s\trrel s'}f(s')+1$ for $s\in S \setminus C$, and $\Phi(f)(s)=0$ otherwise; then one would notice that the following.

\begin{enumerate}
\item For $s\in S$, let $\mathsf{steps}(s)$ be the minimum number of transitions to be made to reach the region $C$ from $s$ (let $\mathsf{steps}(s)=0$ for $s \in C$). then the function $\mathsf{steps}$ is a \emph{fixed point} of $\Phi$, i.e. $\Phi(\mathsf{steps})=\mathsf{steps}$.
\item A function $f\in F(S,\overline\nat)$ is a ranking function if and only if it is a \emph{pre-fixed point} of $\Phi$, i.e. $\Phi(f) \sqsubseteq f$.
\end{enumerate}

One can also show that the function {\sf steps} is in fact the \emph{least} fixed point of $\Phi$. The following fundamental result of lattice theory tells us that those facts imply that any ranking function is above {\sf steps}, and thus Lemma~\ref{rfunction} holds.

\noindent
\begin{minipage}{\textwidth}
\begin{mytheorem}[Knaster--Tarski]\label{thm:fixedPointChar}
 Let $(L,\sqsubseteq)$ be a complete lattice, and $f\colon L\to L$ be a monotone function. Then $f$ has the least fixed point $\mu f$ and the greatest $\nu f$. Moreover, 
      The lfp is the least pre-fixed point:
      $\mu f=\min\{l\in L\mid f(l)\sqsubseteq l \}$. Similarly, the gfp is the greatest post-fixed point:
$\nu f=\max\{l\in L\mid l\sqsubseteq f(l) \}$. 
\end{mytheorem}
\end{minipage}

\noindent
\begin{minipage}{\textwidth}
\begin{mycorollary}\label{cor:fixedPtReasoningPrinciples}
 \begin{itemize}
  \item \textbf{(lfp-KT)} $f(l)\sqsubseteq l$ implies $\mu f\sqsubseteq l$.
  \item \textbf{(gfp-KT)} $l\sqsubseteq f(l)$ implies $l\sqsubseteq \nu f$.
 \end{itemize}
\end{mycorollary}
\end{minipage}

In~\S{}\ref{sec:NNRepSupM}
--\ref{sec:gammaSclSubM} we utilize this idea for the case of pCFG, which is a much larger class than Kripke frames.
We need some additional effort for this because the space $\borel(\config, \overline{\real}_{\geq 0})$, which corresponds to $F(S,\overline\nat)$ in the above case, is not a complete lattice.
In what follows we collect lemmas that are needed for the fixed point characterization of martingale methods.



\begin{myproposition}\label{prop:adapted_KT}
\begin{itemize}
\item Let $L$ be an \emph{$\omega$-cpo}, i.e. any ascending countable chain in $L$ has the supremum. 
Let $F:L \to L$ be \emph{$\omega$-continuous}, i.e. $\bigsqcup_{n<\omega}F(l_n) = F(\bigsqcup_{n<\omega}l_n)$ holds for any $\{l_n\}_{n \in \nat}$. Then $\mu F$ exists, and $Fl \leq l$ implies $\mu F \leq l$.
\item Let $L$ be an \emph{$\omega^{\mathrm{op}}$-cpo}, i.e. any descending countable chain in $L$ has the infimum. 
Let $F:L \to L$ be \emph{$\omega^{\mathrm{op}}$-continuous}, i.e. $\bigsqcap_{n<\omega}F(l_n) = F(\bigsqcap_{n<\omega}l_n)$ holds for any $\{l_n\}_{n \in \nat}$. Then $Fl \leq l$ implies $\mu F \leq l$, assuming $\mu F$ exists.
\end{itemize}
\end{myproposition}
}
%
%
%
Order-theoretic fixed points are central to computer science, for recursive computation, inductive/coinductive datatypes and reasoning and specification of reactive behaviors, etc. In general, a fixed-point equation can have multiple solutions; often we are interested in extremal solutions: least fixed points (lfp's, for liveness, induction, etc.) and greatest ones (gfp's, for safety, coinduction, etc.). The following fundamental results (in a simple setting of complete lattices) give two different characterizations of lfp's and gfp's.

\begin{mytheorem}\label{thm:fixedPointChar}
 Let $(L,\sqsubseteq)$ be a complete lattice, and $f\colon L\to L$ be a monotone function. Then $f$ has the least fixed point $\mu f$ and the greatest $\nu f$. Moreover, 
\begin{enumerate}
 \item\label{item:KnasterTarski} (Knaster--Tarski)
      The lfp is the least pre-fixed point:
      $\mu f=\min\{l\in L\mid f(l)\sqsubseteq l \}$. Similarly, the gfp is the greatest post-fixed point:
$\nu f=\max\{l\in L\mid l\sqsubseteq f(l) \}$. 
 \item \label{item:CousotCousot} (Cousot--Cousot~\cite{CousotC79})
       The (potentially transfinite) ascending chain $\bot\sqsubseteq f(\bot)\sqsubseteq
f^{2}(\bot)\sqsubseteq\cdots$  stabilizes to $\mu f$.  Here $f^{\alpha}(\bot)$ is defined by obvious induction:
	$f^{\alpha+1}(\bot)=f(f^{\alpha}(\bot))$ for a successor
	ordinal; and
	$f^{\alpha}(\bot)=\bigsqcup_{\beta<\alpha}f^{\beta}(\bot)$
	for a limit ordinal.

Similarly, the descending chain 
      $\top\sqsupseteq f(\top)\sqsupseteq\cdots$ stabilizes to $\nu f$. 
\qed
\end{enumerate}
\end{mytheorem}
From these characterizations we can derive the following reasoning principles.

\vspace{.3em}
\noindent
\begin{minipage}{\textwidth}
\begin{mycorollary}\label{cor:fixedPtReasoningPrinciples}
 \begin{itemize}
  \item \textbf{(lfp-KT)} $f(l)\sqsubseteq l$ implies $\mu f\sqsubseteq l$.
  \item \textbf{(gfp-KT)} $l\sqsubseteq f(l)$ implies $l\sqsubseteq \nu f$.
  \item \textbf{(lfp-CC)} For each ordinal $\alpha$, $f^{\alpha}(\bot)\sqsubseteq \mu f$.   \item \textbf{(gfp-CC)} For each ordinal $\alpha$, $\nu f\sqsubseteq f^{\alpha}(\top)$. 
\qed
 \end{itemize}
\end{mycorollary}
\end{minipage}

\vspace{.2em}
 The arguments so far are  symmetric for lfp's and gfp's. However, if one turns to the common proof methods for lfp specifications (termination, reachability, liveness) and those for gfp specifications (safety), a strong contrast emerges. Here is an example. 

\vspace{.3em}
\noindent
\begin{minipage}{\textwidth}
\begin{mylemma}\label{lem:KripkeFrameLivenessSafety}
 Let $(S,\,{\mapsto}\subseteq S\times S)$ be a Kripke frame, $s_{0}\in S$ and $C\subseteq S$. 
\begin{itemize}
 \item \textbf{(Invariant for safety)} Let $I\subseteq S$ be an \emph{invariant}, that is, $I\subseteq \Box I$. Here $\Box I$ is defined by $\Box I=\{s\in S\mid \text{$s\mapsto s'$ implies $s'\in I$}\}$. Assume also that $I\cap C=\emptyset$. Then $s_{0}\in I$ implies that there is no path from $s_{0}$ to $C$.
 \item \textbf{(Ranking function for liveness)} Let $\eta\colon S\to \mathbb{N}\cup\{\infty\}$ be a \emph{ranking function} for $C$. That is, 1) for each $s\in S\setminus C$, there is a successor $s'$ such that $\eta(s)\ge \eta(s')+1$; and 2) for each $s\in S$, $\eta(s)=0$ implies $s\in C$. Then, $\eta(s_{0})\neq\infty$ implies that there is a path from $s_{0}$ to $C$. \qed
\end{itemize}
\end{mylemma}
\end{minipage}

\vspace{.3em}
\begin{wrapfigure}{r}{0pt}
\begin{tabular}{c||c|c}
 & Knaster--Tarski & Cousot--Cousot \\\hline\hline
 lfp& overapprox. & underapprox.\\\hline
 gfp& underapprox. & overapprox.\\
\end{tabular}
\end{wrapfigure}
\noindent 
The difference between the two methods is accounted for by the fact that, in Cor.~\ref{cor:fixedPtReasoningPrinciples},  two items give \emph{under}-approximations while the other two give \emph{over}-approximations. It is clear that the invariant method in Lem.~\ref{lem:KripkeFrameLivenessSafety} comes from (gfp-KT) of Cor.~\ref{cor:fixedPtReasoningPrinciples}. Its dual, (lfp-KT), gives only an overapproximation $l$---it can be used for refutation but not for verification. Similarly, ranking functions come from (lfp-CC)---the  role of well-foundedness of the value domain $\mathbb{N}$ mirrors the structure of ordinals. Its dual (gfp-CC) only gives an overapproximation of $\nu f$. The situation is summarized in the above table. 

The above foundations underpin our technical developments: this is because reachability probabilities and reaching times are characterized as least fixed points.  We note that our semantical domains $L$ in later sections need not be complete lattices. In those cases we exploit the $\omega$- and $\omega^{\mathrm{op}}$-cpo structures, the corresponding continuity of $f$, and the \emph{Kleene theorem}. The last is understood as a  variation of the Cousot--Cousot theorem.


\subsection{Invariants and the Nexttime Operations}
In~\S{}\ref{sec:EpsRepSupM}--\ref{sec:gammaSclSubM} 
the following definitions will be used. 

\vspace{.3em}
\noindent
\begin{minipage}{\textwidth}
\begin{mydefinition}[(pure) invariant for pCFG]\label{def:invpCFG}
Let $\Gamma$ 
be a pCFG.
A measurable set $I\in \mathcal B(L\times \mathbb{R}^V)$ is called a \emph{(pure) invariant}
 for $\Gamma$ if
$(\linit,\xinit) \in I$, and 
for each $(l,\vec{x})\in I$, if $(l',\vec{x}')$ is a successor of $(l,\vec{x})$ then $(l',\vec{x}')\in I$.
\end{mydefinition}
\end{minipage}

\begin{mydefinition}[the ``nexttime'' operation $\upre,\lpre$]\label{pre}
%
%
Let $\Gamma$ be a pCFG, 
$I$ be a pure invariant and $\mathbb K \in \borel(\real)$.
For a measurable $\eta: I \to \mathbb{K}$ 
we define the function $\upre\eta$ of the same type as $\eta$ as follows, provided the right-hand side of each equation is well-defined.
\begin{itemize}
\item For $l \in L_N$, $(\upre\eta)(l,\x) = \max_{l \trrel l'} \eta(l',\x)$.
\item For $l \in L_P$, $(\upre\eta)(l,\x) = \sum_{l \trrel l'} \myPr_l(l') \eta(l',\x)$.
\item For $l \in L_D$, $(\upre\eta)(l,\x) = \eta(l',\x)$ where $l'$ is the unique location s.t.\  $\x \models G(l,l')$.
\item For $l \in L_A$, let $\myUp (l) = (x_j, u)$.
\begin{itemize}
\item $(\upre\eta)(l,\x) = \eta(\mysucc(l),u(\x))$ if $u$ is a measurable function. 
\item $(\upre\eta)(l,\x) = \int_{x \in \mathrm{supp}(u)} \eta(\mysucc(l),\x(x_j \leftarrow x))\mathrm{d}u$ if $u$ is a distribution.
\item $(\upre\eta)(l,\x) = \sup_{x \in u}\eta(\mysucc(l),\x(x_j \leftarrow x))$ if $u$ is a measurable set.
\end{itemize}
\end{itemize}
The function
 $\underline{\pre}\eta: I \to
\mathbb{K}$ 
is defined as above, but 
replacing $\max$ with $\min$ in the first line and $\sup$ with $\inf$ in the last.
\end{mydefinition}

\noindent
\begin{minipage}{\textwidth}
\begin{myproposition}\label{prop:nexttimeIsMeasAndConti}
We define a pointwise partial order $\sqsubseteq$ on $\borel(I,
\mathbb{K})$, i.e.\ $f \sqsubseteq g$ if and only if $f(c) \leq g(c)$ holds for every $c \in I$. Let $\mathbb K$ be a proper closed convex subset of $\real \cup \{\pm \infty\}$. 
Then $\upre\eta$ and $\lpre\eta$ are well-defined for every $\eta \in \borel(I, \mathbb{K})$, and the following hold.
\begin{enumerate}
\item The operators $\upre$ and  $\lpre$ are monotone endofunctions over $\borel(I,
\mathbb{K}
)$. 
In particular, $\upre\eta$ and $\lpre\eta$ are Borel measurable for any $\eta \in \borel(I, 
\mathbb{K}
)$.
\item $\upre$ is $\omega$-continuous, and $\lpre$ is $\omega^{\mathrm{op}}$-continuous. \qed
\end{enumerate}
\end{myproposition}
\end{minipage}


\section{$\varepsilon$-Decreasing Repulsing Supermartingales ($\varepsilon$-RepSupM)}\label{sec:EpsRepSupM}
In~\S{}\ref{sec:EpsRepSupM}--\ref{sec:gammaSclSubM} we will discuss the four martingale-based techniques
in Table~\ref{table:overview}. Here 
we briefly review the notion of $\varepsilon$-decreasing repulsing supermartingale ($\varepsilon$-RepSupM) from~\cite{ChatterjeeNZ17}. 
It is, to the best of our knowledge, the only existing martingale-based notion for 
overapproximating reachability probabilities. 

\vspace{.3em}
\noindent
\begin{minipage}{\textwidth}
\begin{mydefinition}[$\varepsilon$-RepSupM~\cite{ChatterjeeNZ17}]\label{def:EpsRepSupM}
Let $\Gamma$ 
be a pCFG, 
$I
$ be a pure invariant, and $C\subseteq I$ be a Borel set.
An \emph{$\varepsilon$-repulsing supermartingale ($\varepsilon$-RepSupM)} 
for $C$ supported by
 $I$ is a measurable function 
 $\eta\colon I\to\real$
 such that
i) $\eta(c)\ge (\upre\eta)(c)+\varepsilon$ for each $c\in I\setminus C$, and
ii) $\eta(c) \geq 0$ for each $c \in C$. 
\end{mydefinition}
\end{minipage}

\vspace{.3em}
\noindent
\begin{minipage}{\textwidth}
\begin{mytheorem}[soundness,~\cite{ChatterjeeNZ17}]\label{thm:probBoundChatterjee}
Suppose there exists an $\varepsilon$-RepSupM for $C$ supported by $I$
such that $\eta(\linit,\xinit) <0$. Further assume 
that
$\eta$ has \emph{$\kappa$-bounded differences} for some $\kappa >0$, i.e. for each $c \in I$ and its successor $c'$ it holds $|\eta(c) - \eta(c')| \leq \kappa$.
Let $\gamma = e^{-\frac{\varepsilon^2}{2(\kappa+\varepsilon)^2}}$ and $\alpha = e^{\frac{\varepsilon \cdot \eta(\linit,\xinit)}{(\kappa+\varepsilon)^2}}$.

\noindent
\begin{minipage}{\textwidth}
 \begin{enumerate}
 \item\label{item:thm:probBoundChatterjee1}  We have the following inequality:
 \begin{equation}\label{eq:1801301505}
 \upreach_C(\linit,\xinit) \leq \alpha \cdot \frac{\gamma ^{\lceil|\eta(\linit,\xinit)| \slash \kappa \rceil}}{1-\gamma}.
 \end{equation}
 \item\label{item:thm:probBoundChatterjee2} 
 If the right-hand side of $(\ref{eq:1801301505})$ is greater than $1$, still $
 \upreach_C(\linit,\xinit) <1$ holds. \qed
\end{enumerate}\end{minipage}
\end{mytheorem}
\end{minipage}


\vspace{.3em}
We note that for any $\eta \in \borel(I,\real)$ that has $\kappa$-bounded differences, the function $\upre\eta$ is well-defined.
The bound in~(\ref{eq:1801301505}) is derived from \emph{Azuma's concentration inequality}, a well-known martingale concentration lemma that exploits $\kappa$-bounded differences. 
 $\varepsilon$-RepSupM is not complete: there exist a pCFG $\Gamma$ and a set $C$ of configurations such that
$\upreach_{C}<1$ but no $\varepsilon$-RepSupM can prove it. See Fig.~\ref{fig:counterex} below.
\smallskip

\begin{figure}
\begin{center}
\begin{tikzpicture}[x = 2cm]
\node(init) at (0,0) {start};
\node[up] (up1) at (0.5,0) {$l_0$};
\node[det] (det1) at (2.0,0)  {$l_1$};
\node[up] (up2) at (3.2,0) {$l_2$};
\node[ran] (ran1) at (4.0,0) {$l_3$};
\node[det](det2) at (5,0.8) {$l_4$};
\node[det](det3) at (5,-0.8) {$l_5$};
\node(dum) at (3.5,0){};
\draw[tran] (init) -- (up1);
\draw[tran] (up1) to node[auto] {x:={\bf ndet}(0,1)}  (det1);
\draw[tran] (det1) to node[font=\scriptsize,draw, fill=white, 
rectangle,pos=0.5] {$x<1$} (up2);
\draw[tran] (up2) -- (up2|-det2) -- node[auto] {x:=2x}(det1|-det2) -- (det1);
\draw[tran] (det1) -- (det1|-det3) -- node[font=\scriptsize,draw, fill=white, 
rectangle,pos=0.5] {$x\geq1$}
(dum|-det3) -- (dum|-ran1) -- (ran1);
\draw[tran] (ran1) -- node[font=\scriptsize,draw, fill=white, 
rectangle,pos=0.5] {$\frac{1}{2}$}(det2);
\draw[tran] (ran1) -- node[font=\scriptsize,draw, fill=white, 
rectangle,pos=0.5] {$\frac{1}{2}$}(det3);
\draw[tran, loop, looseness = 5, in =25, out = -25] (det2) to (det2);
\draw[tran, loop, looseness = 5, in =25, out = -25] (det3) to (det3);
\end{tikzpicture}
\caption{An example of incompleteness of $\varepsilon$-RepSupM. Probabilistic locations are depicted by circles.
This pCFG satisfies $\upreach_{\{l_5\}\times \real}(l_0,0) = \frac{1}{2}$ but no $\varepsilon$-RepSupM can refute its a.s. reachability. 
Indeed, any $\varepsilon$-RepSupM $\eta$ for $\{l_5\}\times \real$ must satisfy $\lim_{x\to+0}\eta(l_1,x) = \infty$ due to the $\varepsilon$-decreasing condition, but such an $\eta$ cannot have $\kappa$-bounded differences at $(l_0,0)$.
}\label{fig:counterex}
\end{center}
\end{figure}

\section{Nonnegative Repulsing Supermartingales (NNRepSupM)
}
\label{sec:NNRepSupM}
%
We move on to another notion for overapproximating reachability probabilities,
\emph{nonnegative repulsing supermartingale} (NNRepSupM). We believe this is new.
Compared to the notion of $\varepsilon$-RepSupM, NNRepSupM has the following features.
\begin{itemize}
 \item 
 NNRepSupM is derived from the theory of order-theoretic fixed points (\S{}\ref{subsec:fixedPt}), unlike $\varepsilon$-RepSupM that relies on Azuma's martingale concentration lemma.
 \item  Consequently, we can show soundness and completeness of NNRepSupM rather easily, while $\varepsilon$-RepSupM is sound but not complete.
\item We experimentally observe that NNRepSupM often gives better bounds (\S{}\ref{sec:implexp}). 
\end{itemize}

The definition of NNRepSupM resembles \emph{probabilistic barrier certificates} used in control theory~\cite{PrajnaJ04,steinhardtT12IJRR}. Our technical contributions are the following: 
i) we develop the theory of NNRepSupM in the presence of nondeterminism, while the settings in~\cite{PrajnaJ04,steinhardtT12IJRR} are purely probabilistic; and 
ii) we characterize NNRepSupM in the general terms of order-theoretic fixed points (\S{}\ref{subsec:fixedPt}), unlike the previous theory in~\cite{PrajnaJ04,steinhardtT12IJRR} that relies on Markov's martingale concentration lemma.\footnote{We note that the theory of NNRepSupM can also be developed using Markov's lemma.} The latter unveils the mathematical similarity between NNRepSupM and ARnkSupM (\S{}\ref{sec:ARnkSupM}).

The notion comes with upper and lower variants. They are used to overapproximate $\upreach_C$ and $\lpreach_C$, respectively (Def.~\ref{def:reachabilityprob}). In this section we use $\mathbb{K}=[0,\infty]$.

\begin{mydefinition}[NNRepSupM for pCFG]\label{def:NNRepSupMpCFG}
Let $\Gamma$ 
be a pCFG, 
$I
$ be a pure invariant, and $C\subseteq I$ be a Borel set. 
An \emph{upper nonnegative repulsing supermartingale (U-NNRepSupM)} over $\Gamma$ for $C$ supported by $I$ is a function $\eta\in\borel(I,
[0,\infty])$ s.t. 

\noindent
\begin{minipage}{\textwidth}
 \begin{displaymath}
 \begin{array}{l}
 \text{i) $\eta(c) \geq 1$ for each $c \in C$, and}\qquad
 \text{ii) $\eta(c) \geq \upre\eta(c)$ for each $c \in I \setminus C$.}
 \end{array}
 \end{displaymath}
\end{minipage}
The function $\eta$ is 
a \emph{lower nonnegative repulsing supermartingale (L-NNRepSupM)} 
 if it satisfies the above conditions, but with  $\upre$  replaced with $\lpre$.
\end{mydefinition}


We shall prove soundness  and completeness of NNRepSupM, based on the foundations in~\S{}\ref{subsec:fixedPt}. 
The following characterization is fundamental.
\begin{myproposition}\label{prop:reachabilityProbAsFixedPt}
In the setting of Def.~\ref{def:NNRepSupMpCFG}, 
we define endofunctions $\overline{\Phi_{C}}$ and $\underline{\Phi_{C}}$ over $\mathcal B(I, [0,\infty])$ 
as follows:
\[
\overline{\Phi_{C}}(\eta)(x) = 
\begin{cases}
1 & (x \in C) \\
(\upre\eta)(x) & (x \not\in C),
\end{cases}
\quad\quad
\underline{\Phi_{C}}(\eta)(x) = 
\begin{cases}
1 & (x \in C) \\
(\lpre\eta)(x) & (x \not\in C).
\end{cases} 
\]
%
Then 
the upper reachability probability $\upreach_C \colon L \times \mathbb R^V \to [0,\infty]$\footnote{
Precisely it is the restriction of $\upreach_C$ to $I$; in what follows we do this identification for $\upreach_C$, $\lpreach_C$, $\uesteps_C$, and $\lesteps_C$.
} 
is the least fixed point (lfp) of $\overline{\Phi_{C}}$. Similarly, 
$\lpreach_C$ is  the lfp of $\underline{\Phi_{C}}$.
\end{myproposition}
\begin{proof} (Sketch)
We first need to show that  $\upreach_C$ and $\lpreach_C$ are Borel measurable. This is not very easy, 
as they are defined via  supremum or infimum over uncountably many schedulers. We use the technique of \emph{$\varepsilon$-optimal scheduler} known from control theory~\cite{BertsekasS07stochastic}. 

Checking that $\upreach_C$  and $\lpreach_C$ are fixed points is not hard, though laborious. We use $\varepsilon$-optimal schedulers again for interchange between sup./inf.\ and integration.

Finally, the proofs for  minimality differ for $\upreach_C$  and $\lpreach_C$. 
For $\upreach_C$, we first observe that $\overline{\Phi_{C}}$ is $\omega$-continuous
(immediate from Prop.~\ref{prop:nexttimeIsMeasAndConti}). Therefore by the Kleene theorem, the lfp of $\overline{\Phi_{C}}$ is given by $\overline{\Phi_{C}}^{\omega}(\bot)$ (i.e.\ the chain in Thm.~\ref{thm:fixedPointChar} stabilizes after $\omega$ steps). We can check the coincidence between 
$\overline{\Phi_{C}}^{\omega}(\bot)$ and $\upreach_C$ by direct calculation.

For $\lpreach_C$, let $\eta$ be a fixed point of $\underline{\Phi_{C}}$. Then for each $\varepsilon >0$, we can construct a scheduler $\sigma$ such that $\preach_{C,\sigma}(c)\sqsubseteq \eta(c) + \varepsilon$ for each $c$ (at the $n$-th step, $\sigma$ chooses a $\varepsilon/2^{n}$-optimal successor). Since $\lpreach_{C}=\inf_{\sigma}\preach_{C,\sigma}$, this proves $\lpreach_{C}\sqsubseteq \eta$. 
\qed 
\end{proof}




It is easy to see that a U-NNRepSupM $\eta$ is nothing but a pre-fixed point of $\overline{\Phi_{C}}$  (i.e.\ $\overline{\Phi_{C}}(\eta)\sqsubseteq \eta$), and that an L-NNRepSupM $\eta$ is a pre-fixed point of $\underline{\Phi_{C}}$. Therefore,
soundness and completeness of NNRepSupM follow essentially from Cor.~\ref{cor:fixedPtReasoningPrinciples}. 
\begin{mycorollary}\label{cor:NNRepSupMsoundness}
\begin{enumerate}
 \item (Soundness) If $\eta$ is a U-NNRepSupM for $C$ supported by $I$, then for each $c \in I \setminus C$ we have $\upreach_{C} (c) \leq \eta(c)$.

 Similarly, if $\eta$ is an L-NNRepSupM for $C$ supported by $I$, then for each $c \in I \setminus C$ we have $\lpreach_{C} (c) \leq \eta(c)$. This means, concretely, that 
for each $\varepsilon>0$ there is a scheduler $\sigma \in \Sch_\Gamma$ such that, for any $c \in I \setminus C$, we have $\preach_{C,\sigma} (c) \leq \eta(c)+\varepsilon$. 
 \item (Completeness) There exists a U-NNRepSupM $\eta$ that gives the optimal bound for $\upreach_{C}$. The same for L-NNRepSupM. 
 \qed
\end{enumerate}
\end{mycorollary}

\section{Additive Ranking Supermartingales (ARnkSupM)}
\label{sec:ARnkSupM}
We move on to the notion of additive ranking supermartingale (ARnkSupM) in Table~\ref{table:overview}. It is the best-known martingale-based notion for analysis of probabilistic programs and is used for overapproximating the expected reaching time. That its value is finite implies almost-sure reachability, too. We review its theory; the reason is to demonstrate that the same order-theoretic structure (see~\S{}\ref{subsec:fixedPt}) underlies ARnkSupM and NNRepSupM in the previous section. 
The completeness result ((\ref{item:ARnkSupMComplete}) of Cor.~\ref{cor:ARnkSupMSoundComp}) for pCFGs with real-valued variables seems new, too;
See~\S{}\ref{sec:relwork} for a detailed comparison to existing works.
Proofs are done in a much similar manner to the ones in~\S{}\ref{sec:NNRepSupM}. 
In this section we use $\mathbb{K}=[0,\infty]$.

We note that completeness of U-ARnkSupM we state below is the one for \emph{strong} almost-sure reachability~\cite{AvanziniLY18onprob}. U-ARnkSupM is incomplete for \emph{positive} almost-sure reachability~\cite{FioritiH15}, that is, it cannot witness the condition $\forall \sigma. \esteps_{C,\sigma}(c)<\infty$ in general.

\begin{mydefinition}[ARnkSupM for pCFG,~\cite{chakarovS13probprog}]\label{def:ARnkSupMpCFG}
Let $\Gamma$ 
be a pCFG,
$I\in \mathcal B(L\times \mathbb{R}^V)$ be a pure invariant, and $C\subseteq I$ be a Borel set. 
An \emph{upper additive ranking supermartingale (U-ARnkSupM)} over $\Gamma$ for $C$ supported by $I$ is a function $\eta\in\borel(I,
[0,\infty]
)$ 
 that satisfies $\eta(c) \geq 1+\upre\eta(c)$ for each $c \in I \setminus C$.

The function $\eta$ is a \emph{lower additive ranking supermartingale (L-ARnkSupM)} if it satisfies the above conditions, but with  $\upre$  replaced with $\lpre$.
\end{mydefinition}

\begin{myproposition}\label{prop:ExpReachTimeAsFixedPt}
In the setting of Def.~\ref{def:ARnkSupMpCFG}, we define endofunctions $\overline{\Psi_{C}}$ and $\underline{\Psi_{C}}$ over $\mathcal B(I, [0,\infty])$ as follows:
\[
\overline{\Psi_C}(\eta)(x) = 
\begin{cases}
0 & (x \in C) \\
1+ (\upre\eta)(x) & (x \not\in C),
\end{cases}
\quad\quad
\underline{\Psi_C}(\eta)(x) = 
\begin{cases}
0 & (x \in C) \\
1+ (\lpre\eta)(x) & (x \not\in C).
\end{cases} 
\]
%
Then the upper expected reaching time $\uesteps_C\colon \config \to [0,\infty]$ is the lfp of $\overline{\Psi_C}$. Similarly, $\lesteps_C$ is the lfp of $\underline{\Psi_C}$.
\qed
\end{myproposition}

\begin{mycorollary}\label{cor:ARnkSupMSoundComp}
\begin{enumerate}
 \item (Soundness, e.g. \cite{AgrawalC018}) If $\eta$ is a U-ARnkSupM for $C$ supported by $I$, then for each $c \in I \setminus C$ we have $\uesteps_{C} (c) \leq \eta(c)$. In particular, for each $c\in I\setminus C$ that satisfies $\eta(c) < \infty$ we have $\lpreach_C (c) = 1$. 

 Similarly, if $\eta$ is an L-ARnkSupM for $C$ supported by $I$, then for each $c \in I \setminus C$ we have $\lesteps_{C} (c) \leq \eta(c)$. This means, concretely, that 
 for each $\varepsilon>0$ there is a scheduler $\sigma \in \Sch_\Gamma$ such that, for any $c \in I \setminus C$, we have $\esteps_{C,\sigma} (c) \leq \eta(c)+\varepsilon$. 
In particular, for each $c\in I\setminus C$ that satisfies $\eta(c) < \infty$ we have $\upreach_C (c) = 1$.
 \item\label{item:ARnkSupMComplete} (Completeness) There exists a U-ARnkSupM $\eta$ that gives the optimal bound for $\uesteps_{C}$. The same holds for L-ARnkSupM. 
 \qed
\end{enumerate}
\end{mycorollary}



\section{$\gamma$-Scaled Submartingales ($\gamma$-SclSubM)}\label{sec:gammaSclSubM}
Here we present the theory of \emph{$\gamma$-scaled submartingales} ($\gamma$-SclSubM). It is for underapproximating reachability (Table~\ref{table:overview}). Compared to the well-known method of ARnkSupM, the greatest advantage is in \emph{quantitative} reasoning: the value of a $\gamma$-SclSubM is guaranteed to be below the reachability probability (which can be less than $1$), while ARnkSupM is useful only if almost reachability holds. In this section we use $\mathbb{K}=[0,1]$.

The notion of $\gamma$-SclSubM is first introduced in~\cite{urabeHH17LICS}, as an instance of a categorical abstraction of ranking functions. The current paper's  contribution lies in the following: 
i) the theoretical developments about $\gamma$-SclSubM in concrete (non-categorical) terms; 
ii) introduction of nondeterminism (the setting of~\cite{urabeHH17LICS} is purely probabilistic); and 
iii) template-based synthesis of  $\gamma$-SclSubM. 

\vspace{.3em}
\noindent
\begin{minipage}{\textwidth}
 \begin{mydefinition}[$\gamma$-SclSubM for pCFG,~\cite{urabeHH17LICS}]\label{def:GSclSubMpCFG}
 Let 
 $\gamma \in (0,1)$ be given. 
 An \emph{upper $\gamma$-Scaled Submartingale (U-$\gamma$-SclSubM)} over $\Gamma$ for $C$ supported by $I$ is a function $\eta\in\borel(I,[-\infty,1])$ 
that satisfies $\eta(c) \leq \gamma\cdot\upre\eta(c)$
 for each $I\setminus C$.
 A \emph{lower $\gamma$-Scaled Submartingale (L-$\gamma$-SclSubM)} over $\Gamma$ for $C$ supported by $I$ is a function $\eta\in\borel(I,[-\infty,1])$ that satisfies 
 $\eta(c) \leq \gamma\cdot\lpre\eta(c)$ for each $I\setminus C$.
\end{mydefinition}\end{minipage}

\vspace{.3em}
The derivation of  $\gamma$-SclSubM, from a categorical account in~\cite{urabeHH17LICS}, can be described in the following concrete terms. A $\gamma$-SclSubM is a post-fixed point of certain functions (namely $\gamma\cdot\overline{\Phi_C}$ and $\gamma\cdot\underline{\Phi_C}$ below). According to (gfp-KT) in Cor.~\ref{cor:fixedPtReasoningPrinciples}, $\gamma$-SclSubM underapproximates a \emph{greatest} fixed point---but reachability is a \emph{least} fixed point. The trick here is as follows: 1) thanks to the scaling by $\gamma \in (0,1)$, the gfp and lfp of $\gamma\cdot\overline{\Phi_C}$ coincide; and 2) the lfp (hence the gfp) of $\gamma\cdot\overline{\Phi_C}$ is easily seen to be below the lfp of $\overline{\Phi_C}$,
 that is, the reachability probability that we are after. The overall argument signifies the role of the Knaster--Tarski theorem.

\begin{myproposition}\label{prop:fixed_point_char_gamma}
Let $\overline{\Phi_C}$ and $\underline{\Phi_C}$ be as defined in Prop.~\ref{prop:reachabilityProbAsFixedPt}.
%
Define endofunctions
$\gamma\cdot\overline{\Phi_C}$ and 
$\gamma\cdot\underline{\Phi_C}$ over $\mathcal B(I, [0,1])$ as follows:
\begin{math}\small
(\gamma\cdot\overline{\Phi_C})(\eta)(x) = 
\begin{cases}
1 & (x \in C) \\
\gamma\cdot(\upre\eta)(x) & (x \not\in C),
\end{cases}
\end{math}
and
\begin{math}\small
(\gamma\cdot\underline{\Phi_C})(\eta)(x) = 
\begin{cases}
1 & (x \in C) \\
\gamma\cdot(\lpre\eta)(x) & (x \not\in C).
\end{cases} 
\end{math}
Then we have
i) $\mu(\gamma\cdot\overline{\Phi_C}) \sqsubseteq \mu\overline{\Phi_C}$ and $\mu(\gamma\cdot\underline{\Phi_C}) \sqsubseteq \mu\underline{\Phi_C}$ , and 
ii) $\nu(\gamma\cdot\overline{\Phi_C}) = \mu(\gamma\cdot\overline{\Phi_C})$ and $\nu(\gamma\cdot\underline{\Phi_C}) = \mu(\gamma\cdot\underline{\Phi_C})$. \qed
\end{myproposition}


\begin{mycorollary}[soundness]\label{cor:GSclSubMSound} 
If $\eta$ is a U-$\gamma$-SclSubM for $C$ supported by $I$, then for each $c \in I \setminus C$ we have $\upreach_C(c) \geq \eta(c)$.
This means, concretely, that 
 for each $\varepsilon>0$ there is a scheduler $\sigma \in \Sch_\Gamma$ such that, for any $c \in I \setminus C$, we have $\preach_{C,\sigma} (c) \geq \eta(c)-\varepsilon$. 
 
Similarly, if $\eta$ is an L-$\gamma$-SclSubM for $C$ supported by $I$, then for each $c \in I \setminus C$ we have $\lpreach_{C} (c) \geq \eta(c)$.
\end{mycorollary}

\begin{proof}

Just notice that if $\eta$ is an upper- or lower-$\gamma$-SclSubM, then so is $\max\{\mathbf{0},\eta\}$. The rest is as described in the paragraph before Prop.~\ref{prop:fixed_point_char_gamma}. 
\qed 
\end{proof}

\section{Implementation and Experiments}\label{sec:implexp}
We implemented template-based automated synthesis algorithms for 
NNRepSupM~(\S{}\ref{sec:NNRepSupM}) and $\gamma$-SclSubM~(\S{}\ref{sec:gammaSclSubM}),
and present some experimental results. We implemented the following programs: 

\begin{enumerate}
\renewcommand{\theenumi}{\Roman{enumi}}
\item\label{item:impl1} 
 synthesis of a U-NNRepSupM for an APP based on a linear template.

\item\label{item:impl2} 
synthesis of a U-NNRepSupM for a  PPP based on a polynomial template.

\item\label{item:impl3} 
 synthesis of an L-$\gamma$-SclSubM for an APP based on a linear template.



\end{enumerate}
%


Each algorithm first translates given an APP or a PPP to a pCFG $\Gamma$ and a terminal configuration $C$, and
then solves an optimization problem of finding a U-NNRepSupM (L-$\gamma$-SclSubM) over $\Gamma$ for $C$ that gives a small (large) value as possible at the initial configuration. 
Reduction of optimization problems to LP or SDP ones are done in standard ways in the literature; 
we use 
\emph{Farkas' lemma} (see e.g.~\cite{chakarovS13probprog,ChatterjeeNZ17}) 
for the case of APPs, and
\emph{Schm\"udgen's Positivstellensatz} (see e.g.~\cite{ChatterjeeFG16,chakarovVS16TACAS}) 
for PPPs.

We have augmented the syntax of APPs and PPPs (\S{}\ref{sec:pp}) so that we can specify
 an invariant $I$ and a terminal configuration $C$.
The program does not synthesize an invariant nor prove the correctness of the given invariant, and
therefore the user has to provide a correct invariant by hand or by using some algorithm, e.g.~\cite{KatoenMMM10}.

All the programs are implemented in OCaml. 
We have used glpk (v4.63)~\cite{glpk} and SDPT3~\cite{sdpt3} for 
the LP and SDP solvers respectively.
For the implementation of Prog.~\ref{item:impl2}, 
we have also made use of a MATLAB toolbox SOSTOOLS (v3.03)~\cite{sostools}.
We tested our implementations for several APPs and PPPs.
We have used different benchmark sets for Prog.~\ref{item:impl1}--\ref{item:impl2} and Prog.~\ref{item:impl3} because
what is overapproximated by Prog.~\ref{item:impl1}--\ref{item:impl2} ($\upreach_C$) and what is underapproximated by Prog.~\ref{item:impl3} ($\lpreach_C$)  are different. 
The benchmarks implement the following probabilistic processes that are used as benchmarks in the literature.
More details and codes are given in \S{}\ref{sec:ppused}.
\begin{enumerate}
\renewcommand{\theenumi}{(\alph{enumi})}

\item\label{item:experiment2}({\bf Adversarial random walk})
A variation of a random walk, whose analysis is more challenging because of additional adversarial nondeterministic choices~\cite{chatterjeeFNH16algorithmicanalysis}.
We have considered three variants: (a-1) 1D, (a-2) 2D and (a-3) a variant of 2D. 
(a-1) is a random walk over $\mathbb{R}$ modeling a \emph{discrete queuing system}, and is parametrized by $p_1,p_2\in[0,1]$ that determines the distribution of the number of packets that arrive in each round.
(a-2) and (a-3) are random walks over $\mathbb{R}^2$ parametrized by $M_1,M_2\in\mathbb{R}$. They determine the distribution of 
movement distances in each round.
We added a queue size limit 
for (a-1) and a time limit 
for (a-2) and (a-3).
If the queue size exceeds $10$ in (a-1) or $100$ rounds were consumed in (a-2) or (a-3), the program stops, 
and it is not counted as termination.

\item\label{item:experiment3}({\bf Room temperature control})
A model of an air conditioning system for adjacent two rooms~\cite{abateKLP10,chakarovVS16TACAS}.
It is parametrized by real numbers $c$ and $p$: the former determines the power of the air conditioner, and
the latter determines the size of perturbation.
We have also added a time limit of $100$ as in~\ref{item:experiment2} above.

\auxproof{
\item\label{item:experiment4}({\bf Simple pendulum})
A model of a pendulum with perturbations
whose dynamics is modeled by a stochastic difference equation~\cite{steinhardtT12IJRR}. The perturbation follows a normal distribution, and is therefore arbitrarilly big (though with small probabilities). 
The process starts from the initial angle $\theta_0$, and terminates if the angle leaves a specified region 
after $3600$ seconds.
}
\end{enumerate}

We have coded \ref{item:experiment2}--\ref{item:experiment3} as an APP.
\auxproof{
and \ref{item:experiment4} as a PPP.
Hence the latter was fed to only Prog.~\ref{item:impl2}.
}
%
%
Experiments for Prog.~\ref{item:impl1} and \ref{item:impl3} were carried out on 
a  MacBook Pro laptop with a Core i5 processor (2.6 GHz, 2 cores) and 16 GiB RAM.
That for Prog.~\ref{item:impl2} was carried out on 
an Amazon EC2 c4.large instance (May 2018, 2 vCPUs and 3.75 GiB RAM) running Ubuntu 16.04.4 LTS (64 bit).
The results are in Table.~\ref{tab:expresRepSupM}--\ref{tab:expresSclSubM}.
For each program, the first column (``time (s)'') shows the total execution time, 
and the second column  (``bound'') shows the calculated probability bound. 

\noindent
{\bfseries (Applicability of NNRepSupM)}
Table~\ref{tab:expresRepSupM} shows the results for Prog.~\ref{item:impl1}--\ref{item:impl2}; the goal of these experiments is to certify the applicability of NNRepSupM to programs with nondeterminism (a-1).
We have tested them for (a-1) with two combinations of parameters.
Prog.~\ref{item:impl1}--\ref{item:impl2} found a nontrivial bound for the reachability probability when $(p_1,p_2)=(0.2,0.4)$
while it failed to find such a bound when $(p_1,p_2)=(0.8,0.1)$.
Intuitively, the random walk is more ``unfavorable'' in the former case in the sense that 
the opposite direction from a terminal configuration is chosen in higher probabilities.
As expected, a polynomial NNRepSupM gives tighter bound than a linear one, but it took much longer.
The bound was not improved by increasing the degree of the polynomial template.
\auxproof{
Prog.~\ref{item:impl2} also gave a nontrivial bound for the PPP~\ref{item:experiment4}.
}

\noindent
{\bfseries (Applicability of $\gamma$-SclSubM)}
Table~\ref{tab:expresSclSubM} shows the results for Prog.~\ref{item:impl3};
here we wish to certify applicability of our new method $\gamma$-SclSubM.
For each variant of \ref{item:experiment2},
we have tested Prog.~\ref{item:impl3} for two combinations of parameters.
In each variant, Prog.~\ref{item:impl3} gives a nontrivial probability bound for one combination 
and a trivial bound for the other combination.
In fact, all the cases where nontrivial bounds were 
``favorable'' random walks where the direction to a terminal configuration tends to be chosen.
In contrast, the cases where no nontrivial bound was found were ``unfavorable'' random walks.
Note that this is the converse of the results for Prog.~\ref{item:impl1}--\ref{item:impl3}.
Prog.~\ref{item:impl3} also succeeded in giving a nontrivial bound for \ref{item:experiment3}.
However, if we increase the parameter $c$ (i.e.\ if we strengthened the power of air conditioners), it failed to give a nontrivial bound.

\begin{table}[t]
\begin{minipage}{0.58\hsize}
\noindent
\begin{minipage}{\hsize}
\scriptsize
\begin{center}
\scalebox{0.85}{
  \begin{tabular}{|l|l||c|c|c|c|c|c|}\hline
 \multicolumn{2}{|l||}{}  & \multicolumn{2}{c|}{Prog.~\ref{item:impl1} (linear)} & \multicolumn{2}{c|}{Prog.~\ref{item:impl2} (deg.-2 poly.)} & \multicolumn{2}{c|}{Prog.~\ref{item:impl2} (deg.-3 poly.)} \\\hline 
  & param.\ & time (s) & bound & time (s) & bound & time (s) & bound \\\hline\hline
(a-1)&  \scriptsize $\begin{aligned}&p_1=0.2\\[-1.5mm] &p_2=0.4\end{aligned}$& 0.021 & $\leq 0.825$ & 530.298 &$\leq 0.6552$ & 572.393 &$\leq 0.6555$ \\\cline{2-8}  
 &\scriptsize $\begin{aligned}&p_1=0.8\\[-1.5mm] &p_2=0.1\end{aligned}$ & 0.024 & $\leq 1$ & 526.519 & $\leq 1.0$ & 561.327 & $\leq 1.0$ \\\hline   
%
   %
   \auxproof{
  \ref{item:experiment4}& $\theta_0=0.5$ & \multicolumn{1}{|c|}{---} & \multicolumn{1}{|c|}{---} & 636.732 & 
  $\begin{aligned}&\leq 5.785 \\[-2mm]&\;\times 10^{-9}\end{aligned}$& 1031.042 & $ \begin{aligned}&\leq 1.047 \\[-2mm]&\;\times 10^{-8}\end{aligned}$  \\\hline    
  }
  \end{tabular}
  } 
\end{center}  
\vspace{-2mm}
  \caption{Bounds by U-NNRepSupM}
  \label{tab:expresRepSupM}
  \end{minipage}

%
\begin{minipage}{\hsize}
\footnotesize
  \begin{center}
  \scalebox{0.85}{
  \begin{tabular}{|l|c|c|c|}\hline

   & true reachability probability & {\scriptsize U-NNRepSupM} & \scriptsize $1$-RepSupM \\\hline\hline
 (c-1) & $\frac{(0.4/0.6)^5-(0.4/0.6)^{10}}{1-(0.4/0.6)^{10}}\approx 0.116$ & $0.505$ & $<1$ \\\hline  
 (c-2) & $0.5$ & $0.5$ & --- \\\hline  
 (c-3) & $\int^{1}_0(\frac{0.25}{0.75})^{\lceil \log_2(1/x)\rceil}dx\approx 0.2$ & $0.5$ & --- \\\hline
 (c-4)& $(\frac{0.25}{0.75})^{1}\approx 0.333$ & --- & $<1$ \\\hline
  \end{tabular}
    } 
\end{center}  
\addtocounter{table}{1}
\vspace{-1mm}
  \caption{Probabilistic bounds given by U-NNRepSupM and $\varepsilon$-RepSupM}
  \label{tab:expresCompare}
  \vspace{-.6cm}  
  \end{minipage}  
  \end{minipage}
%
\qquad
\begin{minipage}{0.4\hsize}
\scriptsize
\begin{center}
\scalebox{0.85}{
  \begin{tabular}{|l|l||c|c|}\hline
 \multicolumn{2}{|l||}{}   & \multicolumn{2}{c|}{Prog.~\ref{item:impl3} (linear)}  \\\hline 
  & param.\ & time (s) & bound   \\\hline\hline
(a-1)&  \scriptsize $\begin{aligned}&p_1=0.2\\[-1.5mm] &p_2=0.4\end{aligned}$&  0.026 & $\geq 0$ \\\cline{2-4}  
  &\scriptsize $\begin{aligned}&p_1=0.8\\[-1.5mm] &p_2=0.1\end{aligned}$ & 0.022 & $\geq 0.751$\\\hline  
(a-2)  & \scriptsize $\begin{aligned}&M_1=-1\\[-1.5mm]&M_2=2\end{aligned}$&  0.033& $\geq 0$\\\cline{2-4}
    &\scriptsize $\begin{aligned}&M_1=-2\\[-1.5mm]&M_2=1\end{aligned}$ & 0.033 & $\geq 0.767$\\\hline  
(a-3)  & \scriptsize $\begin{aligned}&M_1=-1\\[-1.5mm]&M_2=2\end{aligned}$ & 0.028&$\geq 0$  \\\cline{2-4}
   &\scriptsize $\begin{aligned}&M_1=-2\\[-1.5mm]&M_2=1\end{aligned}$ & 0.040&$\geq 0.801$  \\\hline    
    &\scriptsize $\begin{aligned}&c=0.1\\[-1.5mm]&p=0.5\end{aligned}$ & 0.056&$\geq 0$ \\\cline{2-4}
  \ref{item:experiment3} &\scriptsize $\begin{aligned}&c=0.1\\[-1.5mm]&p=0.1\end{aligned}$ & 0.054&$\geq 0.148$  \\\hline 
  \end{tabular}
  } 
\end{center}  
\addtocounter{table}{-2}
  \caption{Bounds by L-$\gamma$-SclSubM with $\gamma=0.999$}
  \label{tab:expresSclSubM}
  \addtocounter{table}{1}
  \end{minipage}
  \vspace{-.2cm}
\auxproof{
\begin{minipage}{0.9\hsize}
\footnotesize
  \begin{center}
  \begin{tabular}{|l|c|c|c|}\hline

   & true reachability probability & U-NNRepSupM & $1$-RepSupM \\\hline\hline
 (d-1) & $\frac{(0.4/0.6)^5-(0.4/0.6)^{10}}{1-(0.4/0.6)^{10}}\approx 0.116$ & $0.505$ & $<1$ \\\hline  
 (d-2) & $0.5$ & $0.5$ & --- \\\hline  
 (d-3) & $\int^{1}_0(\frac{0.25}{0.75})^{\lceil \log_2(1/x)\rceil}dx\approx 0.2$ & $0.5$ & --- \\\hline
 (d-4)& $(\frac{0.25}{0.75})^{1}\approx 0.333$ & --- & $<1$ \\\hline
  \end{tabular}
\end{center}  

  \caption{Comparison of bounds given by NNRepSupM and $\varepsilon$-RepSupM}
  \label{tab:expresCompare}
  \vspace{-.6cm}  
  \end{minipage}
  }
\end{table}

\noindent
{\bfseries (Comparison between NNRepSupM and $\varepsilon$-RepSupM)}
 Both of NNRepSupM and $\varepsilon$-RepSupM (\S{}\ref{sec:EpsRepSupM})  overapproximate $\upreach_C$.
To compare them, we have also tested them for the following four simple pCFGs:
(c-1) a bounded random walk over $[0,10]$; 
(c-2) a simple system with an infinite branching where $x$ is assigned a value taken from a geometric distribution;
(c-3) a random walk over $[0,1]$ that exhibits geometric behaviors; and 
(c-4) an unbounded random walk.
See \S{}\ref{sec:pcfgused} 
for the concrete definitions of the pCFGs.

The results are shown in Table~\ref{tab:expresCompare}.
The second column shows the true reachability probability obtained by hand calculation.
The third and fourth columns show the probability bounds 
calculated by a linear NNRepSupM and a linear $1$-RepSupM respectively. 

For (c-1), both a linear NNRepSupM and a linear $1$-RepSupM were found.
However, while the NNRepSupM gave a non-trivial bound for the reachability probability,
the probability bound calculated from the $1$-RepSupM as in (\ref{eq:1801301505}) was greater than $1$ and hence trivial 
(cf. Thm.~\ref{thm:probBoundChatterjee}). 
Recall from Thm.~\ref{thm:probBoundChatterjee}.\ref{item:thm:probBoundChatterjee2} 
that the $1$-RepSupM can still refute almost-sure reachability.
%
%
For (c-2) and (c-3), whose almost-sure reachability
 cannot be refuted by $1$-RepSupMs, our algorithm found NNRepSupMs that give non-trivial probability bounds.
In contrast, for (c-4), no NNRepSupM gave non-trivial bound while
a $1$-RepSupM that refutes almost-sure reachability was found. 

\section{Related Work}\label{sec:relwork}
The notion of ranking supermartingale is first proposed by~\cite{chakarovS13probprog} aiming at extending applicability of \emph{quantitative invariants}~\cite{DBLP:conf/psse/McIverM04,McIver:2004:ARP:1036296} to probabilistic programs with real-valued variables, but nondeterminism is not considered.
Soundness of the method under 
demonic nondeterminism is studied in~\cite{AgrawalC018,ChatterjeeFG16,FioritiH15};
among them, \emph{lexicographic ranking supermartingales}~\cite{AgrawalC018} 
can be seen as an extension of our U-ARnkSupM.
Soundness under finite demonic/angelic nondeterminism is shown in~\cite{chatterjeeFNH16algorithmicanalysis}.

Completeness of U-ARnkSupM for strong almost-sure termination~\cite{AvanziniLY18onprob} has previously been shown, but only in  discrete settings~\cite{ChatterjeeF17termnondet,AvanziniLY18onprob}. The closest result to ours is~\cite{ChatterjeeF17termnondet}, where they study pCFGs with demonic nondeterminism but restrict to integer-valued variables. Our proof that also works for real-valued variables utilizes $\varepsilon$-optimal schedulers from control theory~\cite{BertsekasS07stochastic}.



Several under- and overapproximation methods for \emph{expected runtimes} of probabilistic programs, which is defined inductively on its structure rather than on its semantics, is studied in~\cite{KaminskiKMO16}. \emph{Upper invariants of {\tt while}-loops} among them 
is a U-ARnkSupM-like notion in their setting.
In~\cite{KaminskiKMO16} soundness and completeness of the upper invariant technique are derived from order-theoretic considerations. 
They handle probabilistic programs with demonic nondeterminism, but only discrete updates are allowed.


\emph{Probabilistic barrier certificates} are studied in control theory~\cite{PrajnaJ04,steinhardtT12IJRR} as a tool for overapproximating reachability. While it resembles our NNRepSupMs, their setting is purely probabilistic;  we extend applicability of the technique to systems with nondeterminism.

$\varepsilon$-RepSupM~\cite{ChatterjeeNZ17} is also a technique for overapproximating reachability, which is studied for the purpose of synthesizing \emph{stochastic invariants}. 
It is combined with ranking supermartingales to verify the \emph{persistence} property of programs, too~\cite{chakarovVS16TACAS,ChatterjeeNZ17}.
While there are certain similarities between $\varepsilon$-RepSupMs and NNRepSupMs, 
they are technically different because $\varepsilon$-RepSupMs exploit the \emph{$\kappa$-bounded differences condition}, which is not assumed in our case.
Their method is sound for refuting almost-sure reachability but does not provide nontrivial probability bound in general, and is not complete (see Fig.~\ref{fig:counterex}).






 
\section{Conclusions and Future Work}\label{sec:concfut}

We gave a comprehensive and comparative account of martingale-based techniques for approximating reachability probabilities. 
We demonstrated that several different approximation techniques--NNRepSupM, ARnkSupM, and $\gamma$-SclSubM-- 
had a common structure of order-theoretic fixed points in their theory, while they originally arose from different communities.
The key observation was that the reachability probability and the expected reaching time were the least fixed points of certain monotone endofunctions; 
soundness and completeness of the first two techniques are derived as its corollaries, 
and it is the basis for the proof of soundness of $\gamma$-SclSubM.
We also implemented the techniques above and conducted experiments, of which results suggest the advantage of $\gamma$-SclSubM in quantitative reasoning, and the comparative advantage of NNRepSupM over $\varepsilon$-RepSupM in the quality of bounds.

In this paper we have focused on over- and underapproximating (i.e.\ refuting and verifying) reachability probabilities. For future work, we wish to study more complicated specifications such as recurrence ($\mathbf{GF}\varphi$) and persistence ($\mathbf{FG}\varphi$), too. Some martingale-based techniques have already been used  (see e.g.~\cite{chakarovVS16TACAS}); we will investigate the use of \emph{lattice-theoretic progress measures}, introduced in~\cite{HasuoSC16} as a generalization of progress measures for parity games~\cite{Jurdzinski00}, in the probabilistic settings.
 
\paragraph{Acknowledgment.}
We thank Shin-ya Katsumata, Takamasa Okudono and the anonymous referees for useful comments. 
The authors are supported by JST ERATO HASUO Metamathematics for Systems Design Project (No. JPMJER1603), the JSPS-INRIA Bilateral Joint Research Project ``CRECOGI,'' and JSPS KAKENHI Grant Numbers 15KT0012 \& 15K11984. Natsuki Urabe is supported by JSPS KAKENHI Grant Number 16J08157.
\bibliographystyle{splncs04}
\bibliography{myref}

\newpage
\appendix

\section{Omitted Definitions in Section~\ref{prelim}}\label{appendix:pcsigma}

\begin{mydefinition}[scheduler]\label{def:strategy}
 A \emph{scheduler} $\sigma$ of a pCFG is a pair $(\sigma_t, \sigma_a)$ of functions that satisfy the following. 
\begin{itemize}
\item $\sigma_t$ maps each finite path $\pi$ that ends with a nondeterministic  location $l$ (i.e.\  $\pi \in
 (L \times \mathbb R^V)^* (L_{N} \times \mathbb R^V)$) to a probability distribution over $L$.
We require that the support of $\sigma_t(\pi)$ is a subset of ${\trrel_{l}}=\{l' \mid l \trrel l'\}$.
We further require that, for each $l' \in L$,
the function that maps a finite path $\pi$ to $\sigma_t(\pi) (\{l'\})$ is universally  measurable,
 with respect to the  canonical measurable structure on the set of finite paths that extends $\mathcal{B}(\mathbb{R})$.
\item $\sigma_a$ maps each finite path $\pi$ that ends with an assignment location $l\in L_{A}$ for nondeterministic assignment (i.e.\ $\mathcal U (l) \in \mathcal B(\mathbb R^n)$) to a probability distribution over $\mathbb R$. We require that the  support of $\sigma_a(\pi)$ is a subset of $\mathcal U(l)$.
We further require that for each $A \in \mathcal B(\mathbb R)$,
the function
that maps a finite path $\pi$ to $\sigma_a(\pi) (A)$ is universally measurable.
\end{itemize}

\end{mydefinition}

\begin{mydefinition}[the map $\mathcal \mu_{\_}^{\sigma}$]\label{def:trans_prob}
For a given pCFG $\Gamma$, 
a scheduler $\sigma \in \Sch_{\Gamma}$ and a nonempty sequence $\pi\in (L \times \mathbb R^V)^+$ of configurations, we define a probability measure
 $\mathcal \mu_{\pi}^{\sigma}
\in \dist(L \times \mathbb R^V)$   as follows.
\begin{itemize}
\item For any $\pi \in (L \times \mathbb R^V)^*$ and $(l,\x) \in L \times \mathbb R^V$ such that $\pi (l,\x)$ is a finite path,
\begin{itemize}
\item if $l\in L_N$ then $\mu_{\pi(l,\x)}^{\sigma} = \Sigma_{l \trrel l'} \sigma_t(\pi(l,\x))(l') \cdot \delta_{(l',\x)}$,
\item if $l\in L_P$ then $\mu_{\pi(l,\x)}^{\sigma} = \Sigma_{l \trrel l'}\myPr_l(l')\cdot\delta_{(l',\x)}$,
\item if $l\in L_D$ then $\mu_{\pi(l,\x)}^{\sigma} = \delta_{(l',\x)}$, where $l'$ is the unique location that satisfies $\x \models G(l,l')$,
\item if $l\in L_A$, $\mysucc(l) =l'$ and $\myUp(l) = (x_j,u)$, then
\begin{itemize}
\item if $u$ is a measurable function, then $\mu_{\pi(l,\x)}^{\sigma} = \delta_{(l',\x(x_j \leftarrow u(\x)))}$,
\item if $u$ is a distribution, then $\mu_{\pi(l,\x)}^{\sigma}$ is the unique measure that satisfies 
\[
\forall A \in \mathcal B(L \times \mathbb R^V) .\,\bigl(\,\mu_{\pi(l,\x)}^{\sigma}(\{l'\} \times A) = u(A)\,\bigr),
\]
\item if $u$ is a measurable set, then $\mu_{\pi(l,\x)}^{\sigma}$ is the unique measure that satisfies 
\[
\forall A \in \mathcal B(L \times \mathbb R^V) .\,\bigl(\,\mu_{\pi(l,\x)}^{\sigma}(\{l'\} \times A) = \sigma_a(\pi(l,\x))(A)\,\bigr).
\]
\end{itemize}
\end{itemize}
\item For other cases, $\mu_{\pi(l,\x)}^{\sigma} = \delta_{(l,\x)}$.
\end{itemize}
\end{mydefinition}

\section{Formal Definition for Syntax of Probabilistic Programs}\label{appendix:syntaxPP} The following definitions are from~\cite{ChatterjeeNZ17}. 
We fix a countably infinite set $\vars$ of variables.

\begin{mydefinition}[affine probabilistic program]\label{def:app}
An \emph{affine probabilistic program} (APP) is a program defined by the following BNF notation: 
\begin{align}
\pc{stmt}&::=\pc{assgn} \mid \app{skip}\mid 
\pc{stmt} ; \pc{stmt} \notag\\
&\mid \app{if}\;\pc{ndbexpr}\;\app{then}\;\pc{stmt}\;\app{else}\;\pc{stmt}\;\app{f\,\!i} \notag\\
&\mid \app{while}\;\pc{bexpr}\;\app{do}\;\pc{stmt}\;\app{od} \notag\\
\pc{assgn}&::=\pc{pvar}:=\pc{expr}\mid\pc{pvar}:=\sample(\pc{dist}) \notag\\
&\mid\pc{pvar}:=\ndet(\pc{dom})\notag\\
\pc{expr}&::=\pc{constant}\mid\pc{pvar}\mid\pc{constant}\cdot\pc{pvar} \mid\pc{expr}+\pc{expr} \notag\\
&\mid \pc{expr}-\pc{expr} \label{eq:def:app:expr}\\
\pc{dom}&::=
\Real\mid 
\Real[\pc{const},\pc{const}]\mid\pc{dom}\;\app{or}\;\pc{dom}\notag\\
\pc{bexpr}&::=\pc{conjexpr}\mid\pc{conjexpr}\;\app{or}\;\pc{bexpr} \notag\\
\pc{conjexpr}&::=\pc{literal}\mid\pc{literal}\;\app{and}\;\pc{conjexpr} \notag\\
\pc{literal}&::=\pc{expr}\leq\pc{expr}\mid \pc{expr}\geq\pc{expr}\mid\neg\pc{literal} \notag\\
\pc{ndbexpr}&::=\star\mid\prob(p)\mid\pc{bexpr}\notag \qquad\text{(where $p\in [0,1]$)}\\
\pc{pvar}&::=v\in\vars\qquad
\pc{dist}::=d\in \dist(\mathbb{R}) \qquad
\pc{const}::=c\in\mathbb{R}\,.\notag
\end{align}
\end{mydefinition}

\begin{mydefinition}[polynomial probabilistic program]\label{def:ppp}
A \emph{polynomial probabilistic program} (PPP) is a program defined 
in almost the same manner as APP (Def.~\ref{def:app}) except that 
an expression $\pc{expr}$ ((\ref{eq:def:app:expr}) in Def.~\ref{def:app}) is defined by the following BNF notation:
\[
\pc{expr}::=\pc{constant}\mid\pc{pvar}\mid\pc{expr}\cdot\pc{expr}\mid\pc{expr}+\pc{expr}\mid \pc{expr}-\pc{expr}\,.
\]
\end{mydefinition}

\section{Existence of an $\varepsilon$-optimal Scheduler}\label{appendix:optimal_scheduler}

In this section we show existence of $\varepsilon$-optimal schedulers for certain functions. 
For a given function $\varphi$ over pairs of schedulers and configurations, an \emph{$\varepsilon$-optimal scheduler} $\sigma$ is the one for which the function $\varphi(\sigma, \underline{\ \ \ })$ is uniformly $\varepsilon$-close to $\sup_\rho \varphi(\rho, \underline{\ \ \ })$. 
They resolve some measurability issues that arise in the proofs in~\S{}\ref{sec:NNRepSupM}--\ref{sec:gammaSclSubM}. 
We first fix some notations that simplify the description in later sections. 
In what follows, let $\Gamma$ 
be a pCFG, $I\in \borel(\config)$ be an invariant, $C\in\borel(I)$, and $\overline C = I \setminus C$. Also for any $A\in \borel(\config)$, let $\chi_A$ be the characteristic function of $A$.

The \emph{reachability probability $\preachEN_{C,\sigma}(c_0)$ in $N$ steps} from a configuration $c_0\in I$ to a region $C \in \mathcal B(I)$ under a scheduler $\sigma \in \Sch_{\Gamma}$
is defined 
by $\preachEN[0]_{C,\sigma}(c_0)=\chi_C(c_0)$, and for $N\geq 1$,
\[
\preachEN_{C,\sigma}(c_0) =
\chi_{\overline C}(c_0)
\int_{I}\chi_{\overline C}(c_1)  \mathrm{d}\mu^\sigma_{c_0}
\ldots 
\int_{I}\chi_{\overline C}(c_{N-1})\mathrm{d}\mu^\sigma_{c_0\ldots c_{N-2}}
\int_{I}\chi_C(c_{N}) \mathrm{d}\mu^\sigma_{c_0\ldots c_{N-1}}.
\]

Also define the \emph{non-reaching probability $\preachNN_{C,\sigma}(c_0)$ within $N$ steps} from $c_0$ to $C$ under $\sigma$ by $\preachNN[0]_{C,\sigma}(c_0)=\chi_{\overline{C}}(c_0)$, and for $N\geq 1$,
\[
\preachNN_{C,\sigma}(c_0) =
\chi_{\overline C}(c_0)
\int_{I}\chi_{\overline C}(c_1)  \mathrm{d}\mu^\sigma_{c_0}
\ldots 
\int_{I}\chi_{\overline C}(c_{N-1})\mathrm{d}\mu^\sigma_{c_0\ldots c_{N-2}}
\int_{I}\chi_{\overline C}(c_{N}) \mathrm{d}\mu^\sigma_{c_0\ldots c_{N-1}}.
\]

We omit the letter $I$ in descriptions of these functions (namely we do not write e.g. $\preachEN_{I,C,\sigma}$), 
as invariants just determine the domains of functions and do not affect their values at each configuration.


\begin{mydefinition}
	For $\gamma \in (0,1]$, we define the following functions.
\begin{eqnarray*}
	 \preachN_{C,\sigma} &=&  \sum_{i = 0}^N \preachEN[i]_{C,\sigma}, \quad
	 \estepsN_{C,\sigma}= \biggl(\sum_{i = 0}^N i \cdot \preachEN[i]_{C,\sigma}\biggr)+ N \cdot \preachNN_{C,\sigma}, \\
	 \preachN_{C,\gamma,\sigma} &=&  \sum_{i = 0}^N \gamma^i \cdot \preachEN[i]_{C,\sigma}, \quad 
	 \preach_{C,\gamma,\sigma} =  \sum_{i \in \nat} \gamma^i \cdot \preachEN[i]_{C,\sigma}.
\end{eqnarray*}
For each of these functions, let an overline and an underline indicate the supremum and the infimum of the function over all possible configurations, respectively 
(e.g. $\upreachN_{C} = \sup_{\sigma} \preachN_{C,\sigma}$ and $\lpreachN_{C} = \inf_{\sigma} \preachN_{C,\sigma}$).
\end{mydefinition}

By definition $\preachN_{C,1,\sigma}=\preachN_{C,\sigma}$, $\upreachN_{C,1}=\upreachN_{C}$ and $\lpreachN_{C,1}=\lpreachN_{C}$ hold, and similar coincidences also hold in the infinite horizon case. 
Now we are to introduce the definition of an $\varepsilon$-optimal scheduler and the lemma to be proved.

\begin{mydefinition}
Let $\varphi$ be a function from $I \times \Sch_{\Gamma}$ to $\real\cup\{\pm\infty\}$, and $\varepsilon >0$. A scheduler $\sigma$ is \emph{$\varepsilon$-optimal for $\varphi$} if, for every $c\in I$,  
$\varphi(c,\sigma) \geq \sup_{\sigma'} \varphi(c,\sigma') -\varepsilon$ holds when 
$\sup_{\sigma'} \varphi(c,\sigma') < +\infty$
, or 
$\varphi(c,\sigma) \geq \frac{1}{\varepsilon}$ holds otherwise.
\end{mydefinition}

\begin{mylemma}\label{lem:optimal}
Let a pCFG $\Gamma$, an invariant $I$, a set $C \in \borel(I)$, a positive number $\varepsilon >0$ and a natural number $N$ be given. Let $\varphi: I \times \Sch_{\Gamma} \to \real \cup \{\pm\infty\}$ be either of the following:
\begin{enumerate}
\item $\varphi(c,\sigma)= \pm\preach_{C,\gamma,\sigma}(c)$ $(\gamma \in (0,1])$,
\item $\varphi(c,\sigma)= \pm\preachN_{C,\gamma, \sigma}(c)$ $(\gamma \in (0,1])$,
\item $\varphi(c,\sigma)= \pm\esteps_{C,\sigma}(c)$, or
\item $\varphi(c,\sigma)= \pm\estepsN_{C,\sigma}(c)$.
\end{enumerate}
Then an $\varepsilon$-optimal scheduler $\sigma$ for $\varphi$ exists. 
Furthermore, they can be deterministic, i.e. $\sigma_t$ and $\sigma_a$ map any element in their domains to Dirac distributions.
\end{mylemma}

To show this, we translate each pCFG to an \emph{infinite horizon stochastic optimal control model} \cite{BertsekasS07stochastic}, for which existence of an $\varepsilon$-optimal scheduler (a.k.a. $\varepsilon$-optimal \emph{policy}) is well known. 
The following is a slightly modified definition of the one in~\cite{BertsekasS07stochastic}.

\begin{mydefinition}[\cite{BertsekasS07stochastic}] 
An \emph{(infinite horizon) stochastic optimal control model} 
is a 7-tuple $\mathsf{SM} = (S_\mathsf{SM},C_\mathsf{SM},U_\mathsf{SM},W_\mathsf{SM},p,f,\alpha)$ which consists of the following.
\begin{itemize}
\item $S_\mathsf{SM}$ is a \emph{state space}, a nonempty Borel space.
\item $C_\mathsf{SM}$ is a \emph{control space}, a nonempty Borel space.
\item $U_\mathsf{SM}$ is a \emph{control constraint}, a function $U_\mathsf{SM}:S_\mathsf{SM} \to \mathcal P (C_\mathsf{SM}) \backslash \{\phi\}$ such that the set
\[
\mathrm{Graph}(U_\mathsf{SM}) = \{(x,u) \mid x \in S_\mathsf{SM}, u \in U_\mathsf{SM}(x)\}
\]
is analytic in $S_\mathsf{SM} \times C_\mathsf{SM}$.
\item $W_\mathsf{SM}$ is a \emph{disturbance space}, a nonempty Borel space.
\item $p(dw | x,u)$ is a \emph{disturbance kernel}, a Borel measurable stochastic kernel on $W_\mathsf{SM}$ given $S_\mathsf{SM} \times C_\mathsf{SM}$.
\item $f$ is a \emph{system function}, a Borel measurable function $f:S_\mathsf{SM} \times C_\mathsf{SM} \times W_\mathsf{SM} \to S_\mathsf{SM}$.
\item $\alpha$ is a \emph{discount factor} a positive real number.
\end{itemize}

A \emph{policy} for $\mathsf{SM}$ is a sequence $\pi = (\mu_i)_{i \in \mathbb N}$, where $\mu_i(du_k|x_0,u_0,\ldots,u_{i-1},x_i)$ is a universally measurable stochastic kernel on $C_\mathsf{SM}$ given $(S_\mathsf{SM}\times C_\mathsf{SM})^i \times S_\mathsf{SM}$ that satisfies
\[
\mu_i(U_\mathsf{SM}(x_i)|x_0,u_0,\ldots,u_{i-1},x_i) = 1
\]
for each $x_0,u_0,\ldots,u_{i-1},x_i$. The set of all policies is denoted by $\Pi$.
\end{mydefinition}

For each natural number $N$, Borel distribution $p_0$ on $S_\mathsf{SM}$ and policy $\pi$, we have a unique probability measure $r_N(\pi,p_0)$ on $(S_\mathsf{SM}\times C_\mathsf{SM})^N$ with the following condition; for each Borel set $B \subseteq (S_\mathsf{SM}\times C_\mathsf{SM})^N$, the value of $r_N(\pi,p_0)(B)$ is the probability that the SM produces a run in $B$ under the initial distribution $p_0$ and the policy $\pi$. We can uniquely extend the collection $\{r_N(\pi,p_0)\}_{N\in \mathbb N}$ to the Borel measure $r(\pi,p_0)$ on $(S_\mathsf{SM}\times C_\mathsf{SM})^\omega$ such that $r_N(\pi,p_0)(B) = r(\pi,p_0)([B])$ for each $B \in \mathcal B((S_\mathsf{SM}\times C_\mathsf{SM})^N)$. For the concrete definition, see 9.1 in \cite{BertsekasS07stochastic}.

\smallskip

For a given {\sf SM}, we consider a \emph{one-stage cost function} $g$, a lower-semianalytic function $g:\Gamma \to \real\cup\{\pm\infty\}$ that is either nonnegative or nonpositive.

\begin{mydefinition}
Let $\pi$ be a policy for {\sf SM}, and let a cost function $g$ and a natural number $N$ be given. 
The \emph{N-stage cost corresponding to $\pi\in \Pi$ at $x\in S_\mathsf{SM}$} is
\[
J_{N,\pi}(x) = \int \biggl[\sum_{k=0}^N \alpha^k g(x_k, u_k)\biggr]dr(\pi,\delta_x).
\]
The \emph{infinite horizon cost corresponding to $\pi$ at $x$} is
\[
J_\pi(x) = \int \biggl[\sum_{k=0}^\infty \alpha^k g(x_k, u_k)\biggr]dr(\pi,\delta_x).
\]
The \emph{N-stage optimal cost
 at $x$} is
\[
J_N^*(x) = \inf_{\pi\in \Pi} J_\pi(x).
\]
The \emph{infinite horizon optimal cost at $x$} is
\[
J^*(x) = \inf_{\pi\in \Pi} J_\pi(x).
\]
\end{mydefinition}

We give a correspondence between pCFGs and stochastic control models as follows.
Let $L_{AD},L_{AP}$ and $L_{AN}$ be the set of all locations $l \in L_A$ such that $\mathcal U(l)$ is a Borel function, a Borel measure and a Borel set respectively. 
For a pCFG $\Gamma=(L,V,\linit,\xinit,\trrel,\myUp,\myPr,G)$, let $\mathsf{SM}(\Gamma, \alpha)$ be the following stochastic control model: 
\begin{itemize}
\item $S_\mathsf{SM} = L\times \mathbb R^V$ 
\item $C_\mathsf{SM} = W_\mathsf{SM} = L \cup \mathbb R$
\item $U_\mathsf{SM} = ((L_D \cup L_P \cup L_{AD} \cup L_{AP}) \times \{0\}) \cup (L_D \times L) \cup (L_{AN} \times \mathbb R)$
\item $p$ is defined as follows:
\begin{itemize}
\item $p(\{l'\} \mid (l,\x),\xi) = \myPr_l(\{l'\})$ and $p(\mathbb R \mid (l,\x),\xi) = 0$ if $l \in L_P$ and $l' \in L$
\item $p(L \mid (l,\x),\xi) = 0$ and $p(A \mid (l,\x),\xi) = \mathcal U(l)(A)$ if $l \in L_{AP}$ and $A \in \mathcal B(\mathbb R)$
\item $p(dw \mid (l,\x),\xi) = \delta_0$ otherwise
\end{itemize}
\item $f$ is defined as follows:
\begin{itemize}
\item $f((l,\x), \xi,w) = (l', \x)$ if $l \in L_D$ and $l'$ is the unique location such that $\x \models G(l,l')$
\item $f((l,\x), \xi,w) = (w, \x)$  if $l \in L_P$ and $w \in L$
\item $f((l,\x), \xi,w) = (\xi, \x)$  if $l \in L_N$ and $\xi \in L$
\item $f((l,\x), \xi,w) = (l', \x (v_j \leftarrow \mathcal U(l)(\x))$ if $l \in L_{AD}$ and $l \trrel l'$
\item $f((l,\x), \xi,w) = (l', \x(v_j \leftarrow w)$ if $l \in L_{AP}$ and $l \trrel l'$
\item $f((l,\x), \xi,w) = (l', \x(v_j \leftarrow \xi)$ if $l \in L_{AN}$ and $l \trrel l'$
\item $f((l,\x), \xi,w) = (l, \x)$ otherwise
\end{itemize}
\item $\alpha \in (0,1]$.
\end{itemize}

For given $C \in \mathcal B(L \times \mathbb R^V)$, we consider the following cost functions: 
\begin{flalign*}
g_1((l,\x), \xi) = 
\begin{cases}
1 & ((l,\x) \in C) \\
0 & ((l,\x) \not\in C),
\end{cases}
\quad \quad 
g_2((l,\x), \xi) = 
\begin{cases}
-1 & ((l,\x) \in C) \\
0 & ((l,\x) \not\in C),
\end{cases}
\end{flalign*}
\begin{flalign*}
g_3((l,\x), \xi) = 
\begin{cases}
0 & ((l,\x) \in C) \\
1 & ((l,\x) \not\in C),
\end{cases}
\quad \quad 
g_4((l,\x), \xi) =  
\begin{cases}
0 & ((l,\x) \in C) \\
-1 & ((l,\x) \not\in C).
\end{cases}
\end{flalign*}

Without loss of generality we can assume that $\mathsf{SM}(\Gamma,\alpha)$ satisfies the following for any policy. When we use $g_1$ or $g_2$, any run of $\mathsf{SM}(\Gamma,\alpha)$ visits an element of $C$ at most once. 
When we use $g_3$ or $g_4$, any run of $\mathsf{SM}(\Gamma,\alpha)$ stays at the same state forever once it visits an element of $C$.

The Stochastic model $\mathsf{SM}(\Gamma,\alpha)$ is an interpretation of a pCFG $\Gamma$ in the following sense.
The proof is straightforward.
\begin{myproposition}\label{prop:policy-scheduler}
Let a pCFG $\Gamma$, an invariant $I\in \borel(\config)$ and a set $C \in I$ be given, and let $\mathsf{SM}(\Gamma, \alpha)$ be the stochastic control model defined as above. 
Then we have the following: if we take $g_1$ as the cost function, then we have the following.
\begin{enumerate}
\item For any scheduler $\sigma$ and a natural number $N$, there is a policy $\pi_N$ that satisfies $\preachN_{C,\alpha,\sigma} = J_{N,\pi_N}$, and vice versa. Thus in particular, we have $\lpreachN_{C,\alpha} = J_N^*$. 
\item For any scheduler $\sigma$, there is a policy $\pi_\infty$ that satisfies
 $\preach_{C,\alpha,\sigma} = J_{\pi_\infty}$, and vice versa. Thus in particular, we have $\lpreach_{C,\alpha} = J^*$.
\end{enumerate}
In the same vein, we have:
\begin{itemize}
\item $-\upreachN_{C,\alpha} = J_N^*$ and $-\upreach_{C,\gamma} = J^*$ if the cost function is $g_2$;
\item $\lestepsN_{C} = J_N^*$ and $\lesteps_C = J^*$ if $\alpha = 1$ and the cost function is $g_3$; and
\item $-\uestepsN_{C} = J_N^*$ and $-\uesteps_C = J^*$ if $\alpha = 1$ and the cost function is $g_4$. \qed
\end{itemize}
\end{myproposition}


\smallskip
\noindent
{\it Proof of Lem.~\ref{lem:optimal}.} For every case, take an $\varepsilon$-optimal policy $\pi$ for $\mathsf{SM}(\Gamma)$ that is given by [\cite{BertsekasS07stochastic}, Prop.~8.3 and Prop.~9.20], and then take the corresponding scheduler $\sigma$ that is given by Prop.~\ref{prop:policy-scheduler}. \qed

\section{Detail of Proofs in~\S{}\ref{sec:NNRepSupM}--\ref{sec:gammaSclSubM}}

In this section we provide detailed proofs of soundness and completeness of approximation methods in~\S{}\ref{sec:NNRepSupM}--\ref{sec:gammaSclSubM}. 
We first provide a set of lemmas that are used in the proofs. 
Through this section, fix a pCFG $\Gamma$, 
an invariant $I\in \borel(\config)$ and a set $C\in\borel(I)$.

For given $\sigma \in \Sch_{\Gamma}$ and a finite path $\pi \in I^+$, let $\sigma_\pi = ((\sigma_\pi)_t,(\sigma_\pi)_a)$ be any scheduler that satisfies $(\sigma_\pi)_t(\pi') =\sigma_t(\pi\pi')$ and $(\sigma_\pi)_a(\pi') =\sigma_a(\pi\pi')$ for each $\pi'\in I^+$ such that $\pi\pi'$ is a finite path. Notice that for each $\pi,\pi'\in I^+$ such that $\pi\pi'$ is a finite path, $(\sigma_\pi)_{\pi'} = \sigma_{\pi\pi'}$ and  $\mu_{\pi\pi'}^\sigma = \mu_{\pi'}^{\sigma_\pi}$ hold. 
For convenience, we assume $\sigma_\pi = \sigma$ if $\pi$ is an empty sequence.

\begin{mylemma}\label{apnd:pre-uniform}
Let $\mathbb K$ be a proper convex closed subset of $\real \cup \{\pm \infty\}$ and $\eta\in \borel(I, \mathbb K)$. Then for any $c \in I$ we have 
$\upre\eta(c) = \sup_\sigma\int_{I}\eta \mathrm{d}\mu_{c}^\sigma$ and 
$\lpre\eta(c) = \inf_\sigma\int_{I}\eta \mathrm{d}\mu_{c}^\sigma$.
\qed
\end{mylemma}

\begin{mylemma}\label{apnd:preachEN-decomp}
	For $c \in I$, we have 
	$\preachEN[N+1]_{C,\sigma}(c) = 
	\chi_{\overline C}\int_{I} \preachEN_{C,\sigma_c}(c')\mu_c^\sigma(\mathrm{d}c')$.
	\qed
\end{mylemma}


\begin{mylemma}\label{apnd:int-supinf-interchange}
	Let $\mathbb K$ be a proper convex closed subset of $\real \cup \{\pm \infty\}$ and suppose $\varphi:I\times\Sch_\Gamma\to \mathbb K$ satisfies $\varphi(\underline{\ \ \ },\sigma)\in \borel(I, \mathbb K)$ for each $\sigma \in \Sch_\Gamma$.
	Then we have
	$\sup_\sigma\int_{I}\varphi(c',\sigma_c)\mu_{c}^\sigma(\mathrm{d}c') = \sup_\sigma\sup_\rho\int_{I}\varphi(c',\rho)\mu_{c}^\sigma(\mathrm{d}c') $ and 
	$\inf_\sigma\int_{I}\varphi(c',\sigma_c)\mu_{c}^\sigma(\mathrm{d}c') = \inf_\sigma\inf_\rho\int_{I}\varphi(c',\rho)\mu_{c}^\sigma(\mathrm{d}c') $. \qed
\end{mylemma}

\begin{mylemma}\label{apnd:int-supinf-interchange2}
	Let $\mathbb K$ be a proper convex closed subset of $\real \cup \{\pm \infty\}$ and let $\varphi:I\times\Sch_\Gamma\to \mathbb K$ be a function such that $\varphi(\underline{\ \ \ },\sigma)\in \borel(I, \mathbb K)$ for each $\sigma \in \Sch_\Gamma$.
	Further assume that for any $\varepsilon >0$ there is an $\epsilon$-optimal scheduler for $\varphi$. 
	Then $\sup_\sigma\varphi(\underline{\ \ \ },\sigma)\in\borel(I,\mathbb K)$ and for any $\mu\in\dist(\real)$ we have
	$\sup_\sigma\int_{I}\varphi(c,\sigma)\mathrm{d}\mu = \int_{I}\sup_\sigma\varphi(c,\sigma)\mathrm{d}\mu$.
\end{mylemma}

\begin{proof}
	We have $\sup_\sigma\varphi(c,\sigma) = \sup_n\varphi(c,\sigma^{(n)})$, where $\sigma^{(n)}$ is an $2^{-n}$-optimal scheduler for $\varphi$ which can be chosen independently of $c$. This proves $\sup_\sigma\varphi(\underline{\ \ \ },\sigma)\in\borel(I,\mathbb K)$.
	
	(LHS $\leq$ RHS) Take a sequence $\{\sigma^{(n)}\}_{n\in\nat}$ such that $\int_I\varphi(c,\sigma^{(n)})\mathrm{d}\mu \xrightarrow{n\to\infty}\mathrm{LHS}$. Then we have 
	$\mathrm{LHS} = \sup_n\int_I\varphi(c,\sigma^{(n)})\mathrm{d}\mu
	= \int_I\sup_n\varphi(c,\sigma^{(n)})\mathrm{d}\mu \leq \mathrm{RHS}$.
	
	(LHS $\geq$ RHS) For given $\varepsilon > 0$, let $\sigma$ be an $\varepsilon$-optimal scheduler for $\varphi$. Then we have 
	$\int_I\varphi(c,\sigma)\mathrm{d}\mu \geq \int_I\sup_{\sigma'}\varphi(c,\sigma')\mathrm{d}\mu -\epsilon$. This proves LHS $\geq$ RHS. \qed
\end{proof}

Lemma~\ref{apnd:int-supinf-interchange2} also shows that if for any $\varepsilon >0$ there is an $\varepsilon$-optimal scheduler for $-\varphi$ instead, then we have $\inf_\sigma\int_{I}\varphi(c,\sigma)\mathrm{d}\mu = \int_{I}\inf_\sigma\varphi(c,\sigma)\mathrm{d}\mu$.

\begin{myproposition}\label{prop:adapted_KT}
	Let $F:L\to L$ be a monotone endofunction on a lattice $L$. Then the following hold.
		\begin{itemize}
	\item Let $L$ be an \emph{$\omega$-cpo}, i.e. any ascending countable chain in $L$ has the supremum. 
	Let $F:L \to L$ be \emph{$\omega$-continuous}, i.e. $\bigsqcup_{n<\omega}F(l_n) = F(\bigsqcup_{n<\omega}l_n)$ holds for any $\{l_n\}_{n \in \nat}$. Then $\mu F$ exists, and $Fl \leq l$ implies $\mu F \leq l$.
	\item 
	Let $L$ be an \emph{$\omega^{\mathrm{op}}$-cpo}, i.e. any descending countable chain in $L$ has the infimum. 
	Let $F:L \to L$ be \emph{$\omega^{\mathrm{op}}$-continuous}, i.e. $\bigsqcap_{n<\omega}F(l_n) = F(\bigsqcap_{n<\omega}l_n)$ holds for any $\{l_n\}_{n \in \nat}$. Then $Fl \leq l$ implies $\mu F \leq l$, assuming $\mu F$ exists.
		\end{itemize}
\end{myproposition}

\begin{proof}
	For the $\omega$-continuous case, we have $\mu F = \bigsqcup_{n<\omega}F^n(\bot)$ from Kleene fixed point theorem. As we have $F^n(\bot) \leq l$ for any pre-fixed point $l$ of $F$ and $n\in\nat$, we have $\mu F = \bigsqcup_{n<\omega}F^n(\bot) \leq l$.
	
	For the $\omega^{\mathrm{op}}$-continuous case, from the assumption we can derive $\bigsqcap_{n<\omega}F^n l \leq l$, and as $F$ is $\omega^{\mathrm{op}}$-continuous, $\bigsqcap_{n<\omega}F^n l$ is a fixed point of $F$. Thus we have $\mu F \leq \bigsqcap_{n<\omega}F^n l \leq l$. \qed
\end{proof} 

\begin{mycorollary}\label{cor:adapted_KT_nu}
Let $F:L\to L$ be a monotone endofunction on a lattice $L$. Then the following hold.
\begin{itemize}
	\item If $L$ is an $\omega$-cpo and $F$ is $\omega$-continuous, 
	then $Fl \geq l$ implies $\nu F \geq l$, assuming $\nu F$ exists.
	\item If $L$ is an $\omega^{\mathrm{op}}$-cpo and $F$ is $\omega^{\mathrm{op}}$-continuous, then $\nu F$ exists, and $Fl \geq l$ implies $\nu F \geq l$. \qed
\end{itemize}
\end{mycorollary}

\subsection{Soundness/Completeness of UNNRepSupM} 
 \noindent
 {\it Proof of Proposition~\ref{prop:reachabilityProbAsFixedPt}.} 
 By Lemma~\ref{lem:optimal} 
 we have $\upreach_C(c) = \sup_n\preach_{(C,\sigma^{(n)})}(c)$, where $\sigma^{(n)}$ is an $2^{-n}$-optimal scheduler for $\varphi = \bigl((c,\sigma)\mapsto \preach_{(C,\sigma)}(c)\bigr)$ which can be chosen independently of $c$. This proves $\upreach_C$ is Borel measurable, and similarly we can show that $\lpreach_C$, $\upreachN_C$ and $\lpreachN_C$ are Borel measurable for each $N\in \nat$.
 
 %
 %
 %
 
 \medskip
 ($\upreach_C = \mu\overline{\Phi_C}.$)
 We first show that 
 $\overline{\mathbb{P}}^{\mathrm{reach}\leq N+1}_{C}  
 = \overline{\Phi_C}\upreachN_C$ holds for each $N\in\nat$. 
 To show this, observe the following holds for any $c\in I$:
 \begin{eqnarray}
 \mathbb{P}^{\mathrm{reach}\leq N+1}_{C,\sigma} (c)  
 &=& \preachEN[0]_{C,\sigma}(c) + \sum_{i=1}^{N+1}\preachEN[i]_{C,\sigma}(c) \nonumber \\
 &=& \chi_{C}(c) +  
 \sum_{i=0}^N \chi_{\overline C}(c)\int_{I} \preachEN[i]_{C,\sigma_{c}}(c')\mu_{c}^\sigma(\mathrm{d}c') \nonumber \\
 &=& \chi_{C}(c) +  
 \chi_{\overline C}(c)\int_{I} \preachN_{C,\sigma_{c}}(c')\mu_{c}^\sigma(\mathrm{d}c').\label{eq:preachN_inductive}
 \end{eqnarray}
 From Lemma~\ref{lem:optimal}, \ref{apnd:int-supinf-interchange} and~\ref{apnd:int-supinf-interchange2} we can derive 
 \[
 \sup_\sigma\int_I \preachN_{C,\sigma_{c}}(c') \mu^\sigma_{c}(\mathrm{d}c')
 = \sup_\sigma\int_I \upreachN_{C}(c') \mu^\sigma_{c}(\mathrm{d}c').
 \]
 Thus, taking supremum over $\Sch_{\Gamma}$ in~(\ref{eq:preachN_inductive}), we have 
 $\overline{\mathbb{P}}^{\mathrm{reach}\leq N+1}_{C}  
 = \overline{\Phi_C}\upreachN_C$.
 
 Now observe $\overline{\Phi_C}(\bot)=\chi_{C}=\upreachN[0]_{C}$ holds. Then we can inductively show that $\upreachN_{C}=\overline{\Phi_C}^{N+1}(\bot)$ for each $N\in \nat$, and hence $\upreach_{C}=\overline{\Phi_C}^\omega(\bot)$ holds. As $\overline{\Phi_C}$ is $\omega$-continuous, we have  
 $\upreach_C = \mu\overline{\Phi_C}$ via Kleene fixed point theorem.
 
 \medskip
 ($\lpreach_C = \mu\underline{\Phi_C}.$) Taking limit of $N$ in~(\ref{eq:preachN_inductive}) we have 
 \[
 \preach_{C,\sigma}(c)=\chi_{C}(c) + \chi_{\overline C}(c)
 \int_I\preach_{C,\sigma_{c}}(c')\mu_{c}^\sigma(\mathrm{d}c'),
 \]
 and in a similar way to the upper case we obtain $\lpreach_C=\underline{\Phi_C}\lpreach_C$, i.e. $\lpreach_C$ is a fixed point of $\underline{\Phi_C}$.
 
 Now we show that for any fixed point $\eta$ of $\underline{\Phi_C}$ and $\varepsilon > 0$ there is a scheduler $\tilde\sigma\in\Sch_{\Gamma}$ that satisfies $\preach_{C,\tilde\sigma} (c) \leq \eta(c) + \varepsilon$ for each $c\in I$; from this we have $\lpreach_C\sqsubseteq \eta$ and hence $\lpreach_C = \mu\underline{\Phi_C}$.
 For given a fixed point $\eta$ of $\underline{\Phi_C}$ and $\varepsilon > 0$ define $\tilde\sigma$ as follows;
 for each finite path $\pi(l,\x) \in I^* (I\cap L_{N} \times \mathbb R^V)$ we let $\tilde\sigma_t(\pi(l,\x)) = \delta_{\tilde l}$, where $\tilde l$ satisfies $\eta(\tilde l,\x) = \min_{l \trrel l'} \eta(l',\x)$;
 and for each finite path $\pi(l,\x) \in I^* (I \cap L_{AN} \times \mathbb R^V)$ such that $\myUp(l)=(\mysucc(l),u)$ we let $\tilde\sigma_a(\pi(l,\x)) = \delta_{\tilde y}$, where $\tilde y$ satisfies $\eta(\mysucc(l),\x(x_j \leftarrow \tilde y)) \leq \inf_{y \in u} \eta(\mysucc(l),\x(x_j \leftarrow y)) +2^{-(|\pi|+1)}\varepsilon$. 
 Observe that $\tilde\sigma$ is universally (actually Borel) measurable. 
 
 We show the following by induction on $N \in \mathbb N$: for each $N \in \mathbb N$, $\pi \in I^*$ and $c\in I$ such that $\pi c$ is a finite path, we have
 
 \begin{equation}\label{eq:lpreach_induction}
 \preachN_{C,\tilde\sigma_\pi}(c) \leq \eta(c) +\sum_{k=|\pi|}^{|\pi|+N-1}2^{-(k+1)}\varepsilon.
 \end{equation}
 The case of $N=0$ is immediate to show, as $\mathbb{P}^{\mathrm{reach}\leq 0}_{C,\sigma}=\chi_C$ holds for any scheduler $\sigma$. 
 For the step case, observe that 
 $\chi_C(c) + \chi_{\overline C}(c) \int_{I} \eta \mathrm{d}\mu_{c}^{\tilde\sigma_\pi} \leq \eta(c) + 2^{-(|\pi|+1)}\varepsilon $ 
 holds: indeed, 
 if $c \in \overline C \cap \bigl((L_N \cup L_{AN})\times \real^V\bigr)$ then it is derived from the construction of $\tilde\sigma$; otherwise the value is independent of a scheduler and we have 
 $(\mathrm{LHS}) = \underline{\Phi_C}\eta(c) = \eta(c) < (\mathrm{RHS})$.
 Thus we have
 \begin{eqnarray*}
 	\mathbb{P}^{\mathrm{reach}\leq N+1}_{C,\tilde\sigma_\pi}(c) 
 	&=& 
 	\chi_C(c)  + \chi_{\overline C}(c) \int_{I}\preachN_{C,\tilde\sigma_{\pi c}}\mathrm{d}\mu_{c}^{\tilde\sigma_\pi} \\
 	&\leq& \chi_C(c)  + \chi_{\overline C}(c)  \int_{I} \eta \mathrm{d}\mu_{c}^{\tilde\sigma_\pi} +\sum_{k=|\pi|+1}^{|\pi|+N}2^{-(k+1)}\varepsilon \\
 	&\leq& \eta(c) +\sum_{k=|\pi|}^{|\pi|+N}2^{-(k+1)}\varepsilon.
 \end{eqnarray*}

 Now the claim is shown to be true, and in particular we have 
 $ \preachN_{C,\tilde\sigma}(c) \leq \eta(c) +\sum_{k=1}^N 2^{-k}\varepsilon$. 
 Letting $N \to \infty$ 
 we have $\preach_{C,\tilde\sigma} (c) \leq \eta(c) + \varepsilon$. \qed 

\medskip
\noindent
 {\it Proof of Corollary~\ref{cor:NNRepSupMsoundness}.}
 $\eta \in \borel(I,[0,\infty])$ is a UNNRepSupM (LNNRepSupM) if and only if it is a pre-fixed point of $\overline{\Phi_C}$ ($\underline{\Phi_C}$). Soundness follows from Proposition~\ref{prop:adapted_KT}, and completeness is nothing but the fact that $\upreach_C$ ($\lpreach_C$) itself is a UNNRepSupM (LNNRepSupM). \qed

\subsection{Soundness/Completeness of ARnkSupM}  
 \noindent
 {\it Proof of Proposition~\ref{prop:ExpReachTimeAsFixedPt}.} 
 The proof is similar to the one of Proposition~\ref{prop:reachabilityProbAsFixedPt}, so some details are omitted in the proof below.
 First ofserve that, by Lemma~\ref{lem:optimal}, we can show that $\uesteps_C$, $\lesteps_C$, $\uestepsN_C$ and $\lestepsN_C$ are Borel measurable for each $N \in \nat$.
 
 \medskip
 
 ($\uesteps_C = \mu\overline{\Psi_C}.$)
 We first show that 
 $\uestepsN[N+1]_{C}  
 = \overline{\Psi_C}\uestepsN_C$ holds for each $N\in\nat$. To show that, observe that 
$\preachN_{C,\sigma} + \preachNN_{C,\sigma} = \mathbf{1}$  
%
 holds for each $N\in \nat$ and $\sigma\in \Sch_\Gamma$ (intuitively, it asserts that from any configuration in $I$ the pCFG either visits the region $C$ within $N$ steps, or does not visit $C$ within $N$ steps, with probability 1). Thus for any $c \in I$ we have
 \begin{eqnarray}
 \estepsN[N+1]_{C,\sigma}(c)
 &=& 
 \biggl(
 \sum_{i = 0}^{N+1} \preachEN[i]_{C,\sigma}(c)\cdot i
 \biggr) + \preachNN[N+1]_{C,\sigma}(c) \cdot (N+1)  \nonumber \\
 &=& 
 \biggl(
 \sum_{i = 1}^{N+1} \preachEN[i]_{C,\sigma}(c)
 \biggr)
 + \preachNN[N+1]_{C,\sigma}(c)  \nonumber \\
 & &
 +\biggl(
 \sum_{i = 0}^{N} \preachEN[i+1]_{C,\sigma}(c)\cdot i
 \biggr)
 + \preachNN[N+1]_{C,\sigma} (c) \cdot N  \nonumber \\
 &=& \chi_{\overline C}(c)\biggl(
 1 + \int_{I} \estepsN_{C,\sigma_{c}}(c_1)\mu^\sigma_{c}(dc_1)
 \biggr). \label{eq:estepsN_inductive}
 \end{eqnarray}
 Thus, thanks to Lemma~\ref{lem:optimal}, \ref{apnd:int-supinf-interchange} and~\ref{apnd:int-supinf-interchange2}, taking supremum over $\Sch_{\Gamma}$ in~(\ref{eq:estepsN_inductive}) we have  $\uestepsN[N+1]_{C}  = \overline{\Psi_C}\uestepsN_C$.

 Now because $\bot=\uestepsN[0]_{C}$, we can inductively show that $\uestepsN_{C}=\overline{\Psi_C}^{N}(\bot)$ for each $N\in \nat$, and hence $\uesteps_{C}=\overline{\Psi_C}^\omega(\bot)$ holds. As $\overline{\Psi_C}$ is $\omega$-continuous, we have  
 $\uesteps_C = \mu\overline{\Psi_C}$ via Kleene fixed point theorem.
 
 \medskip
 ($\lesteps_C = \mu\underline{\Psi_C}.$) Taking limit of $N$ in~(\ref{eq:estepsN_inductive}) we have 
 \[
 \esteps_{C,\sigma}(c) =
 \chi_{\overline C}(c)
 \biggl(
 1+
 \int_I\esteps_{C,\sigma_{c}}(c')\mu_{c}^\sigma(\mathrm{d}c')
 \biggr),
 \]
 and in a similar way to the upper case we obtain $\lesteps_C=\underline{\Psi_C}\lesteps_C$, i.e. $\lesteps_C$ is a fixed point of $\underline{\Psi_C}$.
 
 It can be shown that for a given fixed point $\eta$ of $\underline{\Psi_C}$ and $\varepsilon > 0$, the scheduler $\tilde\sigma\in\Sch_{\Gamma}$ in Proposition~\ref{prop:reachabilityProbAsFixedPt} satisfies $\esteps_{C,\tilde\sigma} (c) \leq \eta(c) + \varepsilon$ for any $c\in I$; the proof is almost identical to the one in Proposition~\ref{prop:reachabilityProbAsFixedPt}. This proves $\lesteps_C = \mu\underline{\Psi_C}.$ \qed
 
  \medskip
 \noindent 
 {\it Proof of Corollary~\ref{cor:ARnkSupMSoundComp}.} 
 $\eta \in \borel(I,[0,\infty])$ is a UARnkSupM (LARnkSupM) if and only if it is a pre-fixed point of $\overline{\Psi_C}$ ($\underline{\Psi_C}$). Soundness follows from Proposition~\ref{prop:adapted_KT}, and completeness is nothing but the fact that $\uesteps_C$ ($\lesteps_C$) itself is a UARnkSupM (LARnkSupM). \qed

 \subsection{Soundness of $\gamma$-SclSubM} 
 \noindent
 {\it Proof of Proposition~\ref{prop:fixed_point_char_gamma}.} 
 Similar to~(\ref{eq:preachN_inductive}), we have the following equation:
 \begin{eqnarray}
 \mathbb{P}^{\mathrm{reach}\leq N+1}_{C,\sigma,\gamma} (c)  
 &=& \chi_{C}(c) +  
 \chi_{\overline C}(c) \cdot \gamma \int_{I} \preachN_{C,\sigma_{c},\gamma}(c')\mu_{c}^\sigma(\mathrm{d}c').\label{eq:preachNG_inductive}
 \end{eqnarray}
 Thus, thanks to Lemma~\ref{lem:optimal}, \ref{apnd:int-supinf-interchange} and~\ref{apnd:int-supinf-interchange2}, taking supremum and infimum over $\Sch_{\Gamma}$ in~(\ref{eq:preachNG_inductive}) we have  $\upreachN[N+1]_{C,\gamma}  = (\gamma \cdot \overline{\Phi_C})\upreachN_{C,\gamma}$ and $\lpreachN[N+1]_{C,\gamma}  = (\gamma \cdot \underline{\Phi_C})\lpreachN_{C,\gamma}$ respectively. 
 Thus we can inductively show that 
 $\upreachN_{C,\gamma}  = (\gamma\cdot\overline{\Phi_C})^{N+1}(\bot)$ and 
 $\lpreachN_{C,\gamma}  = (\gamma\cdot\underline{\Phi_C})^{N+1}(\bot)$ holds for each $N \in \nat$.
 
 \medskip
 
 ($\mu(\gamma\cdot\overline{\Phi_C}) \sqsubseteq \mu\overline{\Phi_C}$.) 
 We have $\upreach_{C,\gamma} = (\gamma\cdot\overline{\Phi_C})^\omega(\bot)$. 
 As $\gamma\cdot\overline{\Phi_C}$ is $\omega$-continuous, we have $\mu(\gamma\cdot\overline{\Phi_C}) = (\gamma\cdot\overline{\Phi_C})^\omega(\bot)$, and from the definition we have $\upreach_{C,\gamma} \sqsubseteq \upreach_{C}$. This proves $\mu(\gamma\cdot\overline{\Phi_C}) \sqsubseteq \mu\overline{\Phi_C}$.
 
 \medskip
 ($\mu(\gamma\cdot\underline{\Phi_C}) \sqsubseteq \mu\underline{\Phi_C}$.) 
 Taking limit of $N$ in~(\ref{eq:preachNG_inductive}) we have 
 \begin{eqnarray*}
 	\preach_{C,\sigma,\gamma} (c)  
 	&=& \chi_{C}(c) +  
 	\chi_{\overline C}(c) \cdot \gamma \int_{I} \preach_{C,\sigma_{c},\gamma}(c')\mu_{c}^\sigma(\mathrm{d}c'),
 \end{eqnarray*}
 and in a similar way to the finite horizon case we have 
 $\lpreach_{C,\gamma} = (\gamma\cdot\underline{\Phi_C})\lpreach_{C,\gamma}$, 
 i.e. $\lpreach_{C,\gamma}$ is a fixed point of $(\gamma\cdot\underline{\Phi_C})$. 
 In particular we have 
 $\sup_{N<\omega}\lpreachN_{C,\gamma} 
 = (\gamma\cdot\underline{\Phi_C})^\omega(\bot)
 \sqsubseteq \lpreach_{C,\gamma}$. 
 On the other hand, for any $c \in I$ we have
 \begin{eqnarray*}
 	\lpreach_{C,\gamma} (c)
 	&=& 
 	\inf_\sigma \biggl(\sum_{i \in \nat} \gamma^i \cdot \preachEN[i]_{C,\sigma} (c) \biggr) \\
 	&\leq&
 	\inf_\sigma \biggl(\sum_{i =0}^{N-1} \gamma^i \cdot \preachEN[i]_{C,\sigma} (c) \biggr) +\gamma^N 
 	\leq \sup_{M<\omega}\lpreachN[M]_{C,\gamma}  (c) + \gamma^N,
 \end{eqnarray*}
 thus letting $N \to \infty$ we have $\lpreach_{C,\gamma} \sqsubseteq (\gamma\cdot\underline{\Phi_C})^\omega(\bot)$, and hence $(\gamma\cdot\underline{\Phi_C})^\omega(\bot) =\lpreach_{C,\gamma}$ is a fixed point of $\gamma\cdot\underline{\Phi_C}$. 
 This proves $\lpreach_{C,\gamma} = \mu(\gamma\cdot\underline{\Phi_C})$, and by definition 
 $\lpreach_{C,\gamma} \sqsubseteq \lpreach_{C} = \mu\underline{\Phi_C}$.
 
 \medskip
 
 ($\mu(\gamma\cdot\overline{\Phi_C}) = \nu(\gamma\cdot\overline{\Phi_C}).$)
 Let $\eta$ be any fixed point of $\gamma\cdot \overline{\Phi_C}$.
 In what follows we show the following by induction on $N \in \nat$: for any $c\in I$ and $N\in \nat$ we have  
 $\eta (c) 
 \leq
 \upreachN_{C,\gamma} (c) + \gamma^N \cdot \chi_{\overline{C}} (c)$.
 
 For the case of $N=0$ the inequality is true (it just asserts $\eta \sqsubseteq \mathbf{1}$).  
 For the step case, we have
 \begin{eqnarray*}
 	\eta (c)
 	&=& (\gamma\cdot \overline{\Phi_C})\eta(c) \\
 	&\leq& 
 	(\gamma\cdot \overline{\Phi_C}) (\upreachN_{C,\gamma} + \gamma^N \cdot \chi_{\overline{C}}) (c)\\
 	&=&
 	\sup_\sigma \bigl(\chi_C (c) + \chi_{\overline{C}} (c) \cdot \gamma \int \upreachN_{C,\gamma}(c') + \gamma^N \cdot \chi_{\overline{C}}(c') \mu_{c}^\sigma(\mathrm{d}c')\bigr) \\
 	&\leq&
 	\sup_\sigma \bigl(\chi_C(c) + \chi_{\overline{C}}(c) \cdot \gamma \int \upreachN_{C,\gamma}(c')\mu_{c}^\sigma(\mathrm{d}c')\bigr)
 	+ \gamma^{N+1} \cdot \chi_{\overline{C}} (c) \\
 	&=& (\gamma \cdot \overline{\Phi_C})\upreachN_{C,\gamma}(c) + \gamma^{N+1} \cdot \chi_{\overline{C}} (c) \\
 	&=& \upreachN[N+1]_{C,\gamma}(c) + \gamma^{N+1} \cdot \chi_{\overline{C}} (c). 
 \end{eqnarray*}
 This proves the claim, and thus we have $\eta \sqsubseteq \mu (\gamma\cdot \overline{\Phi_C})$. Hence $\mu (\gamma\cdot \overline{\Phi_C})$ is also the greatest fixed point of $\gamma\cdot \overline{\Phi_C}$, which is our claim.
 
 \medskip
 
 ($\mu(\gamma\cdot\underline{\Phi_C}) = \nu(\gamma\cdot\underline{\Phi_C}).$) 
 Similar to the upper case. \qed
 
 \medskip
 \noindent
 {\it Proof of Corollary~\ref{cor:GSclSubMSound}.} 
 We prove the upper case only, as the proof of the lower case is identical to it. 
 Notice that if $\eta$ is a U-$\gamma$-SclSubM, then so is $\eta' = \max\{\mathbf{0},\eta\}$, and $\eta'$ is U-$\gamma$-SclSubM if and only if it is a post-fixed point of $\gamma\cdot\overline{\Phi_C}$.
 By Corollary~\ref{cor:adapted_KT_nu} this implies $\eta' \sqsubseteq \nu(\gamma\cdot\overline{\Phi_C})$, and from Proposition~\ref{prop:fixed_point_char_gamma} and Proposition~\ref{prop:reachabilityProbAsFixedPt} we have 
 $\eta 
 \sqsubseteq \eta' 
 \sqsubseteq \nu (\gamma\cdot\overline{\Phi_C}) 
 = \mu (\gamma\cdot\overline{\Phi_C})
 \sqsubseteq \mu\overline{\Phi_C} 
 = \upreach_{C}$.
 \qed

\section{Automated Synthesis of Martingale-Based Certificates}\label{sec:synthesis}


In this section we discuss  template-based automated synthesis algorithms for NNRepSupM (\S{}\ref{sec:NNRepSupM}) and
$\gamma$-SclSubM (\S{}\ref{sec:gammaSclSubM}) for a pCFG. 
%
%
%

We  use two types of templates: a \emph{linear template} and a \emph{polynomial template}. 
Linear template-based synthesis has been proposed for
$\varepsilon$-RepSupM~\cite{ChatterjeeNZ17}, ARnkSupM~\cite{chakarovS13probprog}, etc. Polynomial template-based synthesis is  found in 
ARnkSupM~\cite{ChatterjeeFG16,chakarovVS16TACAS}. 
Those algorithms can be easily adopted for synthesis of NNRepSupM and $\gamma$-SclSubM. 

We first introduce necessary notions. Let $V=\{x_1,\ldots,x_{|V|}\}$ be a finite set of variables.
\emph{Linear expressions} and \emph{polynomial expressions} over $V$ are defined in the usual manner; see Def.~\ref{def:exp}. 
We follow~\cite{ChatterjeeNZ17} and 
call linear inequalities \emph{linear constraints}. Their conjunctions are \emph{linear conjunctive predicates}, and a disjunction of the latter is called a \emph{linear predicate}. See Def.~\ref{def:linpred}. The polynomial variations are similarly defined.

\begin{mydefinition}[expression]\label{def:exp}
	A \emph{linear expression over $V$} is a formula defined by the following BNF notation:
	$\mathfrak{a}::=r_1\mid x\mid r_2\mathfrak{a}\mid \mathfrak{a}+\mathfrak{a} $ where $x\in V$ and $r_1,r_2\in \mathbb{R}$.
	We inductively define $\sem{\mathfrak{a}}:\mathbb{R}^V\to \mathbb{R}$ by 
	$\sem{r_1}(c):=r_1$,
	$\sem{x}(c):=c(x)$,
	$\sem{r_2\mathfrak{a}_1}(c):=r_2\cdot \sem{\mathfrak{a}_1}(c)$ and
	$\sem{\mathfrak{a}_1+\mathfrak{b}_2}(c):=\sem{\mathfrak{a}_1}(c)+\sem{\mathfrak{a}_2}(c)$.
	A \emph{polynomial expression over $V$} is defined by:
	$\mathfrak{b}::=r_1\mid x\mid \mathfrak{b}\cdot\mathfrak{b} \mid\mathfrak{b}+\mathfrak{b}$. 
	We  define $\sem{\mathfrak{b}}:\mathbb{R}^V\to \mathbb{R}$ similarly to $\sem{\mathfrak{a}}$.
\end{mydefinition}

\begin{mydefinition}[constraint/(conjunctive) predicate]
	\label{def:linpred}
	A \emph{linear constraint over $V$} is a formula 
	$\mathfrak{a}\rhd 0$ where $\rhd\in\{\geq,>\}$ and $\mathfrak{a}$ is a linear expression over $V$. 
	A \emph{linear conjunctive predicate over $V$} is a formula $\mathfrak{r}_1\wedge\cdots\wedge \mathfrak{r}_m$ where each $\mathfrak{r}_i$ is a linear constraint.
	A \emph{linear predicate over $V$} is a formula  $\mathfrak{q}_1\vee \cdots\vee \mathfrak{q}_n$ where each $\mathfrak{q}_i$ is a linear conjunctive predicate.
	We define $\sem{\mathfrak{a}\rhd 0},\sem{\mathfrak{r}_1\wedge\cdots\wedge\mathfrak{r}_m},\sem{\mathfrak{q}_1\vee \cdots\vee \mathfrak{q}_n}\subseteq\mathbb{R}^V$ by 
	$\sem{\mathfrak{a}\rhd 0}:=\{\vec{x}\mid \sem{\mathfrak{a}}(\vec{x})\rhd 0\}$,
	$\sem{\mathfrak{r}_1\wedge\cdots\wedge\mathfrak{r}_m}:=\{\vec{x}\mid \vec{x}\in \sem{\mathfrak{r}_1}\cap\cdots\cap\sem{\mathfrak{r}_m}\}$ and
	$\sem{\mathfrak{q}_1\vee\cdots\vee\mathfrak{q}_m}:=\{v\mid v\in \sem{\mathfrak{q}_1}\cup\cdots\cup\sem{\mathfrak{q}_n}\}$.
	Notions of \emph{polynomial constraint} and \emph{(conjunctive) predicate} are similarly defined.
\end{mydefinition}


\vspace{.3em}
\noindent
\begin{minipage}{\textwidth}
 \begin{mydefinition}[expression/predicate map]\label{def:LEM}
 A \emph{linear expression map} over a pCFG $\Gamma$ is a tuple $\{\mathfrak{f}(l)\}_{l\in L}$ of linear expressions over $V$.
 Its \emph{semantics} 
 $\sem{\mathfrak{f}}:L\times\mathbb{R}^{V}\to \mathbb{R}$ is defined by 
 $\sem{\mathfrak{f}}(l,\vec{x}):=\sem{\mathfrak{f}(l)}(\vec{x})$.
 A \emph{linear predicate map} over $\Gamma$ is a tuple $\mathfrak{P}=\{\mathfrak{P}(l)\}_{l\in L}$ of linear predicates over $V$.
 We write $\sem{\mathfrak{P}}$ for $\{ (l,\vec{x})\mid \vec{x}\in \sem{\mathfrak{P}(l)}\}\subseteq L\times\mathbb{R}^{V}$.
 If each 
 $\mathfrak{P}(l)$ is a linear conjunctive predicate, then it is called a \emph{linear conjunctive predicate map}.
 Notions of polynomial expression map and polynomial (conjunctive) predicate map are similarly defined.
\end{mydefinition}\end{minipage}


\subsection{Linear Template-Based Synthesis  of NNRepSupM}\label{subsec:lt}
Assume we are given a pCFG $\Gamma=(L,V,\linit,\xinit,\trrel,\myPr,G)$,
%
and
linear predicate maps $\mathfrak{C}$ and $\mathfrak{I}$
over $\Gamma$. 
We further assume that $\sem{\mathfrak{C}}\subseteq L\times\mathbb{R}^V$ is an invariant\footnote{Such an invariant can be generated using an existing method, e.g.~\cite{KatoenMMM10}.}.
We consider a linear template-based algorithm that synthesizes a linear expression map $\mathfrak{\mathfrak{f}}=\{\mathfrak{f}(l)\}_{l\in L}$
such that $\sem{\mathfrak{f}}:L\times\mathbb{R}^V\to \mathbb{R}$ is an U-NNRepSupM (Def.~\ref{def:NNRepSupMpCFG}) over $\Gamma$ for $\sem{\mathfrak{C}}$ supported by 
$\mathfrak{I}$. In fact, it can be easily adopted 
from the ones in~\cite{%
chakarovS13probprog,ChatterjeeNZ17}.
We hereby briefly sketch its overview.
See~\S{}\ref{subsec:lintemplate} for a more detailed description.
The algorithm only works for ``linear'' pCFGs in the following sense.

\noindent
\begin{minipage}{\textwidth}
\begin{myassumption}\label{asm:compresL-NNRepSupM}
\begin{enumerate}
\item\label{item:asm:compresL-NNRepSupM1} For each $l\in L_A$ such that $\myUp(l)=(v,u)$,
\begin{enumerate}
\item\label{item:asm:compresL-NNRepSupM11} if $u\in \borel(\mathbb{R}^V,\mathbb{R})$, then a linear expression $\mathfrak{a}$ 
over $V$ s.t.\ $u=\sem{\mathfrak{a}}$ is given;

\item\label{item:asm:compresL-NNRepSupM12} if $u \in \dist(\mathbb{R})$, then its expectation $\mathbb{E}u\in\mathbb{R}$ is known; and

\item\label{item:asm:compresL-NNRepSupM13} if $u \in \borel(\real)$, then a linear predicate $\mathfrak{P}$ over $\{x_v\}$ s.t.\ $u=\sem{\mathfrak{P}}$ is given.
\end{enumerate}

\item\label{item:asm:compresL-NNRepSupM2} For each $l\in L_P$, the set $\{l'\in L\mid l\trrel l'\}$ of successor states is finite.

\item\label{item:asm:compresL-NNRepSupM3} For each 
$l \in L_D$ and $l' \in \trrel_l$,
a linear conjunctive predicate $\mathfrak{b}$ over $V$ s.t.\  $G(l,l')=\sem{\mathfrak{b}}$ is given.

\item\label{item:asm:compresL-NNRepSupM4} $\mathfrak{C}$ is a linear conjunctive predicate map.
\end{enumerate}
\end{myassumption}
\end{minipage}
%
%
%

\vspace{.3em}
The algorithm is almost the same as the ones in~\cite{chakarovS13probprog,ChatterjeeNZ17}:
it first fixes a linear template, a linear expression map $\mathfrak{f}$ whose coefficients are unknown, for an U-NNRepSupM.
We then reduce synthesis of an U-NNRepSupM to a feasibility problem with a conjunction of formulas of
a form $\forall \vec{x}\in\mathbb{R}^{V'}.\;\varphi\Rightarrow \psi$ is satisfied. Here:
$V'$ is a set of variables including $V$; $\varphi$ is a linear conjunctive predicate without unknown coefficients; and
$\psi$ is a linear expression with unknown coefficients.
The key lemma is \emph{Farkas' lemma} (see e.g.~\cite{boydV04ConvexOptimization}):
it reduces the problem to a feasibility problem with a conjunction of linear constraints,
which is solvable in polynomial time as an linear programming (LP) problem using an LP solver.


\vspace{.3em}
\noindent
\begin{minipage}{\textwidth}
 \begin{mytheorem}\label{thm:compresL-NNRepSupM}
 If Asm.~\ref{asm:compresL-NNRepSupM} is satisfied,
 existence of a linear U-NNRepSupM for $\Gamma$ for $\mathfrak{C}$ at $M$ supported by $\mathfrak{I}$ is decidable in polynomial time.
 \end{mytheorem}
\end{minipage}

\noindent
\begin{minipage}{\textwidth}
\begin{myremark}\label{rem:getbetterbound}
By Cor.~\ref{cor:NNRepSupMsoundness}, 
if $\mathfrak{f}$ is an U-NNRepSupM, then $\upreach_{\sem{\mathfrak{C}}}(\linit,\xinit) \leq \sem{\mathfrak{f}}(\linit,\xinit)$.
We naturally want a better bound, 
and
to this end, we have to minimize $\sem{\mathfrak{f}}(\linit,\xinit)$. 
This task is easy in the algorithm above: 
by setting $\sem{\mathfrak{f}}(\linit,\xinit)$, 
which is a linear expression over the set $U$ of unknown coefficients, 
to the objective function of the induced LP problem and ask the LP solver to minimize it, we can minimize the probability bound.
%
%
Similar arguments hold for polynomial U-NNRepSupM and linear \emph{lower} $\gamma$-SclSubM 
 discussed in later sections (although in the latter case, we \emph{maximize} the probability bound).
%
\end{myremark}
\end{minipage}

\vspace{.3em}
\noindent
\begin{minipage}{\textwidth}
\begin{myremark}\label{rem:unwillingtoreachNNRepSupM}
The algorithm described above does not work for \emph{lower} NNRepSupMs because
if we similarly reduce the axioms of L-NNRepSupM, then 
the resulting formula can contain both conjunction and disjunction, 
and therefore it cannot be solved
using an LP solver.
(They are still decidable because of the decidability of the first-order theory over reals (see e.g.~\cite{Tarski51}).)
An analogous discussion for ARnkSupM is found in~\cite{chatterjeeFNH16algorithmicanalysis}.
%
%
%
\end{myremark}
\end{minipage}

\subsection{Polynomial Template-Based Synthesis of NNRepSupM}\label{subsec:PNNRepSupM}
We next fix a polynomial template for an U-NNRepSupM.
In this case, we can relax $\mathfrak{C}$ and $\mathfrak{I}$ in \S{}\ref{subsec:lt} to
\emph{polynomial} predicate maps.
Asm.~\ref{asm:compresL-NNRepSupM} can be also relaxed. 

\vspace{.3em}
\noindent
\begin{minipage}{\textwidth}
\begin{myassumption}\label{asm:compresPNNRepSupM}
%
\begin{itemize}
\item Similar conditions to Asm.~\ref{asm:compresL-NNRepSupM}.\ref{item:asm:compresL-NNRepSupM11}, \ref{item:asm:compresL-NNRepSupM13}, 
\ref{item:asm:compresL-NNRepSupM2}--\ref{item:asm:compresL-NNRepSupM4} where ``linear'' is replaced by ``polynomial'' hold.

\item For each $l\in L_A$ such that $\myUp(l)=(v,u)$ and $u$ is a distribution on $\mathbb{R}$ and
 $n\in\mathbb{N}$, its \emph{$n$-th moment} $\int_{x\in\mathbb{R}}x^n\mathrm{d} u\in\mathbb{R}$ is known.
\end{itemize}
\end{myassumption}
\end{minipage}

\vspace{.3em}
The algorithm is adopted from~\cite{ChatterjeeFG16,chakarovVS16TACAS}.
We hereby sketch it. See \S{}\ref{subsec:polytemplate} for a more detailed description.
It first fixes a polynomial template $\mathfrak{f}$, a polynomial expression map with unknown coefficients.
Then in a similar manner to the linear case, 
 the axioms of U-NNRepSupM are reduced to a feasibility problem.
The differences from the linear case are that we use a theorem called \emph{Positivstellensatz} instead of Farkas' lemma, 
and the resulting feasibility problem is not an LP problem but an \emph{SDP problem}.
Several variants are known for Positivstellensatz. 
Among them, we use a variant called \emph{Schm\"udgen's Positivstellensatz}.

%

\subsection{Linear Template-Based Synthesis of $\gamma$-SclSubM}\label{subsec:ltSubM}
The linear template-based algorithm for synthesizing a $\gamma$-SclSubM
is very similar to the one for U-NNRepSupM in \S{}\ref{subsec:lt}. 
However, in this case, by a similar reason to Rem.~\ref{rem:unwillingtoreachNNRepSupM}, 
we have to focus on \emph{lower} $\gamma$-SclSubM.
%
%
%
%
We also note 
that we first have to fix $\gamma\in(0,1)$.
The bigger $\gamma$ we fix, the better bound we can obtain.

\vspace{.3em}
\noindent
\begin{minipage}{\textwidth}
\begin{mytheorem}\label{thm:compresLgammaSclSubM}
If Asm.~\ref{asm:compresL-NNRepSupM} is satisfied,
existence of a linear lower $\gamma$-SclSubM for $\Gamma$ for $\mathfrak{C}$ at $M$ supported by $\mathfrak{I}$ is decidable in polynomial time.
\end{mytheorem}
\end{minipage}

\section{Algorithm for Finding Sub-/Supermartingales}\label{sec:concreteSM}
\subsection{Linear Template-based Algorithm for lower NNRepSupM and upper $\gamma$-SclSubM}\label{subsec:lintemplate}
We explain the algorithm that synthesize an upper L-NNRepSupM using a linear template.
It is almost the same as the ones in~\cite{%
chakarovS13probprog,ChatterjeeNZ17}.
The algorithm first 
fixes a linear template $\mathfrak{f}$ for an L-NNRepSupM as: $\mathfrak{f}(l)=a^l_1x_1+\cdots+a^l_{|V|}x_{|V|}+b^l$
where we let $V=\{x_1,\ldots,x_{|V|}\}$ and $a^l_1,\ldots,a^l_{|V|},b^l$ are unknown coefficients for each $l\in L$.
Let $U:=\{a^l_1,\ldots,a^l_{|V|},b^l\mid l\in L\}$.
We also fix a fresh variable $x_{|V|+1}$, and let $V':=V\cup\{x_{|V|+1}\}$ (this variable is used to deal with nondeterministic assignment).
We then reduce the axioms of upper NNRepSupM to a predicate
over variables in  $V'$ and unknown coefficients in $U$.

It results in a conjunction of constraints of a form $\forall \vec{x}\in\mathbb{R}^{V'}.\;\varphi\Rightarrow \psi$, where 
\begin{itemize}
\item $\varphi$ is a linear conjunctive predicate 
over $V'$;%
\footnote{Here we are relaxing a strict inequality $\mathfrak{a}>0$ to $\mathfrak{a}\geq 0$.
This relaxation does not affect soundness of NNRepSupM, but it affects completeness.
As was done for ARnkSupM in~\cite{chakarovS13probprog},
if we use a generalized theorem called \emph{Motzkin's transposition theorem} (see e.g.~\cite{MTT}) instead of Farkas' lemma, 
then such a relaxation is not necessary.}

\item $\psi$ is a formula of a form 
$p_1x_1+\cdots+p_{|V'|}x_{|V'|}+q\geq 0$ where $p_1,\ldots,p_{|V'|},q$ are linear expressions over the set $U$ 
of unknown coefficients.
\end{itemize}
The concrete construction of those formulas is as follows:

\begin{mydefinition}[cf.~\cite{chatterjeeFNH16algorithmicanalysis,ChatterjeeNZ17}]\label{def:concreteL-NNRepSupM}
Let $\Gamma=(L,V,\linit,\xinit,\trrel,\myUp,\myPr,G)$ be a pCFG,
$\mathfrak{I}$ be a linear predicate map  and $\mathfrak{C}$ be a linear conjunctive predicate map
over $\Gamma$, and $M>0$.
Assume $V=\{x_1,\ldots,x_{|V|}\}$.
Let $x_{|V|+1}$ be a fresh variable such that $x_{|V|+1}\notin V$, and let $V'=V\cup \{x_{|V|+1}\}$.
We define a set $U$ of unknown variables by $U:=\bigcup_{l\in L}\{a^l_1,\ldots,a^l_{|V|},b^l\}$.
We write $\FarkasInput$ for the set of formulas of a  form
$\varphi\Rightarrow \psi$ where:
\begin{itemize}
\item
$\varphi$ is a linear conjunctive predicate (Definition~\ref{def:linpred}) over $V'$ all of whose inequalities are $\leq$; and 

\item $\psi$ is a formula of a form
$p_1x_1+\cdots+p_{|V'|}x_{|V'|}+q\geq 0$
where $p_1,\ldots,p_{|V'|},q$ are linear expressions over $U$.
\end{itemize}
For such $\psi$, we define $\sem{\psi}\subseteq \mathbb{R}^U\times\mathbb{R}^{V'}$ by $\sem{\psi}:=\{(u,v)\mid \sem{p_1}(u)v(x_1)+\cdots+\sem{p_{|V'|}}(u)v(x_{|V'|})+\sem{q}\geq 0\}$. 
For each $l\in L$, let $\mathfrak{I}(l,\vec{x})=\bigvee_{j=1}^{N^{\mathfrak{I}}_l}\bigwedge_{k=1}^{N^{{\mathfrak{I}}\prime}_{l,j}}(\alpha^l_{j,k}\rhd 0)=\bigvee_{j=1}^{N^{\mathfrak{I}}_l}\bigwedge_{k=1}^{N^{{\mathfrak{I}}\prime}_{l,j}}(c^l_{j,k,1}x_1+\cdots+c^l_{j,k,|V|}x_{|V|}+d\rhd 0)$
and ${\mathfrak{C}}(l,\vec{x})=\bigwedge_{i=1}^{N^{\mathfrak{C}}_l}(\beta^l_{i}\rhd 0)$  
 where  $\rhd\in\{\geq,>\}$ and $\alpha^l_{j,k}$ and $\beta^l_{i}$ are linear expressions for each $l$. 
For each $l\in L$, we define $A^l_1,A^l_2,A^l_3\subseteq \FarkasInput$ as follows:
\begin{itemize}
\item $A^l_1=\Bigl\{\bigwedge_{k=1}^{N^{{\mathfrak{I}}\prime}_{l,j}}(\alpha^l_{j,k}\geq 0)\Rightarrow (a_1^lx_1+\cdots+a_{|V|}^lx_{|V|}+b^l\geq 0)\mid 1\leq j\leq N^{\mathfrak{I}}_l\Bigr\}$.

\item $A^l_2=\Bigl\{\bigwedge_{i=1}^{N^{\mathfrak{C}}_l}(\beta^l_i \geq 0)\Rightarrow (a_1^lx_1+\cdots+a_{|V|}^lx_{|V|}+(b^l-M)\geq 0)\Bigr\}$.

\item $A^l_3$ is defined as follows.
\begin{itemize}
\item If $l\in L_N$, 
$A^l_3=\bigcup_{(l,l')\in\trrel_l}\Bigl\{\bigwedge_{k=1}^{N^{{\mathfrak{I}}\prime}_{l,j}}(\alpha^l_{j,k}\geq 0)\wedge(-\beta^l_i \geq 0)\Rightarrow \bigl((a_1^l-a_1^{l'})x_1+\cdots+(a_{|V|}^l-a_{|V|}^{l'})x_{|V|}+(b^l-b^{l'})\geq 0\bigr)\mid 
1\leq i\leq N^{\mathfrak{C}}_l\}$\,.

\item If $l\in L_P$, 
$A^l_3=\Bigl\{\bigwedge_{k=1}^{N^{{\mathfrak{I}}\prime}_{l,j}}(\alpha^l_{j,k}\geq 0)\wedge(-\beta^l_i \geq 0)\Rightarrow \bigl((a_1^l-\sum_{(l,l')\in\trrel_l}\myPr_l(l,l')\cdot a_1^{l'})x_1+\cdots+(a_{|V|}^l-\sum_{(l,l')\in\trrel_l}\myPr_l(l,l')\cdot a_{|V|}^{l'})x_{|V|}+(b^l-\sum_{(l,l')\in\trrel_l}\myPr_l(l,l')\cdot b^{l'})\geq 0\bigr)\mid 
1\leq i\leq N^{\mathfrak{C}}_l\Bigr\}$\,.

\item Let $l\in L_D$. For each $l'\in\mathrm{succ}(l)$, assume $G(l,l')=\bigvee_{m=1}^{N^G_l}\bigwedge_{n=1}^{N^{G\prime}_{l,m}}G_{m,n}(l,l')$\,. 
We let: $A^l_3=\bigcup_{l'\in\mathrm{succ}(l)}\Bigl\{\bigwedge_{k=1}^{N^{{\mathfrak{I}}\prime}_{l,j}}(\alpha^l_{j,k}\geq 0)\wedge(-\beta^l_i \geq 0)\wedge\bigwedge_{n=1}^{N^{G\prime}_{l,m}}G_{m,n}(l,l') \Rightarrow \bigl((a_1^l-a_1^{l'})x_1+\cdots+(a_{|V|}^l-a_{|V|}^{l'})x_{|V|}+(b^l-b^{l'})\geq 0\bigr)\mid 
1\leq i\leq N^{\mathfrak{C}}_l,1\leq m\leq N^G_l\Bigr\}$\,.

\item Let $l\in L_A$ and $\myUp(l)=(v,u)\in  \{1,\ldots,|V|\}\times\mathcal{U}$.
Note that there uniquely exists $l'\in L$ such that $l\trrel l'$.
\begin{itemize}
\item If $u$ is a  measurable function $\mathbb{R}^V\to \mathbb{R}$ s.t.\ 
$u=\sem{r_1x_1+\cdots+r_{|V|}x_{|V|}+r}$ where $r_1,\ldots,r_{|V|},r\in\mathbb{R}$,
then $A^l_3=\Bigl\{\bigwedge_{k=1}^{N^{{\mathfrak{I}}\prime}_{l,j}}(\alpha^l_{j,k}\geq 0)\wedge(-\beta^l_i \geq 0)
 \Rightarrow \bigl((a_1^l-a_1^{l'}-r_1a_v^{l'})x_1+\cdots+(a_{v-1}^l-a_{v-1}^{l'}-r_{v-1}a_v^{l'})x_{v-1}+(a_v^l-r_va_v^{l'})x_v+(a_{v+1}^l-a_{v+1}^{l'}-r_{v+1}a_v^{l'})x_{v+1}+\cdots+(a_{|V|}^l-a_{|V|}^{l'}-r_{|V|}a_v^{l'})x_{|V|}+(b^l-b^{l'}-ra_v^{l'})\geq 0\bigr)\mid 
1\leq i\leq N^{\mathfrak{C}}_l\Bigr\}$\,.

\item If $u$ is a  distribution on $\mathbb{R}$ s.t.\ $\mathbb{E}u=r\in\mathbb{R}$, then
$A^l_3=\Bigl\{\bigwedge_{k=1}^{N^{{\mathfrak{I}}\prime}_{l,j}}(\alpha^l_{j,k}\geq 0)\wedge(-\beta^l_i \geq 0)\Rightarrow \bigl((a_1^l-a_1^{l'})x_1+\cdots+(a_{v-1}^l-a_{v-1}^{l'})x_1+a_v^lx_1+(a_{v+1}^l-a_{v+1}^{l'})x_{v+1}+\cdots+(a_{|V|}^l-a_{|V|}^{l'})x_{|V|}+
(b^l-b^{l'}-ra_v^l)\geq 0\bigr)\mid 
1\leq i\leq N^{\mathfrak{C}}_l\}$\,.

\item Let $u$ be a set s.t.\ $u=\sem{\mathfrak{P}}$ for a linear predicate $\mathfrak{P}=\bigvee_{t=1}^{T}\bigwedge_{t'=1}^{T'}(b_{t,t'}x_v+b'_{t,t'}\geq 0)$ over $\{x_v\}$ (see Asm.~\ref{asm:compresL-NNRepSupM}). 
%
Then we let 
$A^l_3=\Bigl\{\bigwedge_{k=1}^{N^{{\mathfrak{I}}\prime}_{l,j}}(\alpha^l_{j,k}\geq 0)\wedge(-\beta^l_i \geq 0)\wedge \bigwedge_{t'=1}^{T'}(b_{t,t'}x_{|V|+1}+b'_{t,t'}\geq 0) \Rightarrow \bigl((a_1^l-a_1^{l'})x_1+\cdots+(a_{v-1}^l-a_{v-1}^{l'})x_{v-1}+a_v^lx_v+(a_{v+1}^l-a_{v+1}^{l'})x_{v+1}+\cdots+(a_{|V|}^l-a_{|V|}^{l'})x_{|V|}+(-a_v^{l'})x_{|V|+1}+
(b^l-b^{l'})\geq 0\bigr)\mid 
1\leq i\leq N^{\mathfrak{C}}_l, 1\leq m\leq N^u_l, 1\leq t\leq T \Bigr\}$\,.
\end{itemize}
\end{itemize}
\end{itemize}
\end{mydefinition}

\begin{myproposition}\label{prop:concreteL-NNRepSupM}
We assume the situation in Definition~\ref{def:concreteL-NNRepSupM}.
Then for  $\vec{u}\in\mathbb{R}^U$, we have:
%
\begin{align*}
&\left(\begin{aligned}
\forall \vec{x}\in\mathbb{R}^{V'}.\;
\forall l\in L.\; 
\forall (\varphi\Rightarrow \psi)\in \textstyle{\bigcup_{l\in L}}\; A^l_1\cup A^l_2\cup A^l_3 .\;  \vec{x}\in\sem{\varphi}\Rightarrow(\vec{u},\vec{x})\in\sem{\psi}
\end{aligned}\right)\\
&\Rightarrow\;
\left(\begin{aligned}
&\text{If we let $\mathfrak{f}=\{\vec{u}(a_1^l)\cdot x_1+\cdots+\vec{u}(a_{|V|}^l)\cdot x_{|V|}+b^l\}_{l\in L}$,} \\
&\text{then $\sem{\mathfrak{f}}$ is an upper NNRepSupM
for $\sem{\mathfrak{C}}$ at $M$ supported by $\sem{\mathfrak{I}}$}
\end{aligned}\right).
\end{align*}
\end{myproposition}

Using matrices, we can rewrite formulas of a form $\forall \vec{x}\in\mathbb{R}^{V'}.\;\varphi\Rightarrow \psi$ as follows:
\begin{equation}\label{eq:1801262015}
\forall \vec{x}\in\mathbb{R}^{V'}.\; 
\vec{A}\vec{x}\leq \vec{b}\,\Rightarrow\,
\vec{c}^T\vec{x}\leq d\,.
\end{equation}
Here $\vec{A}\in\mathbb{R}^{V'\times m}$ is a matrix and $\vec{b}\in\mathbb{R}^m$ is a vector whose elements are real numbers, 
while $\vec{c}$ is a vector and $d$ is a scalar  whose elements are linear expressions over $U$.
The goal is to find a valuation $U\to \mathbb{R}$ that makes (\ref{eq:1801262015}) hold.

By Farkas' lemma (see e.g.~\cite{boydV04ConvexOptimization}), if 
$\{\vec{x}\mid \vec{A}\vec{x}\leq \vec{b}\}$ is not empty then 
the inequality (\ref{eq:1801262015}) is satisfied if and only if
there exists a column vector $\vec{y}=(y_1,\ldots,y_m)^T\geq 0$ over $\mathbb{R}$ 
such that $\vec{A}^T\vec{y}=\vec{c}$ and $\vec{b}^T\vec{y}\leq d$.
Hence the problem is reduced to a satisfiability problem of 
linear inequalities over variables $\{y_1,\ldots,y_m\}\cup U$,
which
can be checked 
in polynomial time using a linear programming solver (LP solver).

\subsection{Polynomial Template-based Algorithm for lower NNRepSupM and upper $\gamma$-SclSubM}\label{subsec:polytemplate}
The algorithm almost the same as the ones~\cite{ChatterjeeFG16,chakarovVS16TACAS}.
The key 
is the use of a theorem called \emph{Positivstellensatz} (see e.g.~\cite{scheiderer09guide}).
Several variants are known for Positivstellensatz, and three of them were used in~\cite{ChatterjeeFG16}.
Among them, we use a variant called \emph{Schm\"udgen's Positivstellensatz}.

We first fix a maximum degree $d\in\mathbb{N}$ of the template, and fix
a polynomial template $\mathfrak{f}=\{\mathfrak{f}(l)=\sum_{h\in\mathcal{M}_d}a^l_h\cdot h\}_{l\in L}$ for 
an U-NNRepSupM.
Here 
$\mathcal{M}_d$ denotes the set of monomials whose degrees are no greater than $d$,
and each $a^l_h$ is an unknown coefficient. 
We let $U=\{a^l_h\mid l\in L, h\in\mathcal{M}_d\}$.
In a similar manner to the linear case, 
we can reduce the axioms of U-NNRepSupM into a conjunction of predicates of a form
$\forall \vec{x}\in\mathbb{R}^V.\;\varphi\Rightarrow \psi$,
where $\varphi$ is a polynomial conjunctive predicate $\bigwedge_{i=1}^m \mathfrak{g}_i\geq 0$ over $V':=V+\{x_{|V|+1}\}$ and 
$\psi$ is a formula of a form 
$\mathfrak{g}=p_1h_1+\cdots+p_{t}h_t\geq 0$ with $p_1,\ldots,p_t$ being polynomial expressions over 
$U$, and
$h_1,\ldots,h_t$ are monomials over $V'$.

To find a valuation $U\to\mathbb{R}$ satisfying
$\forall \vec{x}\in\mathbb{R}^{V'}.\;\psi\Rightarrow \varphi$,
we make use of a notion of \emph{sum-of-square polynomial expression}.
It is a polynomial expression of a form 
$\mathfrak{h}=\sum_{i=1}^s\mathfrak{i}_i^2$ for some polynomial expressions 
$\mathfrak{i}_1,\ldots,\mathfrak{i}_s$. 
It is easy to see that $\forall \vec{x}\in\mathbb{R}^V.\;\sem{\mathfrak{h}}(\vec{x})\geq 0$.
Therefore if 
there exists a family $\{\mathfrak{h}_{w_1\ldots w_m}\}_{w_1,\ldots,w_m\in \{0,1\}}$ of
sum-of-square polynomial expressions s.t.\
$\mathfrak{g}=\sum_{w_1,\ldots, w_m\in\{0,1\}}\mathfrak{h}_{w_1\ldots w_m}\cdot \prod_{j=1}^m \mathfrak{g}_i^{w_i}$, then
we have $\forall \vec{x}\in\mathbb{R}^V.\;\varphi\Rightarrow \psi$ (i.e.\ 
$\forall \vec{x}\in\mathbb{R}^V.\;(\bigwedge_{i=1}^m \mathfrak{g}_i\geq 0)\Rightarrow (\mathfrak{g}\geq 0)$).

Schm\"udgen's Positivstellensatz tells us that under certain conditions, its converse holds.
\begin{mytheorem}[\cite{Schmudgen1991}]\label{thm:SchmudgenPSS}
Let $\mathfrak{g},\mathfrak{g}_1,\ldots,\mathfrak{g}_m$ be polynomial expressions over $V$.
If $\sem{\bigwedge_{i=1}^m \mathfrak{g}_i\geq 0}\subseteq\mathbb{R}^V$ is compact,
then 
$\forall\vec{x}\in \sem{\bigwedge_{i=1}^m \mathfrak{g}_i\geq 0}.\;\sem{g}(\vec{x})>0$ iff
there exists a family $\{\mathfrak{h}_w\}_{w\in \{0,1\}^m}$ of sum-of-square polynomial expressions
such that $\mathfrak{g}=\sum_{w\in\{0,1\}^m}\mathfrak{h}_w\cdot \prod_{j=1}^m \mathfrak{g}_i^{w_i}$\,.
\qed
\end{mytheorem}

To ensure that a polynomial expression is sum-of-square, we can use its well-known characterization
(see e.g.~\cite{Horn:2012:MA:2422911}):
a polynomial expression $\mathfrak{h}$ whose degree is no greater than $2k$ is sum-of-square 
iff there exists a positive semidefinite matrix $A\in\mathbb{R}^{|\mathcal{M}_k|\times |\mathcal{M}_k|}$ such that 
$\mathfrak{h}=\vec{y}_k^TA\vec{y}_k$,
where 
$\vec{y}_k$ is the column vector whose components consist of the monomials in $\mathcal{M}_k$. 

To summarize, to check existence of a valuation $U\to\mathbb{R}$
satisfying
$\forall \vec{x}\in\mathbb{R}^V.\;\varphi\Rightarrow \psi$,
it suffices to fix a maximum degree $k\in\mathbb{N}$ for sum-of-square polynomial expressions, 
fix additional unknown parameters $U':=\{a^l_{w,i,j}\mid l\in L, w\in\{0,1\}^m, 1\leq i,j\leq |\mathcal{M}_k|\}$ and
find a valuation $U\cup U'\to\mathbb{R}$ that satisfies the following conditions:
\begin{itemize}
\item for each $w\in\{0,1\}^m$, the following matrix is positive semidefinite:
\[
A^l_w:=\begin{pmatrix}a^l_{w,1,1} & \ldots & a^l_{w,1,|\mathcal{M}_k|+1} \\
\vdots & \ddots & \vdots \\ a^l_{w,|\mathcal{M}_k|+1,1} & \ldots & a^l_{w,|\mathcal{M}_k|+1,|\mathcal{M}_k|+1}
\end{pmatrix}\,.
\]

\item $\mathfrak{g}=\sum_{w\in\{0,1\}^m}\vec{y}_k^TA_w\vec{y}_k\cdot \prod_{j=1}^m \mathfrak{g}_i^{w_i}$ where $\vec{y}_k$ is defined as above.
\end{itemize}
By comparing the coefficients appearing in the equality in the latter condition,
we can obtain a family of linear equalities over $U\cup U'$. 
Therefore polynomial template-based synthesis of an U-NNRepSupM is reduced to a 
\emph{semidefinite programming} (SDP) problem, which is solvable using an SDP solver.

Conditioned polynomial constraints for U-NNRepSupM 
are analogous to Def.~\ref{def:concreteL-NNRepSupM}.
Conditioned linear/polynomial constraints for lower $\gamma$-SclSubM are similar.

\section{Probabilistic Programs Used in Experiments}\label{sec:ppused}
\begin{wrapfigure}[4]{r}{5.4cm}
\vspace{-.4cm}
\hspace*{8mm}
\begin{minipage}{6cm}
\begin{lstlisting}[numbers=left]
{$0 \leq x$} while true do 
{$0 \leq x$}     $\hspace{3mm}$$x := x - 1$
{$0 \leq x$}     $\hspace{3mm}$refute ($x\leq 0$)
$\hphantom{\{0 \leq x\}}$  od
\end{lstlisting}
\end{minipage}
\hspace*{-5mm}
\end{wrapfigure}
We present codes of probabilistic programs used in experiments in \S{}\ref{sec:implexp}.
We have augmented the syntax of APP and PPP (\S{}\ref{sec:pp}) with two components ``$\{...\}$'' and ``$\mathtt{refute}$'' 
for specifying an invariant $I$ and a terminal configuration $C$ respectively.
An example of a program code is shown on the right. 
If we feed this code to Prog.~\ref{item:impl1} for example, then it underapproximates $\upreach_{C}(c)$ where the line $3$ is 
reached when $x\leq 0$, under an assumption that $x\geq 0$ holds in each program location.

\subsubsection{(a-1) (1D adversarial random walk~\cite{chatterjeeFNH16algorithmicanalysis})}\mbox{}
This models a \emph{discrete queuing system} that starts with $2$ packets in the queue, and in each round $0$, $1$ or $2$ packets are queued 
in probability $p_1$, $p_2$ and $1-p_1-p_2$ respectively. Each packet is processed depending on its type that is nondeterministically 
chosen: if it is ``urgent,'' then
the packet is immediately processed in one round in probability $1/8$, but in probability $7/8$ it fails to process the packet and produces another 
packet. If the type is ``standard,'' then it is processed using two rounds in probability $1$. The process terminates if queue 
gets empty, and aborts (it is not counted as termination) if the queue is full.

\begin{lstlisting}[numbers=left]
{ true }               x := 2;
{ 0 <= x and x <= 13 } while x <= 10 do
{ 0 <= x and x <= 10 }   if prob($p_1$) then
{ 0 <= x and x <= 10 }     skip
                         else
{ 0 <= x and x <= 10 }     if prob($\frac{p_2}{1-p_1}$) then
{ 0 <= x and x <= 10 }       x := x + 1
                           else
{ 0 <= x and x <= 10 }       x := x + 2
                           fi
                         fi;
{ 0 <= x and x <= 12 }   if * then
{ 0 <= x and x <= 12 }     if prob(0.875) then
{ 0 <= x and x <= 12 }       x := x - 1
                           else
{ 0 <= x and x <= 12 }       skip
                           fi
                         else
{ 0 <= x and x <= 12 }     if prob($p_1$) then
{ 0 <= x and x <= 12 }       skip
                           else
{ 0 <= x and x <= 12 }       if prob($\frac{p_2}{1-p_1}$) then
{ 0 <= x and x <= 12 }         x := x + 1
                             else
{ 0 <= x and x <= 12 }         x := x + 2
                             fi
                           fi;
{ 0 <= x and x <= 14 }     x := x - 1
                         fi;
{ 0 <= x and x <= 13 }   refute (x <= 0)
                       od
\end{lstlisting}
Here $p_1,p_2\in[0,1]$.

\subsubsection{(a-2) (2D adversarial random walk~\cite{chatterjeeFNH16algorithmicanalysis})}\mbox{}
It is a random walk over $\mathbb{R}^2$ with a time limit of $100$ rounds. 
The process starts from $(x,y)=(2,2)$ and in each 
round, either ``$x$'' or ``$y$'' is nondeterministically chosen, and it is added by $z\in\mathbb{R}$ that is uniformly chosen from $[M_1,M_2]$. The process terminates if $x\leq 0$ or $y\leq 0$.


\begin{lstlisting}[numbers=left]
{ true }                                        
  x := 2;
{ x = 2 }                                       
  y := 2;
{ x = 2 and y = 2 }                             
  t := 0;
{ 0 <= x and 0 <= y and 0 <= t and t <= 101 }   
  while t <= 100 do
{ 0 <= x and 0 <= y and 0 <= t and t <= 100 }   
    if * then
{ 0 <= x and 0 <= y and 0 <= t and t <= 100 }   
      z := Unif ($M_1$,$M_2$);
{ 0 <= x and 0 <= y and $M_1$ <= z and z <= $M_2$ and 0 <= t and t <= 100 }          
      x := x + z
    else
{ 0 <= x and 0 <= y and 0 <= t and t <= 100 }       
      z := Unif ($M_1$,$M_2$);
{ 0 <= x and 0 <= y and $M_1$ <= z and z <= $M_2$ and 0 <= t and t <= 100 }              
      y := y + z
    fi;
{ $M_1$ <= x and $M_1$ <= y and 0 <= t and t <= 100 }   
    t := t + 1;
{ $M_1$ <= x and $M_1$ <= y and 1 <= t and t <= 101 }   
    refute (x <= 0);
{ 0 <= x and $M_1$ <= y and 1 <= t and t <= 101 }    
    refute (y <= 0)
  od
\end{lstlisting}
Here $M_1,M_2\in\mathbb{R}$.

\subsubsection{(a-3) (variant of 2D adversarial random walk~\cite{chatterjeeFNH16algorithmicanalysis})}\mbox{}
It is a random walk over $\mathbb{R}^2$ with a time limit of $100$ rounds. 
The process starts from $(x,y)=(3,2)$, and in each round either of the following is
nondeterministically chosen: (1) $z\in\mathbb{R}$ is uniformly chosen from $[M_1,M_2]$ and added to $x$ (resp.\ $y$) 
in probability $0.7$ (resp.\ $0.3$); or (2)  $z\in\mathbb{R}$ is uniformly chosen from $[-M_2,-M_1]$ and added to $y$ 
(resp.\ $x$) in probability $0.7$ (resp.\ $0.3$).

\begin{lstlisting}[numbers=left]
{ true }                             
  x := 3;
{ x = 3 }                            
  y := 2;
{ x = 3 and y = 2 }                  
  t := 0;
{ x >= y and 0 <= t and t <= 101 }   
  while t <= 100 do
{ x >= y and 0 <= t and t <= 100 }       
    if * then
{ x >= y and 0 <= t and t <= 100 }         
      if prob(0.7) then
{ x >= y and 0 <= t and t <= 100 }           
        z := Unif ($M_1$,$M_2$);
{ x >= y and 0 <= t and t <= 100 and $M_1$ <= z and z <= $M_2$ }            
        x := x + z
      else
{ x >= y and 0 <= t and t <= 100 }           
        z := Unif ($M_1$,$M_2$);
{ x >= y and 0 <= t and t <= 100 and $M_1$ <= z and z <= $M_2$ }            
        y := y + z
      fi
	else
{ x >= y and 0 <= t and t <= 100 }         
      if prob(0.7) then
{ x >= y and 0 <= t and t <= 100 }           
        z := Unif ($-M_2$,$-M_1$);
{ x >= y and 0 <= t and t <= 100 and $-M_2$ <= z and z <= $-M_1$ }          
        y := y + z
	  else
{ x >= y and 0 <= t and t <= 100 }           
        z := Unif ($-M_2$,$-M_1$);
{ x >= y and 0 <= t and t <= 100 and $-M_2$ <= z and z <= $-M_1$ }
        x := x + z
      fi
    fi;
{ x >= y + 2 and 0 <= t and t <= 100 }   
    t := t + 1;
{ x >= y + 2 and 1 <= t and t <= 101 }
    refute (x <= y)
  od
\end{lstlisting}
Here $M_1,M_2\in\mathbb{R}$.

\subsubsection{\ref{item:experiment3} (room temperature control~\cite{abateKLP10,chakarovVS16TACAS})}\mbox{}
The rooms are named Room~1 and Room~2, and we write $x_1$ and $x_2$ for their temperature, with a time limit.
The changes of $x_1$ and $x_2$ are modeled by the following stochastic difference equation:
$x'_i = x_i+ a\cdot (x_j-x_i)+b_i\cdot(x_0-x_i)+c_i\cdot\sigma(x_i)+\nu_i$ where $i,j\in\{1,2\}$ and $j\neq i$.
The second term models the heat exchange between the rooms and 
the third term models the heat exchange with the outside whose temperature is $x_0$.
The forth term models the influence of the air conditioner that is controlled by a function $\sigma:\mathbb{R}\to\mathbb{R}$.
The last term $\nu_i$ is a probabilistic distribution that models the perturbation.
Partially referring to~\cite{chakarovVS16TACAS}, 
we have fixed the parameters as $a=0.0625$, $b_1=0.0375$, $b_2=0.025$,
$c_1=c_2=c$, $\sigma(x)=19.5-x$ and $\nu_1,\nu_2$ independently follow the uniform distribution over $[-p,p]$, where $c,p\in\mathbb{R}$.
The process terminates if $x_1$ or $x_2$ is out of the comfortable range: $[17,22]$ for $x_1$ and $[16,23]$ for $x_2$.

\begin{lstlisting}[numbers=left]
{ true }
  x0 := 6;
{ x0 = 6 }
  x1 := ndet Real[17,22];
{ x0 = 6 and 17 <= x1 and x1 <= 22 }
  x2 := ndet Real[16,23];
{ x0 = 6 and 17 <= x1 and x1 <= 22 and 16 <= x2 and x2 <= 23 }
  t := 0;
{ x0 = 6 and 17 <= x1 and x1 <= 22 and 16 <= x2 
  and x2 <= 23 and 0 <= t and t <= 101 }
  while t <= 100 do
{ x0 = 6 and 17 <= x1 and x1 <= 22 and 16 <= x2 
  and x2 <= 23 and 0 <= t and t <= 100 }
    controller1 := 19.5 - x1;
{ x0 = 6 and 17 <= x1 and x1 <= 22 and 16 <= x2 
  and x2 <= 23 and -2.5 <= controller1 
  and controller1 <= 2.5 and 0 <= t and t <= 100 }
    controller2 := 19.5 - x2;
{ x0 = 6 and 17 <= x1 and x1 <= 22 and 16 <= x2 
  and x2 <= 23 and -2.5 <= controller1 
  and controller1 <= 2.5 and -3.5 <= controller2 
  and controller2 <= 3.5 and 0 <= t and t <= 100 }
    noise1 := Unif(-p,p);
{ x0 = 6 and 17 <= x1 and x1 <= 22 and 16 <= x2 and x2 <= 23 
  and -2.5 <= controller1 and controller1 <= 2.5 and -3.5 <= controller2 
  and controller2 <= 3.5 and -p <= noise1 
  and noise1 <= p and 0 <= t and t <= 100 }
    noise2 := Unif(-p,p);
{ x0 = 6 and 17 <= x1 and x1 <= 22 and 16 <= x2 and x2 <= 23 
  and -2.5 <= controller1 and controller1 <= 2.5 and -3.5 <= controller2 
  and controller2 <= 3.5 and -p <= noise1 and noise1 <= p 
  and -p <= noise2 and noise2 <= p and 0 <= t and t <= 100 }
    x1 := x1 + 0.0375 * x0 - 0.0375 * x1 + 0.0625 * x2 
           - 0.0625 * x1 + c * controller1 + noise1;
{ x0 = 6 and 15 <= x1 and x1 <= 24 and 16 <= x2 and x2 <= 23 
  and -3.5 <= controller2 and controller2 <= 3.5 and -p <= noise2 
  and noise2 <= p and 0 <= t and t <= 100 }
    x2 := x2 + 0.025 * x0 - 0.025  * x2 + 0.0625 * x1 
           - 0.0625 * x2 + c * controller2 + noise2; 
{ x0 = 6 and 14 <= x1 and x1 <= 25 and 13 <= x2 
  and x2 <= 26 and 0 <= t and t <= 100 }
    t := t + 1;
{ x0 = 6 and 14 <= x1 and x1 <= 25 and 13 <= x2 
  and x2 <= 26 and 0 <= t and t <= 101 }
    refute (x1 < 17);
{ x0 = 6 and 17 <= x1 and x1 <= 25 and 13 <= x2 
  and x2 <= 26 and 0 <= t and t <= 101 }
    refute (x1 > 22);
{ x0 = 6 and 17 <= x1 and x1 <= 22 and 13 <= x2 
  and x2 <= 26 and 0 <= t and t <= 101 }
    refute (x2 < 16);
{ x0 = 6 and 17 <= x1 and x1 <= 22 and 16 <= x2 
  and x2 <= 26 and 0 <= t and t <= 101 }
    refute (x2 > 23)
  od
\end{lstlisting}
Here $c,p\in\mathbb{R}$.

 \auxproof{
\subsubsection{\ref{item:experiment4} (simple pendulum~\cite{steinhardtT12IJRR})}\mbox{}
The model has two variables $\theta$ and $\dot{\theta}$ that represent the angle of the pendulum and
its time derivative respectively.
The dynamics is modeled by the following stochastic difference equation. 
\[
\begin{pmatrix} \theta' \\ \dot{\theta}' \end{pmatrix}
=
\begin{pmatrix} \theta + 0.01 \cdot\dot{\theta} \\ -0.0167\cdot \theta^3-0.3\cdot\theta+0.97\cdot \dot{\theta}\end{pmatrix}
+
\begin{pmatrix} 0.01\cdot w_1 \\ 0.05\cdot w_2\end{pmatrix}
\]
Here $w_1$ and $w_2$ follow the Normalized distribution $\mathcal{N}(0,1)$
The process terminates if the angle exceeds $\frac{\pi}{6}$
after $3600$ rounds are consumed.
We fixed no invariant.

\begin{lstlisting}[numbers=left]
theta1 := $\theta$;
dt_theta := 0;
t := 0;
while true do
  w1 := Norm (0,1);
  theta1 := theta1 + 0.01 * dt_theta + 0.01 * w1;
  w2 := Norm (0,1);
  dt_theta := -0.0167 * theta1^3 - 0.3 * theta1 
              + 0.97 * dt_theta + 0.05 * w2;
  t := t + 0.01;
  refute (theta1 >= 0.52359877559 and t >= 3600)
od
\end{lstlisting}
Here $\theta\in\mathbb{R}$.
 }

\section{pCFGs Used in Experiments}\label{sec:pcfgused}
The pCFGs $\mathcal{C}=(L,V,\linit,\xinit,\trrel,\myUp,\myPr,G)$ used to compare NNRepSupM and $\varepsilon$-RepSupM were
as follows:
\subsubsection{(d-1) (bounded random walk over $[0,10]$)}\mbox{}
\begin{itemize}
\item $L_N=\emptyset$, $L_P=\{l_1\}$, $L_D=\{l_0\}$ and $L_A=\{l_2,l_3,l_4\}$.

\item $V=\{x\}$.

\item $\linit=l_0$

\item $\xinit=[x=5]$.

\item $\trrel=\{(l_0,l_1),(l_0,l_4),(l_1,l_2),(l_1,l_3),(l_2,l_0),(l_3,l_0),(l_4,l_4)\}$.

\item $\myUp(l_2)=(x,\lambda x.\,x-1)$,
$\myUp(l_3)=(x,\lambda x.\, x+1)$ and
$\myUp(l_4)=(x,\lambda x.\, x+1)$.

\item $\myPr(l_1,l_2)=0.1$, $\myPr(l_1,l_3)=0.9$.

\item $G(l_0,l_1)=(x<10)$ and $G(l_0,l_4)=(x\geq 10)$.
\end{itemize}
Moreover, let  $C:=\{(l_1,[x\mapsto r])\mid r\leq 0\}\subseteq L\times \mathbb{R}^V$, and let $\mathfrak{C}$ be a linear predicate map representing $C$.
The pCFG
represents a \emph{bounded random walk},
and can be illustrated as follows (for simplicity, we are omitting some configurations).
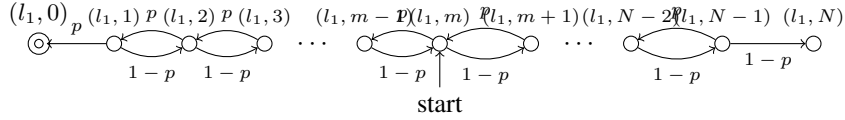
\[
\begin{tikzpicture}[node distance=1.0cm,  bend angle=30, accepting/.style={double distance=2.0pt}, state/.style={circle, draw, inner sep=2pt, minimum size=0pt}]
\node [state, accepting, label={above:\small$(l_1,0)$}, minimum size=0pt] (x_0)       {};
\node [state, label={above:\scriptsize$(l_1,1)$}, minimum size=0pt] (x_1) [right of = x_0] {};
\node [state, label={above:\scriptsize$(l_1,2)$}, minimum size=0pt] (x_2) [right of = x_1] {};
\node [state, label={above:\scriptsize$(l_1,3)$}, minimum size=0pt] (x_3) [right of = x_2] {};
\node        (x_dots) [right=0.2cm of x_3] {$\cdots$}; 
\node [state, label={above :\scriptsize$(l_1,m-1)$}, minimum size=0pt] (x_m0) [right =0.2cm  of x_dots] {};
\node [state, initial below, label={above:\scriptsize$(l_1,m)$}, minimum size=0pt] (x_m1) [right of = x_m0] {};
\node [state, label={above :\scriptsize$(l_1,m+1)$}, minimum size=0pt] (x_m2) [right =of x_m1] {};
\node        (x_dots2) [right=0.2cm of x_m2] {$\cdots$}; 
\node [state, label={above :\scriptsize$(l_1,N-2)$}, minimum size=0pt] (x_N2) [right =0.2cm of x_dots2] {};
\node [state, label={above :\scriptsize$(l_1,N-1)$}, minimum size=0pt] (x_N1) [right = of x_N2] {};
\node [state, label={above :\scriptsize$(l_1,N)$}, minimum size=0pt] (x_N) [right = of x_N1] {};
\path [->] (x_1) edge             node             [above]  {\scriptsize$p$}     (x_0)
           (x_1) edge [bend right] node             [below]  {\scriptsize$1-p$}   (x_2)
           (x_2) edge [bend right] node             [above]  {\scriptsize$p$}     (x_1)
           (x_2) edge [bend right] node             [below]  {\scriptsize$1-p$}   (x_3)
           (x_3) edge [bend right] node             [above]  {\scriptsize$p$}     (x_2)
           (x_m0) edge [bend right] node             [below]  {\scriptsize$1-p$}   (x_m1)
           (x_m1) edge [bend right] node             [above]  {\scriptsize$p$}     (x_m0)
           (x_m1) edge [bend right] node             [below]  {\scriptsize$1-p$}   (x_m2)
           (x_m2) edge [bend right] node             [above]  {\scriptsize$p$}     (x_m1)
           (x_N2) edge [bend right] node             [below]  {\scriptsize$1-p$}   (x_N1)
           (x_N1) edge [bend right] node             [above]  {\scriptsize$p$}     (x_N2)           
           (x_N1) edge             node             [below]  {\scriptsize$1-p$}     (x_N);
\end{tikzpicture}\,.
\]

\subsubsection{(d-2) (simple system with infinite branching)}\mbox{}
\begin{itemize}
\item $L_N=\emptyset$, $L_P=\{l_4,l_5\}$, $L_D=\{l_2\}$ and $L_A=\{l_1,l_3,l_6\}$.

\item $V=\{x\}$.

\item $\linit=l_1$

\item $\xinit=[x=0]$.

\item $\trrel=\{(l_1,l_2),(l_2,l_3),(l_2,l_4),(l_3,l_2),(l_4,l_5),(l_4,l_6),(l_5,l_5),(l_6,l_6)\}$.

\item $\myUp(l_1)=(x,\mathrm{Geometric}(0.5))$ and
$\myUp(l_3)=\myUp(l_6)=(x,\lambda x.\, x-1)$
 where $\mathrm{Geometric}(0.5)$ denotes a geometric distribution that takes a value $i\in\mathbb{N}\setminus\{0\}$ in a probability $0.5^i$.

\item $\myPr(l_4,l_5)=\myPr(l_4,l_6)=0.5$, $\myPr(l_5,l_5)=1$.

\item $G(l_2,l_3)=(x\geq 1)$ and $G(l_2,l_4)=(x< 1)$.  
\end{itemize}
Moreover, let $C=\{(l_5,[x\mapsto r])\mid r\in\mathbb{R}\}\subseteq L\times\mathbb{R}^V$, and assume that $\mathfrak{C}$ is a linear predicate map representing $C$.
It can be illustrated as follows (for simplicity, we are omitting or merging some configurations).
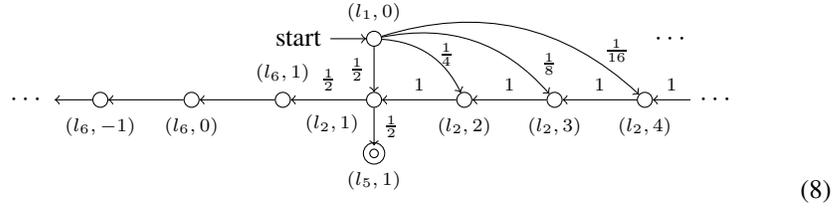
\begin{equation}\label{eq:1801311401}
\begin{tikzpicture}[node distance=1.2cm,  bend angle=30, accepting/.style={double distance=2.0pt}, state/.style={circle, draw, inner sep=2pt, minimum size=0pt}]
\node [state, initial left, label={above:\scriptsize$(l_1,0)$}] (a_1) {};
\node        (c_dots) [right=3.5cm of a_1] {$\cdots$}; 
\node [state, label={below left:\scriptsize$(l_2,1)$}, minimum size=0pt] [below = 6mm of a_1] (x_0) {};
\node [state, label={above:\scriptsize$(l_6,1)$}, minimum size=0pt] [left = 10mm of x_0] (x_100) {};
\node [state, label={below:\scriptsize$(l_6,0)$}, minimum size=0pt] [left = 10mm of x_100] (x_101) {};
\node [state, label={below:\scriptsize$(l_6,-1)$}, minimum size=0pt] [left = 10mm of x_101] (x_102) {};
\node        (x_dots2) [left=0.5cm of x_102] {$\cdots$}; 
\node [state, label={below:\scriptsize$(l_2,2)$}, minimum size=0pt] (x_1) [right of = x_0] {};
\node [state, label={below:\scriptsize$(l_2,3)$}, minimum size=0pt] (x_2) [right of = x_1] {};
\node [state, label={below:\scriptsize$(l_2,4)$}, minimum size=0pt] (x_3) [right of = x_2] {};
\node        (x_dots) [right=0.5cm of x_3] {$\cdots$}; 
%
\node [state, accepting, label={below:\scriptsize$(l_5,1)$}, minimum size=0pt] [below = 5mm of x_0] (y) {};
\path [->] (a_1) edge node [left] {\scriptsize$\frac{1}{2}$}     (x_0)
           (a_1) edge [bend left] node [right] {\scriptsize$\frac{1}{4}$}     (x_1)
           (a_1) edge [bend left] node [xshift=1.0cm, yshift=-0.3cm] {\scriptsize$\frac{1}{8}$}     (x_2)
           (a_1) edge [bend left] node [xshift=1.3cm, yshift=-0.3cm] {\scriptsize$\frac{1}{16}$}     (x_3)
           (x_0) edge node [right] {\scriptsize$\frac{1}{2}$}     (y)
           (x_0) edge  node             [above]  {\scriptsize$\frac{1}{2}$}   (x_100)
           (x_100) edge  node             [above]  {\scriptsize}   (x_101)
           (x_101) edge  node             [above]  {\scriptsize}   (x_102)  
          (x_102) edge  node             [above]  {\scriptsize}   (x_dots2)                      
           (x_1) edge  node             [above]  {\scriptsize$1$}   (x_0)
           (x_2) edge  node             [above]  {\scriptsize$1$}   (x_1)
           (x_3) edge  node             [above]  {\scriptsize$1$}   (x_2)           
           (x_dots) edge  node             [above]  {\scriptsize$1$}   (x_3);
\end{tikzpicture}
\end{equation}

\begin{myremark}\label{rem:noRepSM2}
In fact, there exists no RepSupM that can refute almost-sure termination.
Let $\eta:L\times\mathbb{R}^V\to\mathbb{R}$ be an $\varepsilon$-RepSupM for $C$
that has $\kappa$-bounded differences for some $\kappa>0$.
Assume that $f(l_1,0)=A<0$.
By Def.~\ref{def:EpsRepSupM}, we have:
\[
\eta(l_2,1)\leq \eta(l_3,2)-\varepsilon\leq \eta(l_2,2)-2\varepsilon \leq \eta(l_3,3)-3\varepsilon\leq \eta(l_2,3)-4\varepsilon
\leq\cdots\,.
\]
Hence $\eta(l_2,n)\geq \eta(l_2,1)+2(n-1)\varepsilon$, and therefore $\lim_{n\to\infty}\eta(l_2,n)=\infty$.
This contradicts to the $\kappa$-bounded condition.

In contrast, 
if we define a linear expression map $\mathfrak{f}$ and a linear predicate map $\mathfrak{I}$ by 
$\mathfrak{f}(l_1)=\mathfrak{f}(l_2)=\mathfrak{f}(l_3)=\mathfrak{f}(l_4)=\frac{1}{2}$, 
$\mathfrak{f}(l_5)=1$, $\mathfrak{f}(l_6)=0$ and
$\mathfrak{I}(l)=\true$ for each $l\in L$, then $\mathfrak{f}$ is a linear U-NNRepSupM at $\mathfrak{C}$ and $1$ supported by $\mathfrak{I}$.
\end{myremark}

\subsubsection{(d-3) (random walk over $[0,1]$ that exhibits geometric behaviors)}\mbox{}
\begin{itemize}
\item $L_N=\emptyset$, $L_P=\{l_3\}$, $L_D=\{l_2\}$ and $L_A=\{l_1,l_4,l_5,l_6\}$.

\item $V=\{x\}$.

\item $\linit=l_1$

\item $\xinit=[x=0]$.

\item $\trrel=\{(l_1,l_2),(l_2,l_3),(l_2,l_6),(l_3,l_4),(l_3,l_5),(l_4,l_2),(l_5,l_2),(l_6,l_6)\}$.

\item $\myUp(l_1)=(x,\mathrm{Uniform}[0,1])$,
$\myUp(l_4)=(x,\lambda x.\, 2*x)$,
$\myUp(l_5)=(x,\lambda x.\, 0.5*x)$ and $\myUp(l_6)=(x,\lambda x.\, x)$
 where $\mathrm{Uniform}[0,1]$ is the uniform distribution over $[0,1]$.

\item $\myPr(l_3,l_4)=0.25$, $\myPr(l_3,l_5)=0.75$.

\item $G(l_2,l_3)=(x<1)$ and $G(l_2,l_6)=(x\geq 1)$.  
\end{itemize}
Moreover, let $C=\{(l_6,[x\mapsto r])\mid r\in\mathbb{R}\}\subseteq L\times\mathbb{R}^V$, and let $\mathfrak{C}$ be a linear predicate map representing $C$.

\begin{myremark}\label{rem:noRepSM3}
In fact, no $\varepsilon$-RepSupM can refute almost-sure reachability to $\sem{\mathfrak{C}}$.
Let $\eta:L\times\mathbb{R}^V\to\mathbb{R}$ be an $\varepsilon$-RepSupM for $C$ 
that has $\kappa$-bounded differences for some $\kappa>0$.
Assume $\eta(l_1,0)=A<0$.
Then we have:
\begin{math}
A=\eta(l_1,0)\geq \int_{x\in[0,1]}\eta(l_2,x)+\varepsilon\,.
\end{math}
Hence there exists $x>0$ such that $\eta(l_2,x)\leq A<0$.

Let $N\in\mathbb{N}$ be the minimum number such that $2^N  x\geq 1$.
As $(l_6,2^N  x)\in C$, we have
\begin{math}
\eta(l_2,2^N  x)\geq \eta(l_6,2^N  x)+\varepsilon\geq 0\,,
\end{math}
which means that there exists $n'\in\{0,\ldots,N-1\}$ such that 
$\eta(l_2,2^{n'}  x)-\eta(l_2,2^{n'+1}  x)>0$.

For each $n\in\mathbb{Z}$ such that $n\leq N-1$, we have:
\begin{align*}
\eta(l_2,2^n  x)&\geq \eta(l_3,2^n  x)+\varepsilon \\
&\geq 0.25\cdot  \eta(l_4,2^{n+1}  x)+0.75\cdot  \eta(l_5,2^{n-1}  x)+2 \varepsilon \\
&\geq 0.25\cdot  \eta(l_2,2^{n+1}  x)+0.75\cdot  \eta(l_2,2^{n-1}  x)+3 \varepsilon\,.
\end{align*}
This implies:
\[
\eta(l_2,2^{n-1}  x)-\eta(l_2,2^n  x)\leq \frac{0.25}{0.75} \bigl(\eta(l_2,2^n  x)-\eta(l_2,2^{n+1}  x)\bigr)-\frac{0.25}{0.75}\cdot 3 \varepsilon\,.
\]

By the two discussions above, we can inductively prove $\eta(l_2,2^n x)-\eta(l_2,2^{n+1}  x)\leq -\frac{1}{3}\cdot 3(n-n')\varepsilon$ for each $n\leq n'$.
Hence $\lim_{n\to\infty}f(l_2,2^nx)-f(l_1,0)=-\infty$, and it contradicts to the $\kappa$-bounded condition.

In contrast, 
if we define a linear expression map $\mathfrak{f}$ and a linear predicate map $\mathfrak{I}$ by 
$\mathfrak{f}(l_1)=\frac{1}{2}$, $\mathfrak{f}(l_2)=\mathfrak{f}(l_3)=x$, $\mathfrak{f}(l_4)=2x$, $\mathfrak{f}(l_5)=\frac{1}{2}x$ and
$\mathfrak{I}(l)=\bigl(x\geq 0\bigr)$ for each $l\in L$, then $\mathfrak{f}$ is a linear NNRepSupM for $\mathfrak{C}$ at $1$ supported by $\mathfrak{I}$.
\end{myremark}

\subsubsection{(d-4) (unbounded random walk)}\mbox{}
\begin{itemize}
\item $L_N=\emptyset$, $L_P=\{l_1\}$, $L_D=\emptyset$ and $L_A=\{l_2,l_3\}$.

\item $V=\{x\}$.

\item $\linit=l_1$

\item $\xinit=[x=1]$.

\item $\trrel=\{(l_1,l_2),(l_1,l_3),(l_2,l_1),(l_3,l_1)\}$.

\item $\myUp(l_2)=(x,\lambda x.\,x-1)$ and
$\myUp(l_3)=(x,\lambda x.\, x+1)$.

\item $\myPr(l_1,l_2)=0.25$, $\myPr(l_1,l_2)=0.75$.

\item $G$ is the empty function.
\end{itemize}
We define $C\subseteq L\times \mathbb{R}^V$ by $C:=\{(l_1,[x\mapsto r])\mid r\leq 0\}$, and let $\mathfrak{C}$ be a linear predicate map representing $C$.

\end{document}